\documentclass{aa}

\usepackage[varg]{txfonts}
\usepackage{graphicx}
\usepackage{array, booktabs, makecell}
\usepackage{siunitx}
\usepackage[version=4]{mhchem}
\usepackage{textcomp, gensymb}
\usepackage{tablefootnote}
\usepackage{xcolor}
\usepackage{amsmath}
\usepackage{amsfonts}
\usepackage{comment}
\usepackage{multicol}
\usepackage[utf8]{inputenc}
\usepackage[english]{babel}
\usepackage{ulem}
\usepackage{natbib}
\bibpunct{(}{)}{;}{a}{}{,}
\usepackage[hidelinks,colorlinks=true,linkcolor=blue,citecolor=blue]{hyperref}

\usepackage{orcidlink}

\begin{document}

\title{Mapping the emission and spectral properties of the FRI radio galaxy 3C~449 with LOFAR and the VLA}

\titlerunning{Mapping the emission and spectral properties of 3C~449}
\authorrunning{L.\,Ricci, et al.}

\author{Luca Ricci\orcidlink{0000-0002-4175-3194}\inst{1,2,3},
Luisa Ostorero\orcidlink{0000-0003-3983-5980}\inst{3,4},
Raffaella Morganti\orcidlink{0000-0002-9482-6844}\inst{5,6},
Judith H.\,Croston\orcidlink{0000-0003-2402-9003}\inst{7},
Martin J.\,Hardcastle\orcidlink{0000-0003-4223-1117}\inst{8}, and
Timothy W.\,Shimwell\orcidlink{0000-0001-5648-9069}\inst{5,9}
}

\institute{Julius-Maximilians-Universität Würzburg, Fakultät für Physik und 
Astronomie, Institut für Theoretische Physik und Astrophysik, 
Lehrstuhl für Astronomie, Emil-Fischer-Str. 31, D-97074 Würzburg, Germany
\and
Max-Planck-Institut für Radioastronomie, Auf dem Hügel, 69, D-53121 Bonn, Germany
\and
Dipartimento di Fisica, Università di Torino, Via P.\ Giuria 1, I-10125 Torino, Italy
\and
Istituto Nazionale di Fisica Nucleare (INFN), Sezione di Torino, Via P. Giuria 1, I-10125 Torino, Italy
\and
ASTRON, the Netherlands Institute for Radio Astronomy, Oude Hoogeveensedijk 4, 7991 PD, Dwingeloo, the Netherlands
\and
Kapteyn Astronomical Institute, University of Groningen, Postbus 800, 9700 AV Groningen, The Netherlands
\and
School of Physical Sciences, Open University, Walton Hall, MK7 6AA, UK
\and
Centre for Astrophysics Research, University of Hertfordshire, College Lane, Hatfield AL10 9AB, UK
\and
Leiden Observatory, Leiden University, PO Box 9513, 2300 RA Leiden, The Netherlands
}

\date{Received / Accepted}

\abstract{
The jets and lobes of nearby, extended radio galaxies are ideal laboratories to explore the spectral and dynamical evolution of the radio emitting plasma that emanates from active galactic nuclei and propagates through the ambient medium.
Here, we present a deep, high-resolution radio continuum study of the low-redshift ($z=0.01713$), Fanaroff-Riley class I (FRI) radio galaxy 3C~449 performed with a combination of radio data at 145 MHz acquired with the  LOw Frequency ARray (LOFAR) and archival Very Large Array (VLA) data at 1365, 1485, 4985, and 8485 MHz. Our LOFAR maps of the source have angular resolution 20\arcsec$\times$ 20\arcsec~ (7.2 kpc $\times$ 7.2 kpc) and 6.0\arcsec$\times$ 6.0\arcsec~ (2.2 kpc $\times$ 2.2 kpc), and show the full extent of the known radio emission  ($\approx 22\arcmin$, i.e. $\approx 480$ kpc) at the highest angular resolution to date.
Our spectral index maps show, for the first time, the high-resolution distribution of the spectrum in the 145-8485 MHz frequency range over a source region that extends beyond $2.5\arcmin$ (54 kpc). The average 145-8485 MHz source spectrum is consistent with a single power law and stays approximately constant over the inner $\approx 50\arcsec$ ($\approx 18$ kpc) of both source jets. Beyond $\approx 50\arcsec$, both on the northern and on the southern source sides, the higher-frequency spectrum steepens and the spectral break frequency lowers with increasing distance from the radio core, indicating the absence of relevant sites of particle acceleration beyond those distances.
In our $145-1365$ MHz spectral index map, we detect a flatter spectrum spine surrounded by a steeper spectrum sheath in the inner $\approx 2.5\arcmin$ (54 kpc) and  $\approx 3\arcmin$ (65 kpc) of the southern and northern jet, respectively; beyond $\approx 1\arcmin$, the spine-sheath structure is also detected in the 1365-8485 MHz frequency range, confirming previous findings.
The steep spectrum sheath may be the signature of interaction between the jet and the ambient medium.
By modelling the spectral index maps under the assumption of equipartition and of a constant magnetic field across the source, we derive maps of the highest radiative age of the particles all over the source in a standard ageing scenario. From the oldest radiating plasma, located in the southern radio lobe, we estimate the source spectral age as $\tau\simeq 200$~Myr; at the outer edges of both the northern tail and southern lobe, we estimate a spectral age $\tau_{\rm sp} \simeq 150$ Myr. 
%{\color{red} [Aggiungere risultato di ``robust upper limit on the spectral age throughout the source'']?}
If the latter age were representative of the dynamical source age, the average expansion speed of both jets during the source lifetime would be supersonic, with $M\simeq 4.1$ and $M\simeq 2.8$ for the northern and southern jet, respectively. 
Because numerical magneto-hydrodynamical simulations of FRI jets suggest that the source's current expansion is subsonic, the high average Mach numbers might arise either from the source being highly supersonic for a small fraction of its lifetime or 
from a severe underestimation of the spectral age due either to particle acceleration on scales of hundreds of kpc, not detected in our observations, or to the presence of a non-homogeneous magnetic field with complex structure.}

\keywords{galaxies: individual: 3C 449 (B2 2229$+$39) – galaxies: active – galaxies: jets - radio continuum: galaxies}
\maketitle

%------------
\section{Introduction} 
%------------
Radio galaxies extending on spatial scales of a few hundred kpc or greater are thought to sample the evolved radio galaxy population.
Based on their radio morphology, they can broadly be grouped into \cite{Faranoff_Riley_1974} class I (FRI) and class II (FRII) objects, even though additional classifications were introduced to account for hybrid morphology radio sources and sources without a straight radio structure 
%discovered in the last decades 
\citep[see, e.g.,][for a recent review]{Hardcastle_2020}.

FRII sources are known as ``edge brightened sources'': their jets are thought to be highly relativistic out to large distances from the radio core. 
When detected, FRII jets mostly show a one-sided or two-sided,  highly asymmetric structure, which are found to terminate in a hot spot at their edges, which is the primary site of particle acceleration. 
Terminal hot spots can be detected on both sides even when only the approaching jet is visible \citep[e.g.,][]{Mullin_2006}.
Conversely, FRI sources are ``edge darkened''  (or ``centre brightened''): they possess twin jets that dominate the surface brightness, and diffuse plumes; the jets are thought to be launched with highly relativistic speeds; however, on kpc scales, they experience a deceleration to non-relativistic speeds \citep{Laing_2014}, which may be the primary cause of particle acceleration \citep{Laing_2013}, directly observed in the inner parts of the jets through X-ray and optical synchrotron radiation.

The FRI/FRII dichotomy may be linked to jet dynamics. Originally, the FRI/FRII morphological divide was found to correspond to a radio luminosity break, $L_{\rm 150 MHz}\simeq  10^{26}~$W~Hz$^{-1}$ \citep{Faranoff_Riley_1974}, 
that appeared to increase with the luminosity of the host elliptical galaxy \citep{Owen_1994,Ledlow_1996}. 
This suggested that, for a given radio luminosity, the distinction between FRI and FRII morphology may be a consequence of both the luminosity and the core density of galaxies, with the FRI morphology occurring in the denser cores of more optically luminous hosts \citep[e.g.,][]{DeYoung_1993,Bicknell_1995,Mingo_2019}. 
As a result, FRI and FRII sources would clearly be different in the locations of particle acceleration and the particle contents on large scales \citep[e.g.,][]{Croston_2018}. 
Even though recent studies enabled to show that the radio luminosities of FRI's and FRII's span a wide range and display a significant overlap, it remains true that FRII's are on average more luminous than FRI's \citep[e.g.,][and references therein]{Clews_2025}. In addition, despite the relation between radio luminosity and host luminosity \citep{Ledlow_1996} disappearing when selection effects are properly accounted for, radio galaxies near the FR luminosity divide preferentially show FRI morphology if they are hosted by more massive galaxies, supporting the evidence that the inner environment plays a role in determining jet disruption \citep{Clews_2025}.

A possible additional, extrinsic cause of the FRI/FRII dichotomy is the large scale environment of the host galaxies, with the inter-galactic medium (IGM) density being the differentiating factor in the resistance experienced by the jets \citep[e.g.,][]{Prestage_1988}. Even though there is a significant overlap in environment between the two classes,  FRI sources are found, on average, in higher density environments than FRII sources \citep[e.g.,][]{Zirbel_1997,Gendre_2013,Croston_2019}.
Finally, the role of the accretion mode, in turn linked to the large scale environment,  as a supplementary, intrinsic cause in shaping the jet morphology,  is still under debate \citep[e.g.,][]{Best_Heckman_2012,Gendre_2013,Mingo_2014,Ineson_2015, Tadhunter_2016,Hardcastle_2007,Czerny_2016,Hardcastle_2018}.

Despite the absence of a unique factor separating the FRI and FRII sources, the above evidence has long suggested that the characteristic features of FRI sources could be mostly ascribed to deceleration and disruption of their (typically lower-power) jets by a denser surrounding medium, as originally proposed by \citet{Simon_1978}.
These processes may be triggered by mass entrainment via a mixing, turbulent layer at the jet surface \citep[][]{Bicknell_1984,Bicknell_1986,DeYoung_1993,Laing_2002, Wang_2009}, promoted by various types of jet instabilities \citep[e.g.,][]{Perucho_2005, Perucho_2007, Meliani_2007, Meliani_2009, Matsumoto_2013, Millas_2017, Tchekhovskoy_2016, Rossi_2020, Rossi_2024}, by mass loading from stellar winds within the jet volume \citep[][]{Komissarov_1994}, or by stars penetrating and exiting the jets and triggering instabilities that favour mixing \citep{Perucho_2020}.  The efficiency of all these mechanisms is higher on galaxy scales.

As a consequence of the mass entrainment on galaxy scale, FRI jets develop a boundary layer (sheath) that transfers momentum to the ambient gas and propagates at lower speed than the central spine \citep[e.g.,][]{Bicknell_1984,Bicknell_1986,Rosen_2000,Laing_Bridle_2002}. This spine-sheath structure may persist on larger scales, where the jet interacts with the ICM \citep[e.g.,][]{Loken_1995,Massaglia_2019}.
%, as seen, e.g., by \citet{Lal_2020} in NGC~4869.

The role of the jet magnetic field may also be important for shaping the radio morphology, after the deceleration phase: three-dimensional, magneto-hydrodynamical (3D MHD) numerical simulations of the propagation of plasma jets in a stratified ambient medium 
\citep[e.g.,][]{Massaglia_2019, Massaglia_2022} show that, while low- (high-)power jets lead to FRI (FRII) morphology radio galaxies, intermediate-power jets preferentially lead to FRI morphologies if they are highly magnetised.

Both the magnetic field distribution and the source substructures created by the jet-ambient interaction are expected to determine the surface brightness and spectral index distribution of a radio source.
Indeed, in radio sources, energetic electrons are expected to be accelerated mostly by strong shocks inside the jet and its cocoon via diffusive shock acceleration -  or the Fermi first order process 
 \citep[][]{Blandford_1978, Drury_1983,Blandford_1987},  although  other  microphysical processes, such as 
 Fermi second order acceleration and reconnection at shear layers  \citep[][]{Rieger_2007a, Rieger_2007b,Sironi_2021}, may be at work. 
After an acceleration event, the electrons are expected to suffer cooling due to synchrotron and inverse-Compton losses.
Therefore, testing dynamical models of FRI sources against observations requires the availability of reliable simulations of the surface brightness distribution and corresponding high resolution radio observations at different frequencies. 
In MHD simulations, the surface brightness distribution is often inferred from the simulated density distribution and the pressure behaviour of the jets \citep[e.g.,][]{Rosen_2000,Massaglia_2019,Massaglia_2022}.
On the other hand, semi-analytical calculations can predict the evolution of the energy spectrum of non-thermal electron populations moving in a fluid flow \citep{Kardashev_1962, Jaffe_Perola_1973, Murgia_1999, Hardcastle_2013}, however with simplifying assumptions.
A more complete and self-consistent approach requires following the evolution of the population of relativistic radiating particles in 3D relativistic MHD (RMHD) numerical simulations of the source dynamics \citep[][]{Fromm_2016, Turner_2018, Vaidya_2018, Davelaar2020, Mukherjee_2021}. Even though these simulations are still computationally very expensive, the first results on jets unstable to MHD instabilities, as may be the low-power jets of FRI's, show that the complex shock structures created by the jet-ambient interaction on scales of $\lesssim 10$ kpc are indeed capable of accelerating particles \citep{Mukherjee_2021}. These results are encouraging in showing the possibility to identify the main sites and mechanisms of particle acceleration, and, in the long run, to simultaneously follow the dynamical and spectral evolution of the full radio source.
%}

The LOw Frequency ARray (LOFAR) Telescope \citep{van_Haarlem_2013} is the ideal instrument to perform studies of the surface brightness distribution of extended radio galaxies of the FRI type. With its ability to observe the sky at low frequencies, high sensitivity, and high angular resolution, LOFAR allows to image the low frequency emission of these sources in great detail.
The combination of LOFAR data with data at higher radio frequencies such as, e.g., those acquired with the Very Large Array (VLA), offers the unique opportunity to map the distribution of the spectral properties of these sources. 
In this context, we carried out a study of the radio continuum emission of the FR I radio galaxy 3C~449 (B2~2229+39).
Thanks to the combination of its vicinity, luminosity, as well as apparent and intrinsic large size, this source is one of the best targets to investigate the distribution of the surface brightness and spectral index on scales from few kpc up to hundreds of kpc.

\subsection{Target characteristics and aim of the work}
3C~449 is an FRI radio galaxy \citep{Faranoff_Riley_1974,Perley_1979,Pearson_1988}. It is hosted by the nearby \citep[$z = 0.01708$;][]{De_Vauc_1991} elliptical galaxy UGC~12064, which is the most prominent member, classified as cD by \citet{Wyndham_1966}, of the poor galaxy cluster Zw~2231.2+3732. 
%and has a bright , elliptical companion $37\arcsec$ (13 kpc) in projection to the north, whose halo is connected to the halo of UGC~12064 by a low surface brightness bridge \citep{Martel_1999}.
UGC~12064 has a bright, elliptical companion $37\arcsec$ (13 kpc) in projection to the north; the haloes of the two galaxies are connected by a low surface brightness bridge  \citep{Martel_1999}.
%{\bf Even though \cite{capetti1994} detected, in Hubble Space Telescope images of the source, two blue knots and proposed them to be a possible optical synchrotron counterpart of the radio jet, subsequent HST observations by \citep[][]{Martel_1999} did not confirm 3C\,449 as a source with optical jets. 
The source is not among those with confirmed synchrotron optical jets, acccording to \citet{Martel_1999} \citep[see, however,][]{Capetti1994}. 

From VLA observations at 1.4 GHz, the source appears to have a total extent of $\approx 20\arcmin$ ($\approx 430$ kpc): two opposite jets start from an unresolved core, and appear to be fairly symmetrical on scales of $\approx 1\arcmin$ ($\approx 21.6$ kpc); they then merge into the inner lobes, and subsequently follow a structure that resembles a helix in projection \citep{Perley_1979,Feretti_1999}.
From dynamical studies, the jets were found to be almost aligned with the plane of the sky, with an angle to the line of sight of $\theta >75 \degree$ \citep{Feretti_1999}.
The radio source expands within a distribution of X-ray emitting gas 
whose surface brightness peak corresponds to the position of the radio core \citep{Hardcastle_1998,Croston_2003,Lal_2013}. 
The high galaxy number density and X-ray gas density that characterize the environment of 3C~449 are consistent with its classification as an FRI radio galaxy.
The hot gas distribution that embeds the source is not spherically symmetric, with the large-scale radio emission that avoids the regions of higher X-ray surface brightness. A tunnel-like feature and an X-ray cavity are detected in correspondence of the southern radio jet and inner lobe, respectively  \citep{Hardcastle_1998,Croston_2003,Sun_2009,Lal_2013}.
Even though the orbital motion of the two companion galaxies was initially proposed to be responsible for the wiggles of the jets \citep{Lupton_1982,Hardee_1994}, the fact that both of the inner jets are bent to the west as they enter their respective lobes led \citet{Lal_2013} to suggest that the ambient gas is rather pushing them westward;
moreover, the inner jets flare at approximately the position of a ``sloshing'' gas cold front ascribed to a group merger occurred $\lesssim 1.3-1.6$ Gyr ago, suggesting that the jet is entraining and thermalizing some of the hot gas as it crosses the front, as predicted by \citet{Loken_1995}.

The radio spectral properties of 3C~449 have been extensively studied by means of both single-dish and interferometric observations.
Early, moderate angular resolution (RA$\times$DEC$=29\arcsec \times 46\arcsec$) interferometric observations with the Westerbork Synthesis Radio Telescope at 608 MHz and 1400 MHz enabled  \cite{Jaegers_1987} to show that the 600-1400 MHz %spectral index progressively increases (and the spectrum steepens) 
spectrum progressively steepens with core distance out to $\sim 9'$ in the north and out to $\sim 6'$ in the south; a spectral flattening appears beyond $\sim 6'$ in the southern lobe.
With single-dish, low angular resolution ($2.5'$ and $4.5'$) observations acquired with both the Bologna Northern Cross and Effelsberg radio telescopes at 400, 2700 and 4750 MHz, and 10.7 and 32 GHz, \cite{Andernach_1992} showed that the spectrum progressively steepens with core distance out to $>9'$ in each pair of frequencies in the 400 MHz$-$10.7 GHz range, and out to  $\sim 5'$ at 10.7$-$32 GHz; some spectral flattening is observed at the southernmost edge of the source in the 400-2700 MHz range, consistently with the findings of \cite{Jaegers_1987}. 
%From early interferometric observations at $\sim 4\arcsec$ angular resolution performed with the VLA at 1465 and 4885 MHz, \cite{Perley_1979} showed that the spectral index slightly varies along the inner $\sim 1'$ jets, and displays some steepening at the end of each jet, with the northern jet being steeper than the southern one.
High resolution interferometric observations were performed with the VLA by \cite{Perley_1979} at 1465$-$4885 MHz with $\sim 4\arcsec$ resolution, by \cite{Katz-Stone_Rudnick_1997}  at 330, 1445, and 4835 MHz with $\sim 3.6\arcsec$ resolution, and by \cite{Feretti_1999} at 4985 and 8385 MHz with $1.25\arcsec-5\arcsec$ resolution.
The investigations by \cite{Katz-Stone_Rudnick_1997} and \cite{Feretti_1999} showed that the radio spectrum is roughly constant, on both source sides, out to $\approx 1\arcmin$ from the core, confirming earlier findings by \cite{Perley_1979}. Beyond $1\arcmin$, the overall spectrum steepens with increasing distance from the core out to $\approx 2.5\arcmin$, with an asymmetric behaviour on the two source sides, with the southern, inner lobe that appears to be steeper than the northern, inner lobe.
While \citet{Katz-Stone_Rudnick_1997} found evidence, on both source sides, for a ``flat jet'' whose spectrum remains roughly constant with the core distance, and a ``sheath'' that appears beyond $1\arcmin$ from the core and is responsible for the steepening of the overall spectrum with increasing core distance and for most of the observed source widening, \citet{Feretti_1999} did not distinguish between flat jet and sheath for the northern part of the source, and only confirmed the separation between a flat jet and a surrounding, steeper region in the southern part of the source.
The southern, steeper spectrum radio emitting sheath was more recently associated to the X-ray sheath around the X-ray tunnel by \citet{Lal_2013}.
Both radio spectral studies by \citet{Katz-Stone_Rudnick_1997} and \citet{Feretti_1999} are limited to within $\approx 2.5\arcmin$ ($\approx 54$~kpc) from the core, namely to the source region within the inner lobes.

%\textbf{\citet{Capetti1994}, using Hubble Space Telescope (HST) observations, reported two blue knots possibly associated with the optical synchrotron counterpart to the radio jets.
%Nonetheless, subsequent HST observations by \citep[][]{Martel_1999} did not confirm 3C\,449 as a source with optical jets.}

In this work, we present new LOFAR observations of 3C~449 at 145 MHz, and we combine them with reanalysed archival VLA observations at 1365, 1485, 4985, and 8485 MHz, with the aim of obtaining, for the first time, high-resolution spectral index maps of the source that extend well beyond $2.5\arcmin$. 
\footnote{In this work, we did not use the 330 MHz VLA observations, due to their poor quality and therefore low S/N of the extended emission.}
%These maps are instrumental 
We used these maps to look for signatures of particle acceleration processes occurring in the jets and lobes, to explore the interaction between expanding jet and ambient medium, to estimate the source spectral age and constrain the dynamics of the source expansion.  
The paper is organized as follows. 
In Sect.~\ref{sec:data}, we describe the LOFAR-HBA observations and the data reduction procedure, as well as the reanalysis of the VLA archival data. 
In Sect.~\ref{sec:maps}, we present the source morphology as seen by LOFAR (Sect.~\ref{sec:morphology}), the maps of radio spectral index obtained from the combination of LOFAR and VLA maps, and the profiles of the spectral index along and across the radio structures (Sect.~\ref{sec:spec_index}).
In Sect.~\ref{sec:spectral_age}, we perform the spectral 
analysis: by modelling the combination of the surface brightness images of the source at different frequencies, we map the spectral age of the radio emitting particles and infer the spectral age of the source. 
In Sect.~\ref{sec:discussion} we discuss our results in terms of constraints to the source dynamics (Sect.~\ref{sec:mach_number}) and of particle populations and acceleration mechanisms (Sect.~\ref{sec:particle_population}). We conclude in Sect.~\ref{sec:conclusions}.

We adopt J2000.0 as the equinox of coordinates. 
We adopt a cosmological model that assumes a flat Universe with $\Omega_{\rm M} = 0.308$ and $H_0 = 67.8$ km s$^{-1}$ Mpc$^{-1}$ \citep{Planck_2016}. In this model,
%$\Omega_{\rm Lambda} = 0.692$
$1 \arcsec$ corresponds to 0.360 kpc at the source redshift, $z=0.01713$.

%------------
\section{Observations and data reduction} \label{sec:data}
%------------

%------------
 \subsection{LOFAR HBA data} \label{sec:lofar_data}
%------------

3C~449 was observed twice with the LOFAR high band antenna (HBA). %between 2015 and 2020. 
The first observation was 
%carried out during Cycle 4, in 
a dedicated ten-hour run on August 16, 2015 (project LC4\_028; observation ID L368474; PI: V. Heesen), with the LOFAR telescope operating in the HBA\_DUAL\_INNER configuration, resulting in a field of view (FOV) of approximately $8\degree$  with baseline lengths of up to 85 km. 
The quality of the image obtained with this observation was not good enough to perform a spectral analysis, due to severe artifacts remaining after self-calibration; therefore, we will not discuss it further in this work.

The source was then observed as part of the ongoing LOFAR Two Meter Sky Survey \citep[LoTSS;][]{Shimwell_2017, Shimwell_2019,Shimwell2022} project and we have used the pointing P337+38 observed on March 07, 2020.
In this pointing, 3C~449 is located 1.10 degrees from the field center. The observations were carried out with the standard LoTSS survey setup \citep{Shimwell_2019,Shimwell2022}, i.e., 8 h on-source observations bookended on either side with a flux density calibrator, in this case 3C~295. 
The observation used 51 antennas (24 core, 14 remote, 13 international).
The observations were recorded with an integration time of 1~s, a 48~MHz bandwidth centred at 144.6~MHz and a channel width of 3.05~kHz. 
The details of this observation are reported in Table \ref{tab:LOFAR_obs}. 
The data from the international stations, although recorded, are not used in the present paper. 
Full details of the observational setup and processing of these data will be given as part of the forthcoming LoTSS DR3 release (Shimwell et al in prep.).

The data were passed through the standard LOFAR pre-processing pipeline \citep{Heald_2010} which performed the flagging of the radio frequency interference (RFI) using the AOflagger \citep{Offringa_2012} and averaged down to a channel width of 12.2~kHz. Direction independent calibration was then performed using the PREFACTOR1 pipeline \citep{de_Gasperin_2019,Van_Weeren_2016,Williams_2016}.

The output datasets were then processed with DDF pipeline\footnote{\url{https://github.com/mhardcastle/ddf-pipeline}}, which performs direction dependent (DD) calibration and imaging.  
This pipeline makes use of kMS and DDFacet for calibration and imaging respectively. The processing was performed as part of the standard LoTSS-DR3 processing and uses v3.0 of DDF-pipeline. DDF pipeline is described in detail in \cite{Tasse_2021} and \citet{Shimwell2022}.
%
%The direction dependent self-calibration of the $6\arcsec$ and $20\arcsec$ resolution Dutch array images were made through the standard LoTSS processing infrastructure. This uses the latest version DDF-pipeline (v3.0)\footnote{https://github.com/mhardcastle/ddf-pipeline} which is described in \citet{Tasse_2021} and \citet{Shimwell2022}.

For the analysis presented in this paper, we used images of the field at both $6\arcsec \times 6\arcsec$ and $20.0\arcsec \times 20.0\arcsec$  resolution to trace the faint large-scale emission, while also being able to map the details in the large-scale structure. 
The images at both resolutions are shown in Fig.~\ref{fig:LOFAR_maps}. 
Their characteristics are listed in Table \ref{tab:maps}. 

The flux density scale by \citet{Scaife_Heald_2012} is used for the 
%amplitude calibration, 
calibration of the calibrator source in PREFACTOR1 and TGSS-ADR1 sky models of the target fields are used for an initial phase calibration.  
However, density scale errors arise when transferring the calibration solutions to the target field (possibly related to uncertainties in the LOFAR beam model and the differing elevations of the target and calibrator fields); for this reason, in LoTSS, the flux density scale of the target field is further refined during processing. This is done by deriving a scaling factor to match the flux density scale with that of NVSS \citep{Hardcastle2016}. For this field, a flux scaling factor of 1.161 has been applied.
In addition, the error associated to the measured flux densities is dominated by the flux calibration uncertainty, that is typically assumed to be between 10 and 20\% \citep{Shimwell2022}, and that we evaluated to be $\simeq 14\%$ (see Sect.~\ref{sec:flux_scale} for details).

\begin{table}
\caption{Summary of LOFAR HBA observations.} 
\label{tab:LOFAR_obs}      
\begin{tabular}{lc}         
\hline\hline                        
   LoTSS project code & P337+38  \\       
   Central frequency & 144.627  MHz \\
   Bandwidth  &  48 MHz (120-168 MHz) \\
   %Sample integration time & 1 s \\
   Time on source & 8 h \\    
   Flux calibrator & 3C~295 \\
   %Secondary calibrator & ... \\
   Observation date & 2020-03-07\\
   \hline                                             
\end{tabular}
\end{table}

%-----------
\subsection{VLA data} \label{sec:vla_data}
%------------
%In this work, we also 
We retrieved archival Very Large Array (VLA) data of 3C~449 acquired
by \cite{Katz-Stone_Rudnick_1997} at 1365 MHz, 1485 MHz, 4895 MHz and 8485 MHz (i.e., in the L, C, and X bands), and by \cite{Feretti_1999} at 8485 MHz (i.e.,  in the X band).
The details of the VLA data sets are reported in Table \ref{tab:VLA_data}. 
We reprocessed the data using the Common Astronomy Software Applications (CASA) package.
%The procedure we performed was the standard approach ...
After the online flagging, the data were manually flagged and calibrated using the flux scale of \cite{Perley_Butler_2013}, which is consistent with the scale of \cite{Scaife_Heald_2012} at low frequencies (30-300 MHz).
Phase self-calibration runs were performed for each map. 
A single amplitude self-calibration run was performed when needed.
The primary beam correction was applied to all the VLA images.
The flux calibrators were 3C~286 and 3C~48, which are unresolved at the resolution of our observations. 
The phase calibrator was the nearby source 2200+420.

\begin{table*}[h]
\caption{Summary of the characteristics of the LOFAR and VLA maps obtained and re-obtained, respectively, in this work. }
\label{tab:maps}      
\centering                                    
\begin{tabular}{c c c c c c}          
\hline \hline      
Telescope & Frequency & Configuration & Calibrators & Angular resolution & rms noise \\
& ($\mathrm{MHz}$) & (VLA) & (flux, phase) & (arcsec) & (mJy beam$^{-1}$) \\
\hline %\hline
LOFAR & 145 & // & ... &$6.0\arcsec \times 6.0\arcsec$ &  0.142 \\
LOFAR & 145 & // & ... &$20.0\arcsec \times 20.0\arcsec$ & 0.214 \\
VLA & 1365 & B,C,D & 3C 286, 2200+420 & $6.0\arcsec \times 6.0\arcsec$ & 0.038 \\
VLA & 1365 & C,D  & 3C 286, 2200+420 & $20.0\arcsec \times 20.0\arcsec$ & 0.030 \\
VLA & 1485 & B,D  & 3C 286, 2200+420 & $20.0\arcsec \times 20.0\arcsec$ & 0.054 \\
VLA & 4985 & C,D  & 3C 286, 2200+420 &  $6.0\arcsec \times 6.0\arcsec$ & 0.016 \\
%VLA & 4985 & $20.0\arcsec \times 20.0\arcsec$ & 3C 286, 2200+420 &  0.08 \\
VLA & 8485 &  C,D  & 3C 286, 2200+420 & $6.0\arcsec \times 6.0\arcsec$ & 0.018\\
\hline %\hline    ctr. A&A template
\end{tabular}
\begin{flushleft} 
\textbf{Notes.} Column 1: array used for the observation; Column 2: frequency in MHz of the observation; Column 3: VLA configuration used; Column 4: flux and phase calibrators used for the data reduction; Column 5: angular resolution of the final image in arcsec; Column 6: thermal noise of the final image in mJy $\mathrm{beam}^{-1}$.
\end{flushleft}
\end{table*}

\begin{table*}[h]
\caption{Summary of the VLA data reprocessed in this work.}
\label{tab:VLA_data}    
\centering              
\begin{tabular}{lcc}    
\hline\hline
    Observation ID         & AK0319   & AF0241 \\
%    Proposal PI            & R.\ A.\  Perley & L.\ Feretti\\
    Bands                  &  L, C, X       &  X\\
    Frequency (MHz)        & a) 1365 & 8485   \\
                           & b) 1485 &            \\
                           & c) 4985 &            \\
                           & d) 8485 &            \\
    VLA configurations     & B, C, D     & C       \\
    Observation time (h)   & a) 5.2 (B), 1.0 (C) 0.6 (D)  & 5.5 (C) \\
                           & b) 5.2 (B), 0.6 (D)   &        \\
                           & c) 8.7 (C), 1.8 (D)   &            \\
                           & d) 1.8 (D)   &            \\
    Primary calibrators    & 3C~286   & 3C 286           \\
                           & 3C 48    &         \\
    Secondary calibrator   & 2200+420 & 2200+420           \\
    Observation dates      &  1993/04 -- 1994/01 & 1993/07 \\ 
\hline
\end{tabular}
\end{table*}

The combination of VLA and LOFAR maps to study the spatial distribution of the source radio spectral index requires the same angular resolution for the combined maps.
In order to obtain VLA maps that match the angular resolution of the LOFAR maps at 145 MHz (i.e., $6.0\arcsec \times 6.0\arcsec$ and $20\arcsec \times 20\arcsec$) as much as possible without applying a heavy smoothing of the beam, we proceeded as follows.
We first combined data acquired in different VLA configurations (see Table~\ref{tab:VLA_data}, entry no.~4); subsequently, to achieve exactly the same angular resolutions of the LOFAR images, we restored each pair of maps to have the same $(u,v)$-coverage\footnote{We also produced the spectral index maps without restricting them to the same $(u,v)$-coverage and no important differences were seen.} and then restored the VLA images using the {\tt restoringbeam} command during the imaging process. 
As the short baselines are crucial to recover the extended emission, for each pair of images the $(u,v)$-range has been cut to the shortest common baseline, in order to produce accurate spectral index maps.
Our procedure is similar to the procedure followed by \citet{Heesen_2018} to combine LOFAR and VLA data of 3C~31.
Our reanalysed VLA maps are displayed in Appendix \ref{app:vlamaps}.

\subsection{General map properties} \label{sec:map_properties}

A summary of the characteristics of the LOFAR and VLA maps is reported in Table~\ref{tab:maps},  which lists the observation frequencies, the array configuration used for the VLA, the angular resolution of the maps, the calibrators, and the map rms noise level.

At high resolution (i.e., $6.0\arcsec \times 6.0\arcsec$), the quality of our reprocessed VLA images is comparable to the quality of the original images presented in \citet{Katz-Stone_Rudnick_1997} and in \citet{Feretti_1999}. For example, \citet{Feretti_1999} achieve rms noise levels, $\sigma_{\rm rms}$, of 0.035 mJy beam$^{-1}$ at 1365  MHz, 0.018 mJy beam$^{-1}$ at 4985 MHz, and 0.011 mJy beam$^{-1}$ at 8485 MHz, comparable to our rms noise levels, listed in Table~\ref{tab:maps}.
Furthermore, \citet{Feretti_1999} estimate core flux densities of 37.0 mJy at 5 GHz and 45.0 mJy at 8.4 GHz, both with observations in C configuration, which are consistent with our core flux densities of $38.0 \pm 0.8$ mJy $44.0 \pm 0.9$ mJy, respectively. 
\citet{Katz-Stone_Rudnick_1997} report a surface brightness of 18.42 mJy beam$^{-1}$ for the core region at 20 cm, comparable to our estimate of 23.45 mJy beam$^{-1}$. 
The slightly higher core surface brightness in our VLA images is likely due to the fact that our beam ($6.0\arcsec \times 6.0\arcsec$) is larger than theirs ($3.6\arcsec \times 3.6\arcsec$).
At low resolution (i.e., $20\arcsec \times 20\arcsec$), the quality of our maps in the L band (i.e., at 1365 and 1485 MHz) is comparable to their higher angular resolution counterparts.

%------------
\subsection{Flux scale and uncertainties} \label{sec:flux_scale} 
%------------

Studying the spectral properties of the radio emission requires an accurate flux density scale.
In LoTSS-DR2, the systematic overall flux density scale error is less than 10\%, but there is a further up to 10\% random error; both these errors vary across the sky. 
Assuming these errors are independent, we added them in quadrature to give an error of
%for the total error on the flux density scale of a given source, this overall scaling uncertainty should be added to the 10\% error due to positional variations \citep{Shimwell2022}.yes 
%The combination of these uncertainties yields an upper limit of 
$\simeq 14\%$ to the flux density scale error. 

To further check the accuracy of the flux density scale of our LOFAR image of 3C~449, we assembled the radio spectrum of 3C 449 by combining the integrated flux densities of the source that we derived from the LOFAR and VLA maps analysed in this work with the flux densities taken from the literature. In our images, the integrated flux densities were measured by using $3\sigma_{\rm rms}$ contours as reference.

The total uncertainties on the LOFAR and VLA flux densities derived in this work were computed by combining in quadrature the rms noise, averaged over an area of 10 beams, and the flux density scale error.
%\footnote{To convert the  measured surface brightness noise in Jy beam$^{-1}$ to a flux density uncertainty  in Jy, the surface brightness noise has to be multiplied by the beam size, $\pi \ \theta_{\mathrm{max}}  \theta_{\mathrm{min}} / {(4 \ \mathrm{ln}2)}$, expressed in units of sr.} averaged over an area of 10 beams, and the flux density scale error.
The 10-beam areas considered for the rms noise were outside the radio source and did not show any artifact.
We considered the uncertainty on the flux density scale to be the above-mentioned $14\%$ for the LOFAR maps \citep{Shimwell2022} and $2\%$ for the VLA maps \citep{Perley_Butler_2013}. For both LOFAR and VLA observations, the flux density scale uncertainty is the dominant error. 
The rms noise uncertainty is always $\sigma_{\rm rms} \sim 10^{-4} \, \mathrm{Jy}$.
%, and can thus be neglected.

\begin{figure}
 \resizebox{\hsize}{!}{\includegraphics{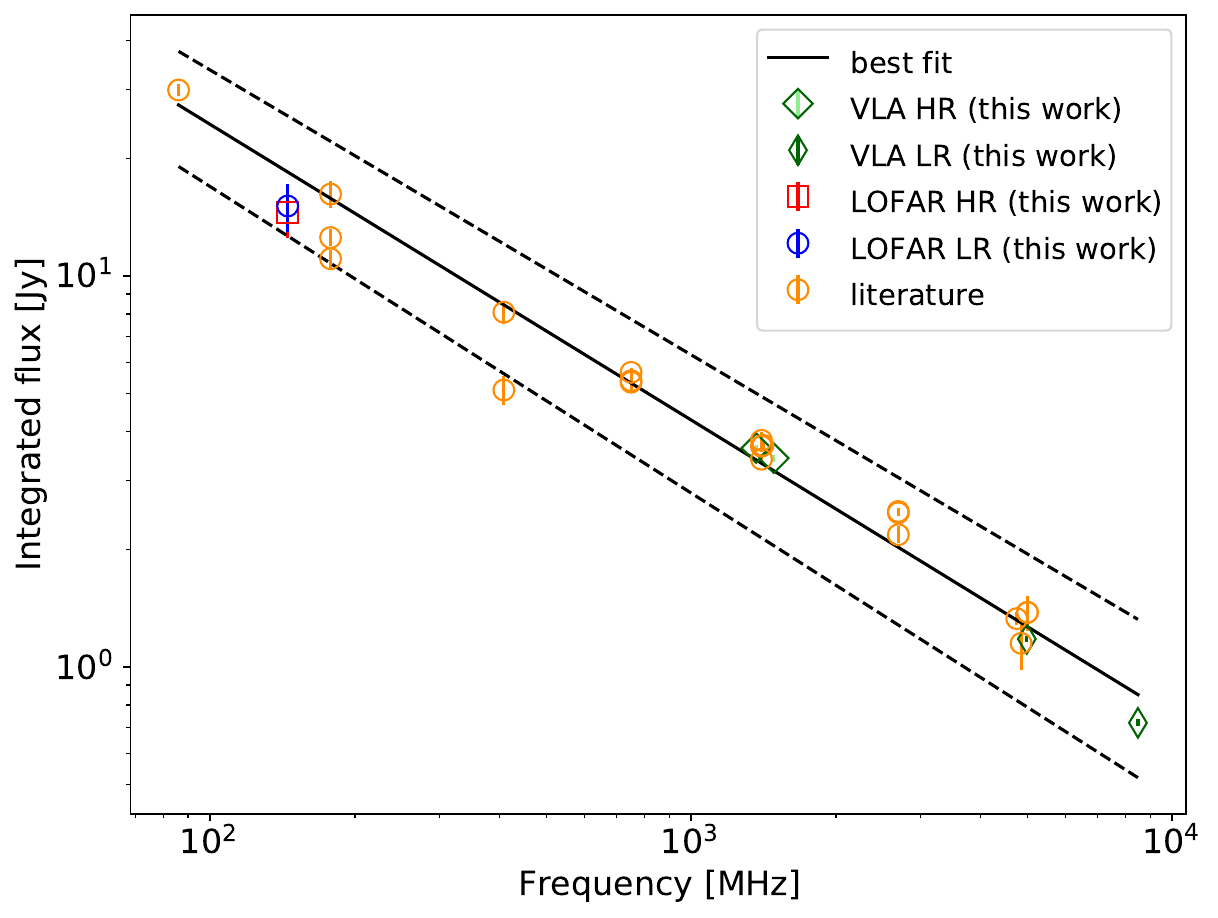}}
  \caption{Integrated radio spectrum of 3C 449 in the frequency range $86-8485$ MHz. Symbols are as follows. Red square: 145 MHz flux density derived from the LOFAR map at $6.0\arcsec \times 6.0\arcsec$; blue circle: 145 MHz flux density derived from the LOFAR map at $20.0\arcsec \times 20.0\arcsec$; small green diamonds: flux densities at 1365 and 1485 MHz derived from the VLA maps at $20.0\arcsec \times 20.0\arcsec$; large green diamonds: flux densities at 4985 and 8485 MHz, derived from the VLA map at $6.0\arcsec \times 6.0\arcsec$; orange circles: data from the literature. More details on the data are listed in Appendix \ref{app:fluxes}.
  The black, solid line represents the best power-law fit to the data. 
  The black, dotted lines enclose the range of variation of the best-fit line according to the uncertainties in the best-fit parameters.}
  \label{fig:integrated_spectrum}
\end{figure}

Fig.~\ref{fig:integrated_spectrum} shows the integrated radio spectrum of 3C 449 in the frequency range $86-8485$ MHz.
The values of the integrated flux densities and their uncertainties are reported in Appendix \ref{app:fluxes}.
Assuming that the radio spectrum is described by a power law of the form $S_\nu = K \nu^{-\alpha}$, we fit a straight line, that takes the form 
$\log S_{\nu} = \log K - \alpha \log \nu $, to the logarithmic spectrum, without including the LOFAR points in the analysis.
The regression line, represented with a solid, black line in Fig.~\ref{fig:integrated_spectrum},  
has the following parameters: $\alpha = 0.76 \pm 0.03$ and $\log K = 795 \pm 169$. 
The black, dotted lines in Fig.~\ref{fig:integrated_spectrum} enclose the range of variation of the best fit according to the 1$\sigma$ uncertainties on the best-fit parameters. The integrated flux densities at 145 MHz derived from the LOFAR maps are consistent with the flux density at 145 MHz expected for a single power-law spectrum at the 1$\sigma$ level.

%------------
\section{Morphology and spectral index distribution} \label{sec:maps}
%------------

%------------
\subsection{Morphology}\label{sec:morphology}
%------------

\begin{figure*}[h]
    \centering
\begin{multicols}{2}
   \includegraphics[width=0.9\linewidth]{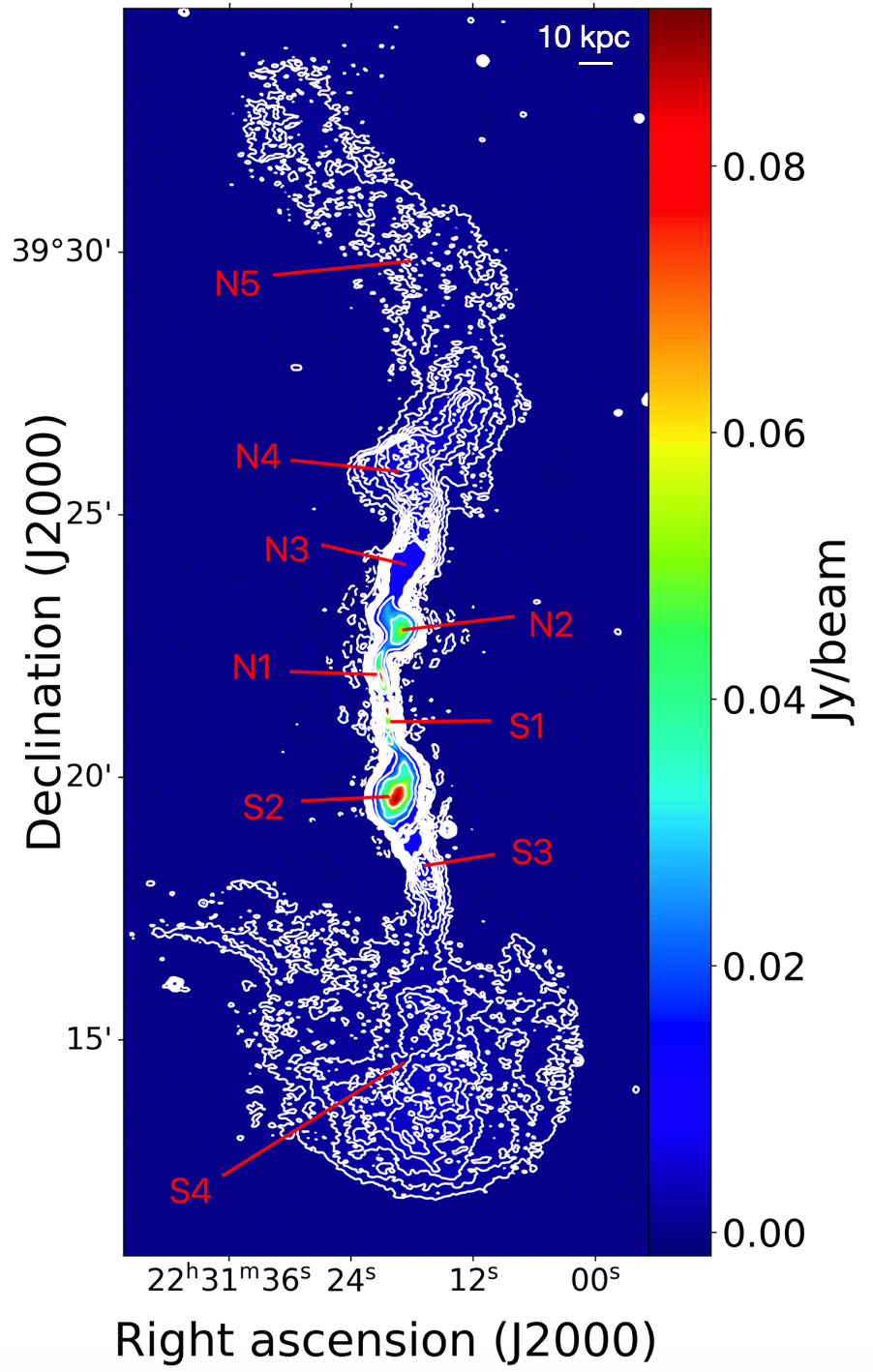}\par
   \includegraphics[width=1.05\linewidth]{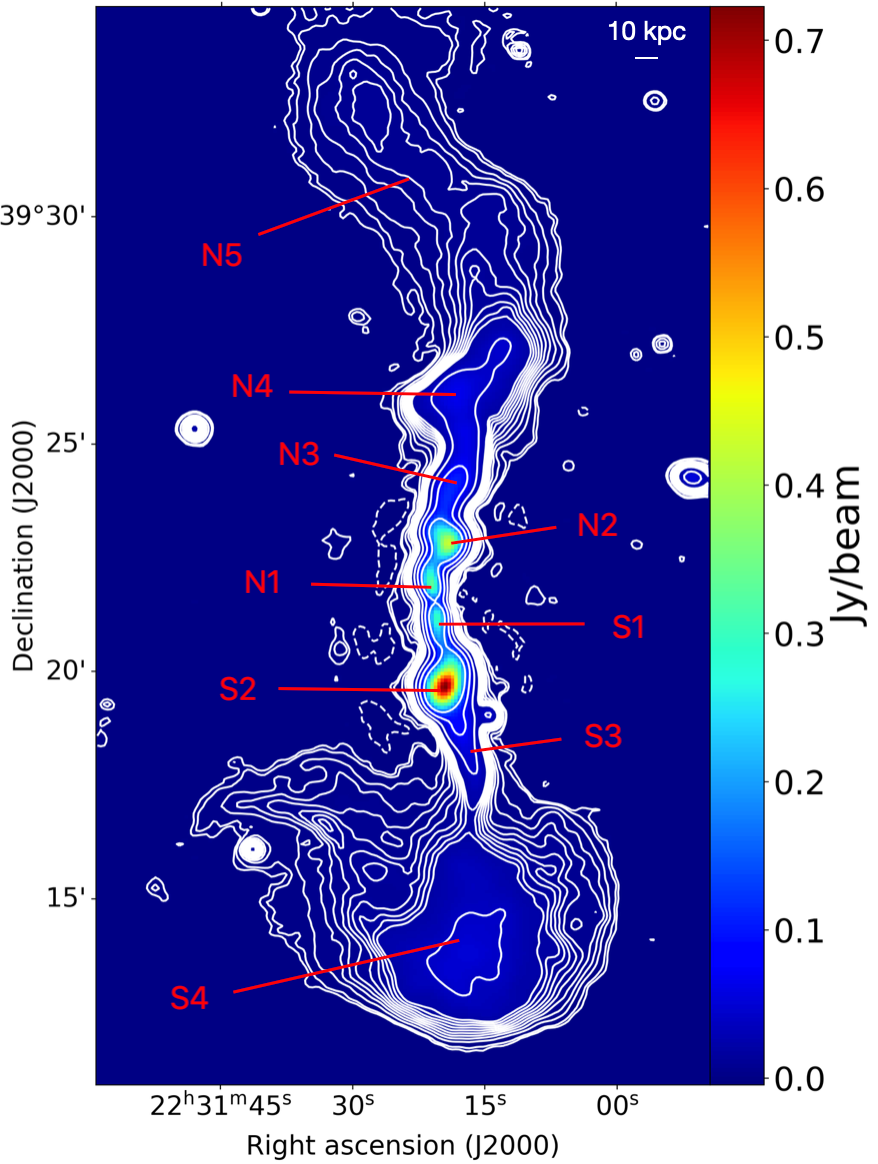}\par
\end{multicols}
  \caption{LOFAR maps of 3C~449 at 145 MHz. Left panel: angular resolution of $6.0\arcsec \times 6.0\arcsec$. The contour levels are $0.006 \, \sqrt{2} \, \times$ [-1.0, 1.0, 2.0, 3.0, 4.0, 5.0, 6.0, 7.0, 8.0] $\times$ the flux peak of 0.09 Jy beam$^{-1}$.  The rms noise of the map is 0.1 mJy beam$^{-1}$. Right panel: angular resolution of $20.0\arcsec \times 20.0\arcsec$. The contour levels are $0.002 \, \sqrt{2} \, \times$ [-1.0, 1.0, 2.0, 3.0, 4.0, 5.0, 6.0, 7.0, 8.0, 9.0, 10.0, 20.0, 40.0, 80.0] $\times$ the flux peak of 0.7 Jy beam$^{-1}$. The rms noise of the map is 0.2 mJy beam$^{-1}$.}
  \label{fig:LOFAR_maps}
  \end{figure*}

Our two LOFAR maps, displayed in Fig.~\ref{fig:LOFAR_maps}, show the source morphology at 145 MHz at the different angular resolutions listed in Table \ref{tab:maps}. 
The map in the left panel has a resolution of $6.0\arcsec \times 6.0\arcsec$ (2.2 kpc $\times$ 2.2 kpc), 
while the map in the right panel has a resolution of $20.0 \arcsec\times20.0 \arcsec$ (7.2 kpc $\times$ 7.2 kpc).
For ease of description of the results and comparison with the literature, in Fig.~\ref{fig:LOFAR_maps} we label the different regions of the radio source as in \cite{Feretti_1999}.

In our LOFAR maps, the source has a total extension out to 3$\sigma$  of $22.2\arcmin$ (479.5 kpc) at $6.0\arcsec \times 6.0\arcsec$ resolution, and of $22.3\arcmin$ (481.7 kpc) at $20.0\arcsec \times 20.0\arcsec$ resolution. 
For comparison, our reanalysed VLA maps at $6.0\arcsec \times 6.0\arcsec$ resolution show a source full extension of $\approx 22\arcmin$ at 1365 MHz, $\approx 11\arcmin$ at 4985 MHz, and $\approx 7\arcmin$ at 8485 MHz, whereas the maps at  $20.0\arcsec \times 20.0\arcsec$ resolution have a total extension of 
$\approx 20\arcmin$ at 1365 MHz and $\approx 22\arcmin$ at 1485 MHz. 

Thanks to the combination of low frequency and high sensitivity, our higher angular resolution ($6.0\arcsec \times 6.0\arcsec$) LOFAR map reveals additional, faint emission previously undetected by the VLA map at 1365 MHz with the same angular resolution (see Appendix \ref{app:vlamaps}).
In particular, in the northern tail beyond region N5, the $6.0\arcsec \times 6.0\arcsec$ LOFAR map clearly shows emission with a high average signal to noise ratio (S/N $\simeq 7$)  out to $\delta \simeq 39\degree \ 33\arcmin$; the emission of this source region was %much more noisy (S/N $\simeq 3$)
only marginally detected in the VLA observations at 1365 MHz with the same angular resolution.
On the other hand, at lower angular resolution ($20.0\arcsec \times 20.0\arcsec$),
emission out to that distance from the core is detected  in both our LOFAR map  at 145 MHz (Fig.~\ref{fig:LOFAR_maps}, right panel) and the VLA maps at 1365 MHz and 1485 MHz.
However, the LOFAR map reveals an additional detail with respect to the VLA maps: beyond region N5, the radio emitting plasma bends from the North-East direction to the North direction at a declination $\delta \simeq 39\degree \ 32\arcmin$; this feature was not clearly seen before, even though hints for a change in direction of the northern tail emerge in our reanalysed archival VLA map at 1485 MHz.
A bending of the radio structure about $\delta \simeq 39\degree \ 32\arcmin$ is also suggested by the higher angular resolution LOFAR map (Fig.~\ref{fig:LOFAR_maps}, left panel), despite the low S/N of the emission in this region.
%high noise level of the map in this region.

In the southern lobe (region S4), the higher angular resolution ($6.0\arcsec \times 6.0\arcsec$) LOFAR map (Fig.~\ref{fig:LOFAR_maps}, left panel) reveals $\approx 2\arcmin$ ($\approx 43$ kpc) of additional faint emission along the East-West direction with respect to the VLA map at 1365 MHz with the same angular resolution.
Overall, region S4 in the LOFAR maps at 145 MHz appears to have an area $\approx 50\%$ larger than in previous VLA maps at 1365 MHz with the same angular resolution. 
Furthermore, region S4 displays a protuberance on the eastern side which extends for $\approx 2'$ ($\approx 33$ kpc) in the East-West direction, and which was partly detected only in our reanalised VLA map at 1365 MHz and 1485 MHz with resolution of $20.0 \arcsec\times20.0 \arcsec$; it was not deteced by the VLA at $6.0\arcsec \times 6.0\arcsec$ resolution.

We highlight that the southern additional, extended emission of 3C~449 recovered by LOFAR at both angular resolutions and by the VLA at $20.0 \arcsec\times20.0 \arcsec$ was actually detected in previous observations performed with the Westerbork Synthesis Radio Telescope (WSRT) 
%at 608 MHz \citep{Perley_1979} 
at 608 MHz \footnote{The WSRT image is available in the online 3CRR Atlas (\url{https://www.jb.man.ac.uk/atlas/}), which presents radio images and other data for the nearest 85 DRAGNS (radio galaxies and related objects) in the so-called 3CRR sample of \citet{Laing_1983}.}.
%\citep[such observations are part of the 3CRR sample,][]{Laing1983}.
However, the larger beam ($48.0 \arcsec \times 30.0 \arcsec$) of the WSRT map makes this map unsuitable for a comparison with the LOFAR and VLA maps presented in this work.

%------------
\subsection{Spectral index maps and profiles} \label{sec:spec_index}
%------------

%------------
\subsubsection{Map characteristics}
\label{sec:spec_index_maps}
%------------

\begin{figure*}
  \centering
\begin{multicols}{3}
   \includegraphics[width=1.058\linewidth]{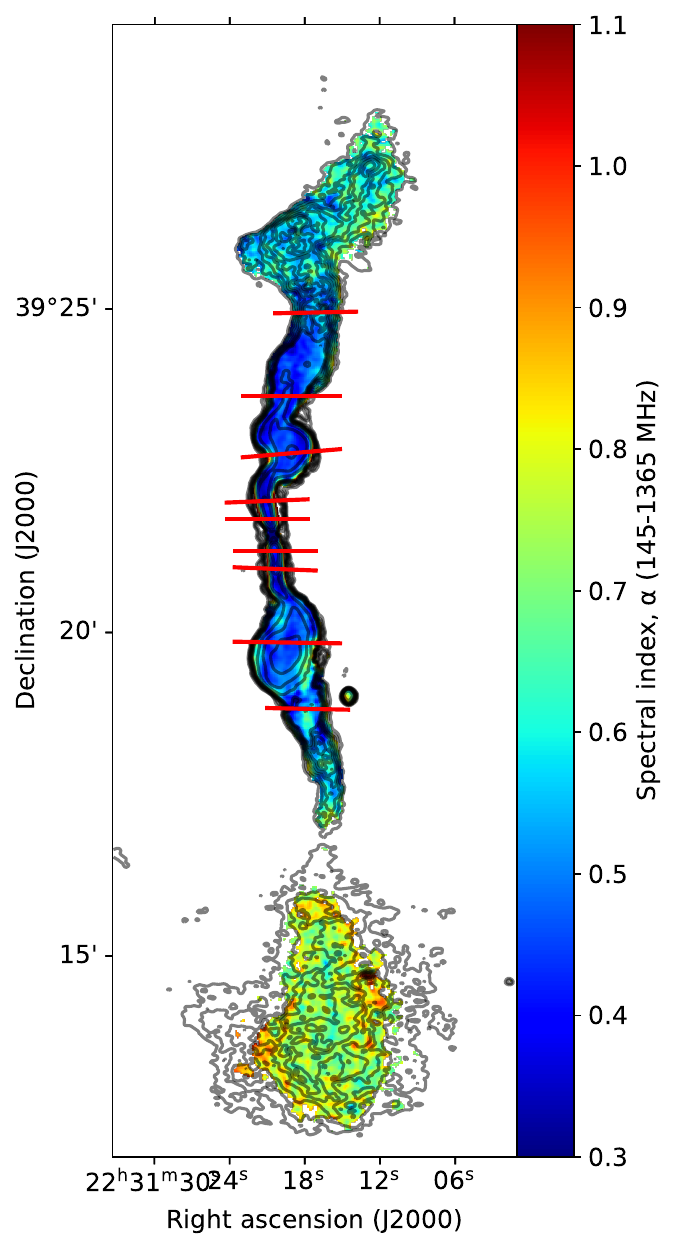}\par
   \includegraphics[width=0.95\linewidth]{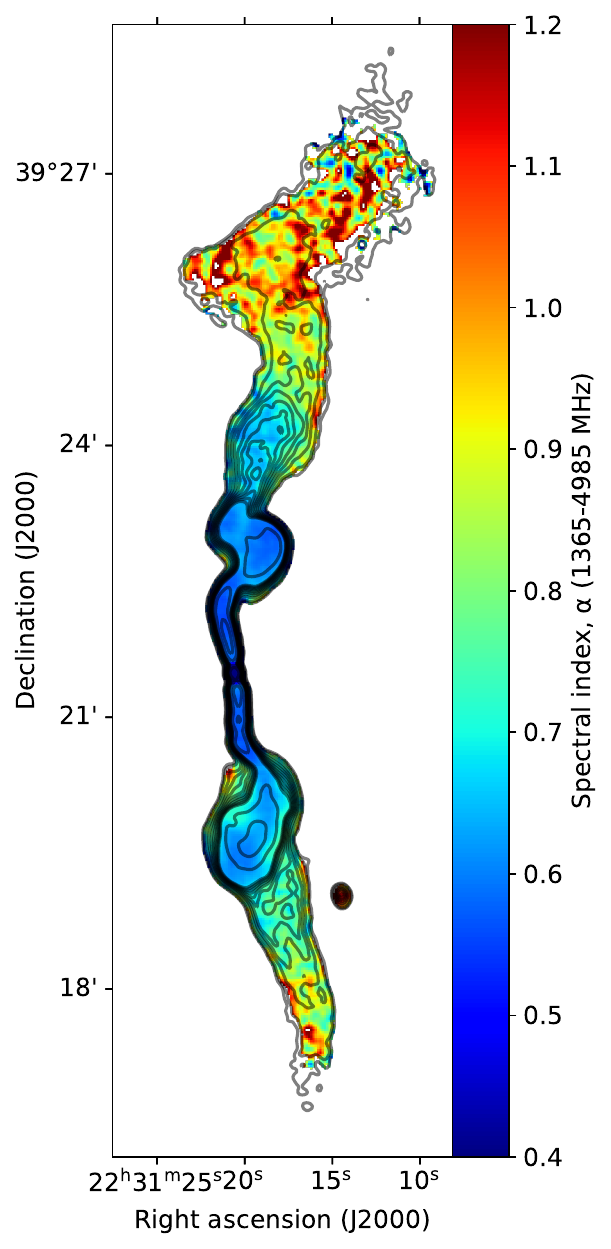}\par
   \includegraphics[width=1.018\linewidth]{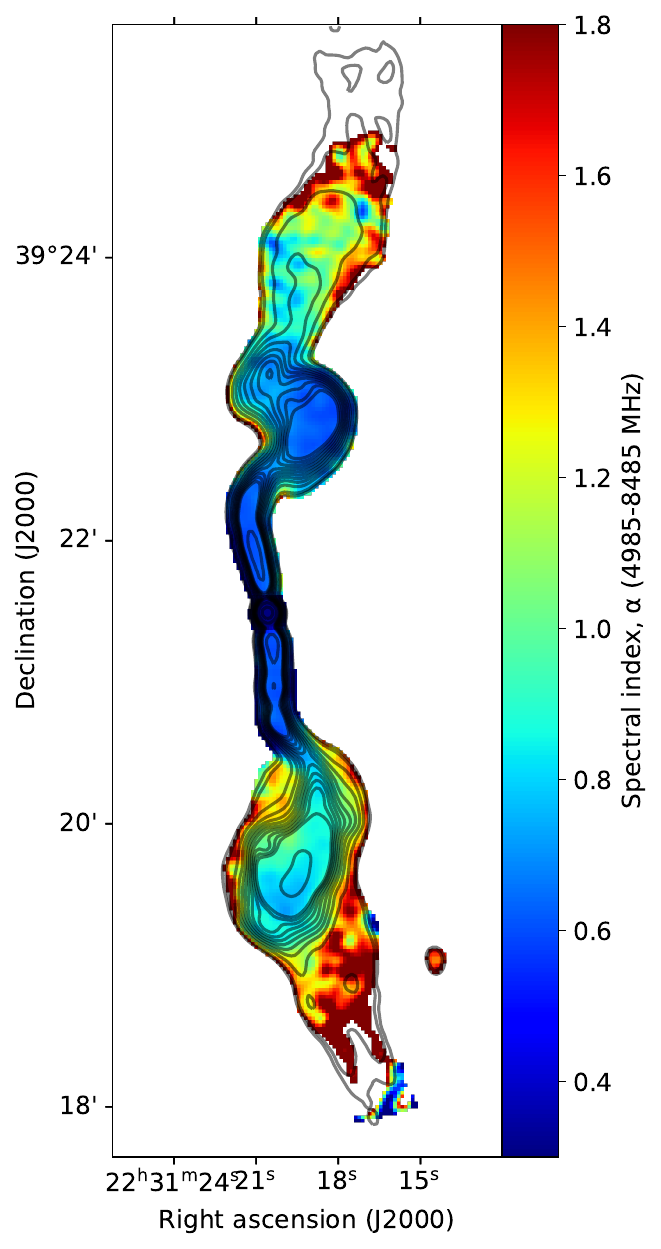}\par
\end{multicols}
\begin{multicols}{3}
   \includegraphics[width=1.06\linewidth]{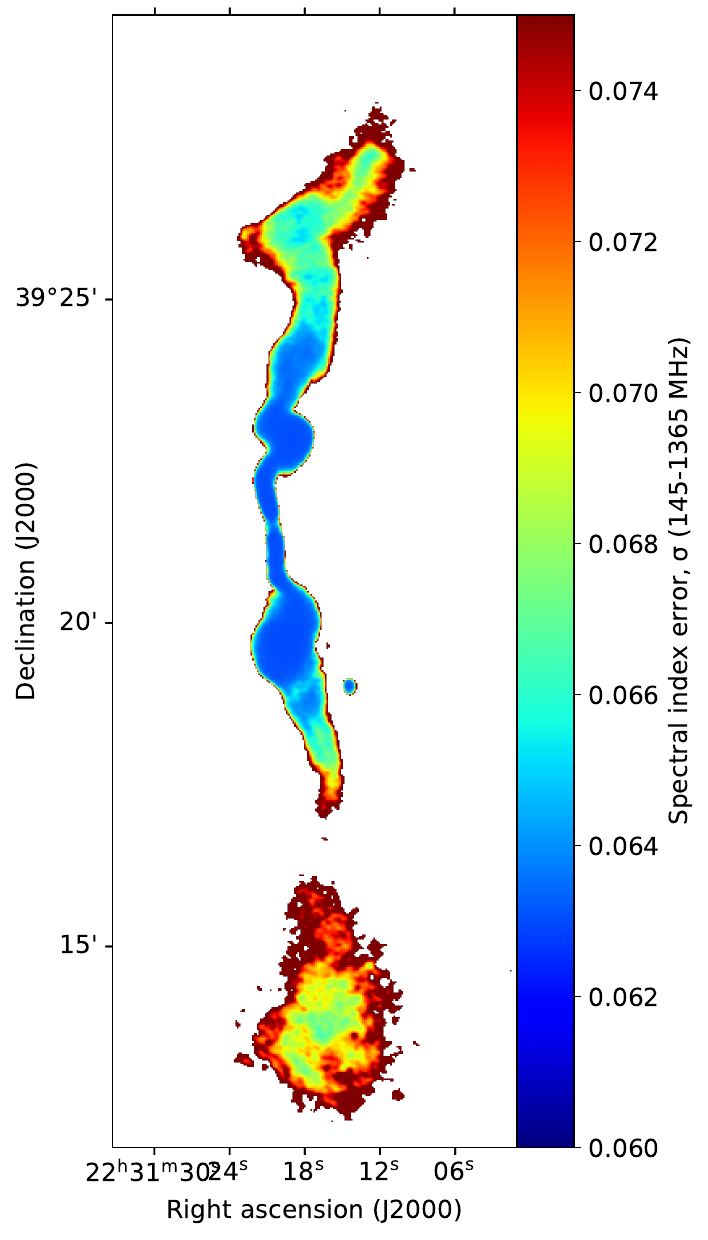}\par
   \includegraphics[width=0.95\linewidth]{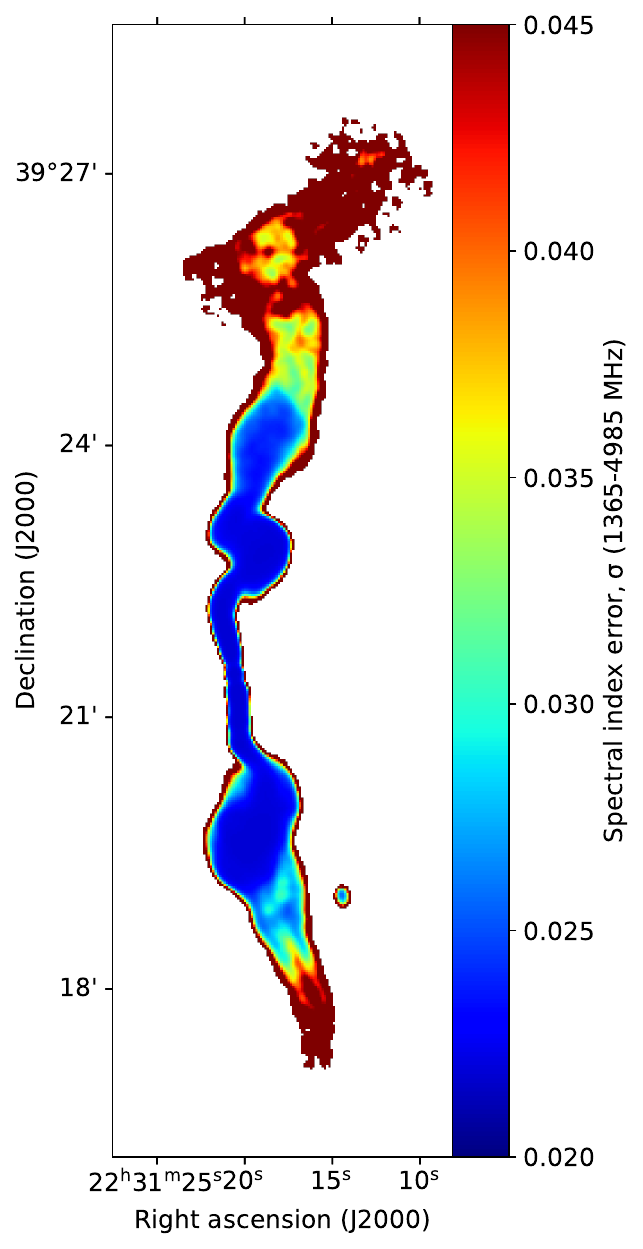}\par
   \includegraphics[width=\linewidth]{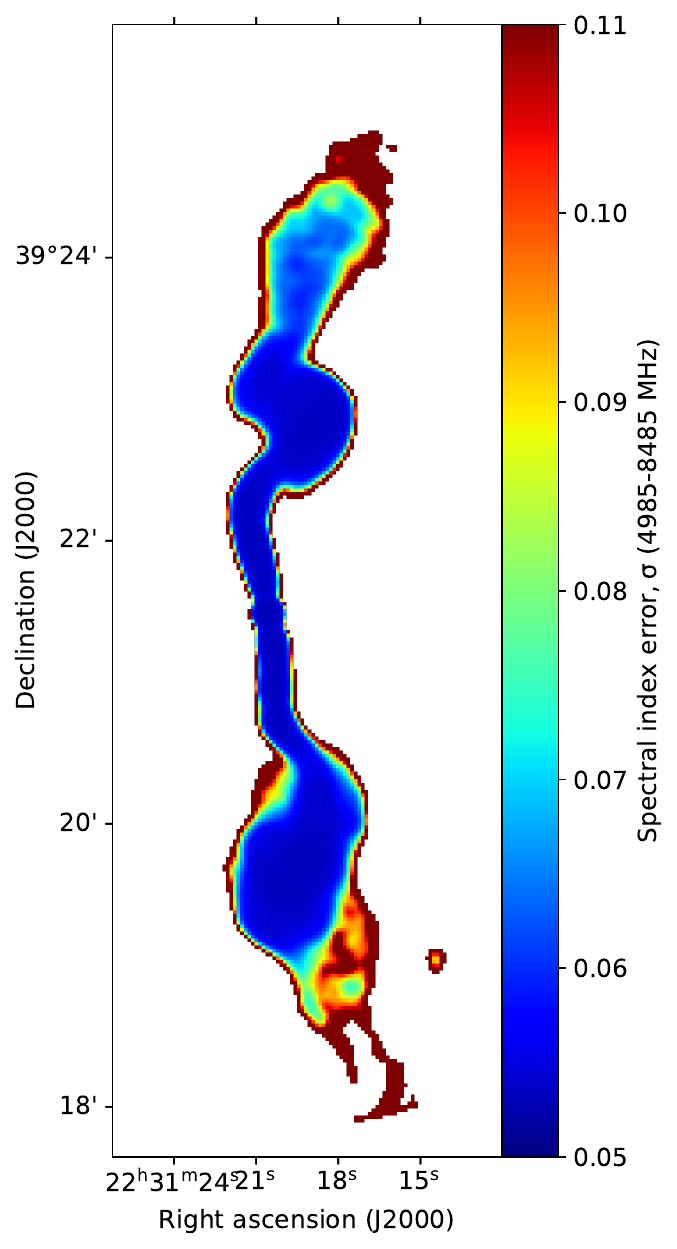}\par
   
\end{multicols}
  \vspace{-0.5cm}
  \caption{Spectral index maps at $6.0\arcsec \times 6.0\arcsec$. Upper panels: maps of the radio spectral index of 3C 449 between the frequencies of 145 and 1365 MHz (left panel), 1365 and 4985 MHz (middle panel), and 4985 and 8485 MHz (right panel). 
  Each map is overlaid with surface brightness contours of the lower frequency map used to compute the spectral index.
  Lower panels: maps of the $1\sigma$ uncertainties on the radio spectral index of 3C 449 between the frequencies of 145 and 1365 MHz (left panel), 1365 and 4985 MHz (middle panel), and 4985 and 8485 MHz (right panel). The red segments on top of the map in the upper, left panel represent the jet sections selected for the analysis of Sect.~\ref{sec:transverse_profiles}.}
  \label{fig:spectral_combined}
\end{figure*}

\begin{figure*}
\begin{multicols}{2}
   \includegraphics[width=0.98\linewidth]{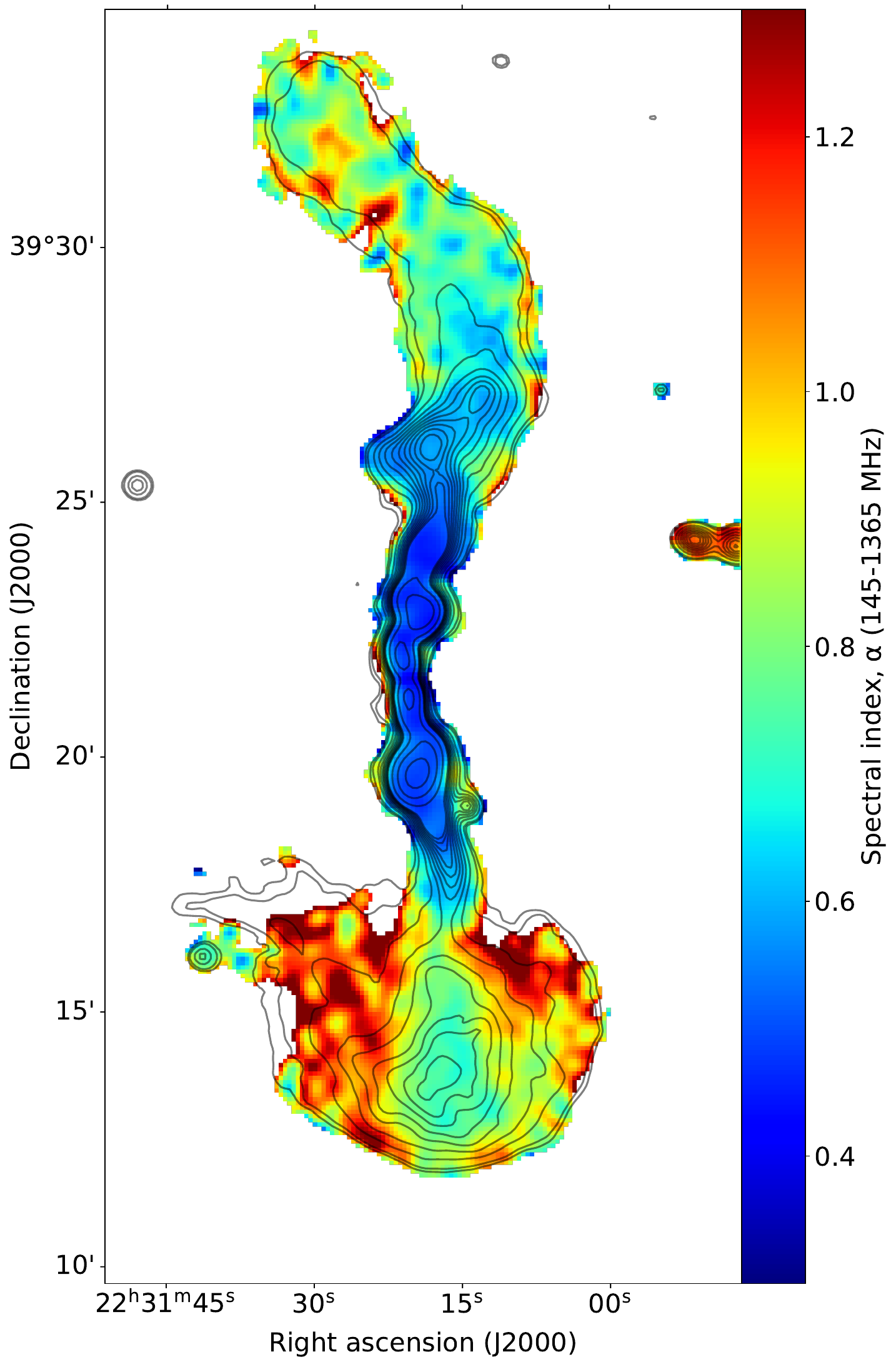}\par
   \includegraphics[width=\linewidth]{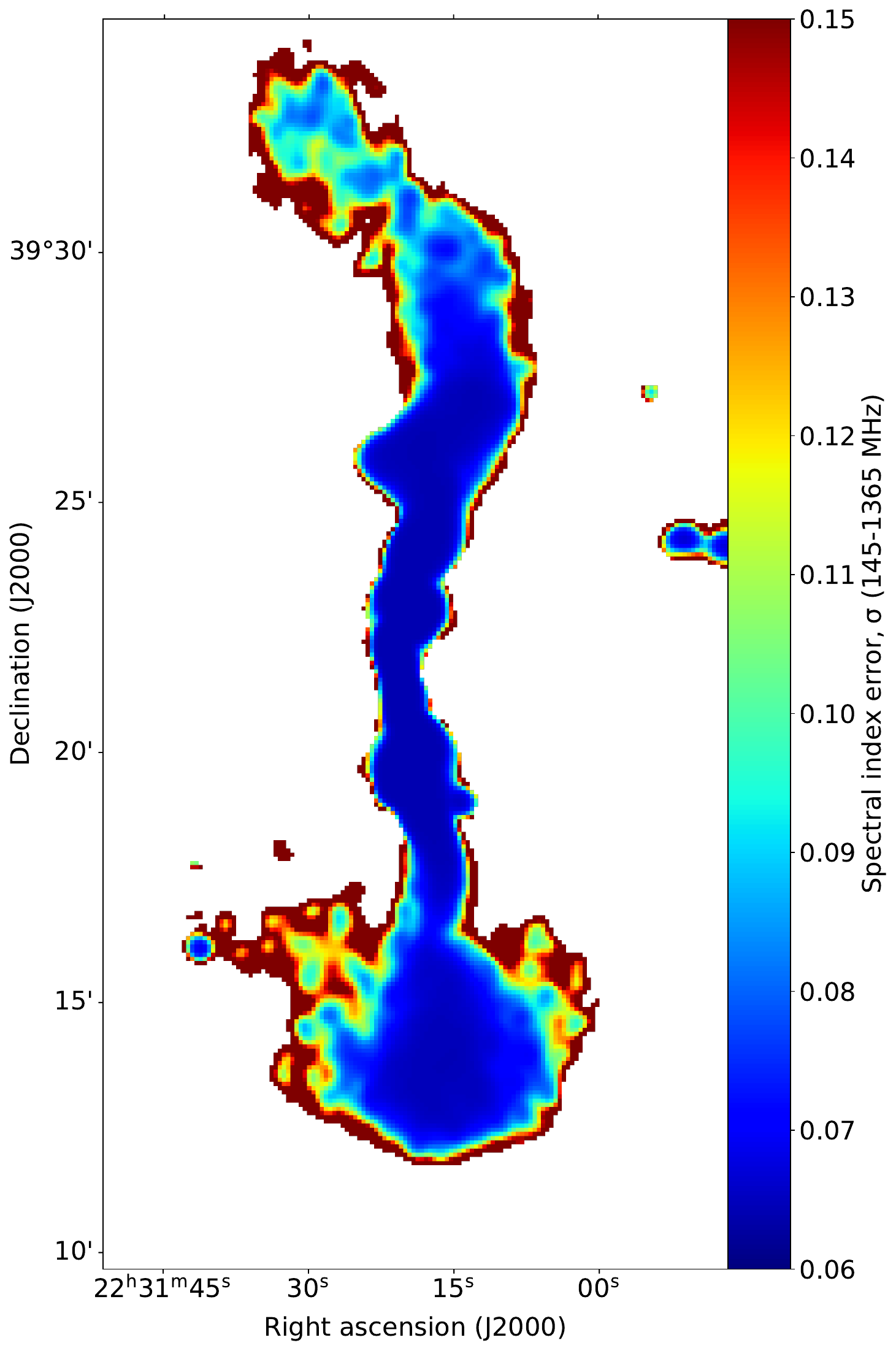}\par
\end{multicols}
  \caption{Radio spectral index of 3C 449 between 145 MHz and 1365 MHz, with the angular resolution of $20.0\arcsec \times 20.0\arcsec$. Left panel: spectral index map overlaid with 145 MHz surface brightness contours. Right panel: map of the 1$\sigma$ uncertainty.}
  \label{fig:Spec_index_low}
\end{figure*}

Following the procedure described in Sect.~\ref{sec:vla_data} to make the spatial resolution of the VLA maps match the resolution of our LOFAR maps, we derived four different maps of the radio spectral index $\alpha$ by combining in pairs the four maps obtained at the frequencies of 145, 1365, 4985, and 8485 MHz.
As mentioned in Sect.\,\ref{sec:vla_data}, the data have been restricted to have the same ($u,v$) range.
Specifically, for the $145-1365$ MHz spectral index map, the ($u,v$) range is ($150, 51000$)\,$\lambda$, for the $1365-4985$ MHz spectral index map ($u,v$) = ($600, 51000$)\,$\lambda$, and for the $4985-8485$ MHz spectral index map ($u,v$) = ($900, 56300$)\,$\lambda$.
Three out of four spectral index maps 
%(in the frequency ranges 145$-$1365, 1365$-$4985, and 4985$-$8485 MHz) 
have an angular resolution of $6.0\arcsec \times 6.0\arcsec$;
%, and are shown in Fig.~\ref{fig:spectral_combined}; 
the remaining map 
%(in the frequency range 145$-$1365 MHz) 
has an angular resolution of $20.0\arcsec \times 20.0\arcsec$.
%, and is displayed in Fig.~\ref{fig:Spec_index_low}.
To build these spectral index maps, in each single-frequency map we selected only pixels whose surface brightness was above $10\sigma_{\rm rms}$ (with $\sigma_{\rm rms}$ the rms noise level of the map).

Fig.~\ref{fig:spectral_combined} shows the spectral index maps at $6.0\arcsec \times 6.0\arcsec$ angular resolution: the upper left, middle, and right panels display 
the spatial distribution of the spectral index, $\alpha$, in the $145-1365$ MHz, $1365-4985$ MHz, and $4985-8485$ MHz frequency ranges, respectively; the bottom row shows the corresponding  maps of the 1$\sigma$ uncertainty on the spectral index, $\sigma_{\alpha}$.
In the panels of the upper row, source regions with lower values of the spectral index, i.e.,\ with flatter spectra, are represented in blue colours, while source regions with steeper spectra are displayed with green-red colours. 
%The full range of $\alpha$ represented in the maps is $\alpha \simeq 0.4-1.5$.
In the bottom row, bluer (redder) colours represent lower (higher) values of $\sigma_{\alpha}$.
Fig.~\ref{fig:Spec_index_low} shows the spectral index maps at $20.0\arcsec \times 20.0\arcsec$ angular resolution (left panel), with the corresponding map of $\sigma_{\alpha}$ (right panel). 

We note that the three spectral index maps of Fig.~\ref{fig:spectral_combined} have different spatial extensions: while the map on the left panel has a full extension of $\approx 15 \arcmin$ ($\approx 324$ kpc),
the maps in the middle and right panels have a full extension of $\approx 11\arcmin$ ($\approx 238$ kpc) and $\approx 7\arcmin$ ($\approx 151$ kpc), respectively.
The smaller extension of the latter two spectral index maps is the result of the smaller extension of the source in the VLA maps at 4985 MHz and 8485 MHz (see Sect.~\ref{sec:morphology}). 
The spectral index map of Fig.~\ref{fig:Spec_index_low} has a full extension of $\approx 20\arcmin$ ($\approx 432$ kpc).

{\subsubsection{Radial variations of the spectral index}
\label{sec:radial_profiles}}

A visual inspection of the $6.0\arcsec \times 6.0\arcsec$ angular  resolution spectral index maps of Fig.~\ref{fig:spectral_combined} 
shows that, in all of the three frequency ranges, the spectral index $\alpha$ on average stays roughly constant ($\alpha \approx 0.6$) along the straight, inner jet, out to 
%$\approx 1'$ from the core (
regions N1 and S1; starting from the inner lobes (regions N2 and S2), it then gradually increases with the distance to the radio core out to the source edges,
%$\gtrsim 10'$, 
implying that the average source spectrum steepens as the radio-emitting plasma propagates outwards, as expected from particle ageing in absence of significant reacceleration processes. 
This behaviour is also apparent in the $20\arcsec \times 20\arcsec$ angular  resolution spectral index map at 145-1365 MHz, shown in Fig.~\ref{fig:Spec_index_low} (left panel).

\begin{figure}[t]
\centering
  {\includegraphics[width=\linewidth]{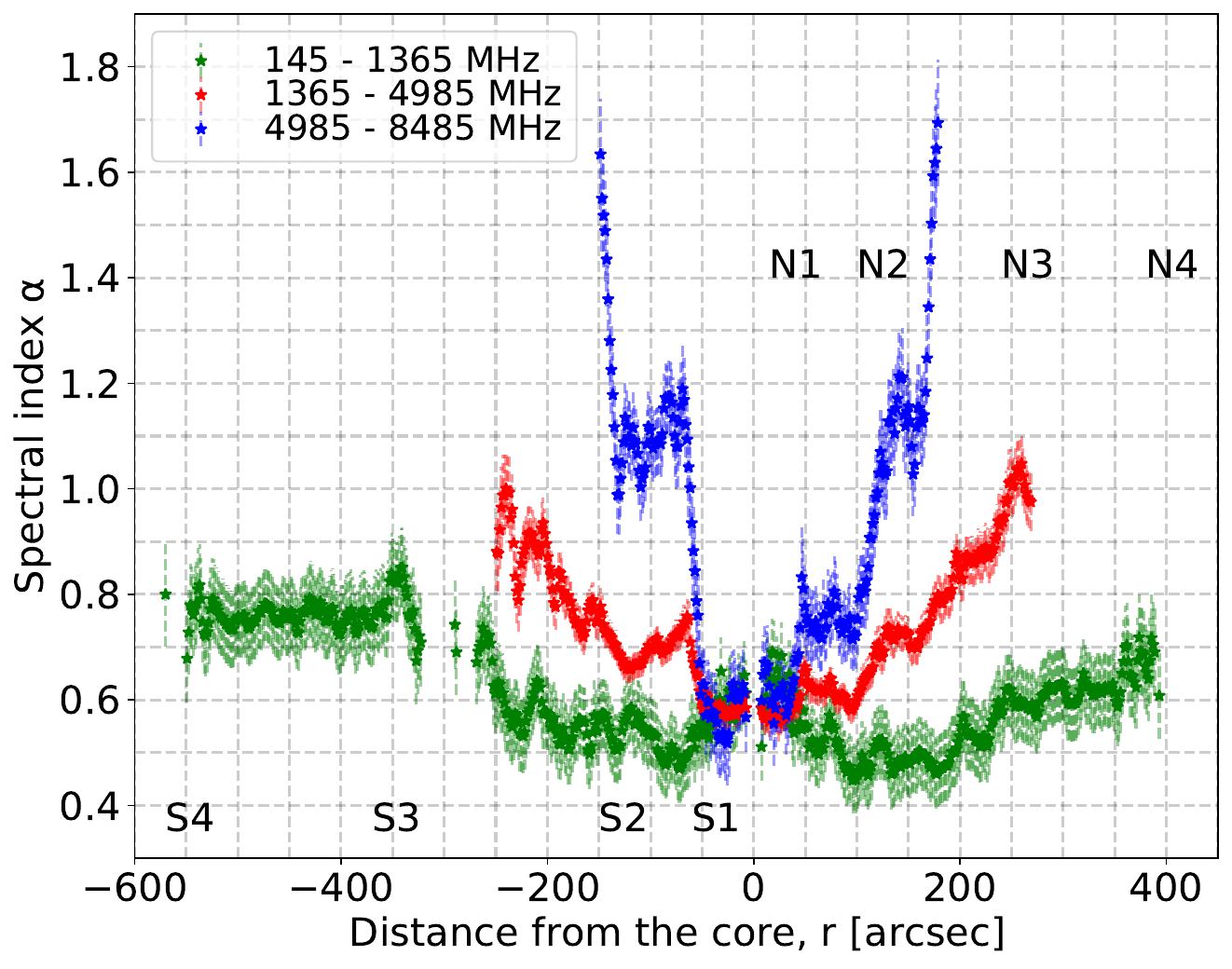}}
  \caption{Radio spectral index, $\alpha$, in different frequency ranges, as a function of the projected distance from the radio core, $r$ (see text for details), for the northern and southern jets of the source. 
  The position of the core corresponds to $r=0$; positive distances are for the northern jet, and negative distances for the southern one. The spectral indices of the core are not included in the plot.
  The frequency ranges and the corresponding colours are listed in the inset. The overall steepening of the spectrum (i.e.,  an increase of the radio spectral index $\alpha$) downstream of the jet can be appreciated in all three spectral index profiles.}
  \label{fig:SM_distribution}
\end{figure}

The evolution of the average radio spectrum with increasing distance from the radio core can be better appreciated in the spectral index radial profiles shown in Fig.~\ref{fig:SM_distribution}, derived from the $6\arcsec \times 6 \arcsec$ angular resolution maps of Fig.~\ref{fig:spectral_combined}.
Here, each spectral index profile shows the average value of the spectral index, $\alpha$, as a function of the projected distance from the radio core, $r$ in a given frequency range.
The average value of $\alpha$ is computed over a source slice arranged along the East-West direction; the slice thickness  $\Delta \delta$ (with $\delta$ the declination) is of one pixel, namely $1.5 \arcsec$; $r$ increases from the core outwards along the direction of the declination, roughly corresponding to the jet direction.

Incidentally, we note that, at any given $r$, the average value of the spectral index shown in each of the profiles of Fig.~\ref{fig:SM_distribution} may be higher than the value of the spectral index that appears to be dominant in the maps of Fig.~\ref{fig:spectral_combined}. 
This is the result of averaging the spectral indices over a jet slice that displays a flatter spectrum in its central part, and a significant spectral steepening in its external part (see Sect.~\ref{sec:transverse_profiles} for details).

Each of the radial profiles of Fig.~\ref{fig:SM_distribution} shows that the spectral index value stays roughly constant within $\approx 40-50\arcsec$ from the core (i.e., in regions N1 and S1),  despite the small-scale fluctuations, and then significantly increases with increasing distance from the core in both jets.
This confirms that, in any given frequency range, the spectrum steepens with the distance from the core beyond regions N1 and S1.

The behaviour we observed in 3C~449 confirms previous findings by  \citet{Katz-Stone_Rudnick_1997} and \citet{Feretti_1999} on this source: they showed that the radio spectrum is roughly constant out to $\approx 1'$ from the core on both sides, and then steepens with increasing distance out to $\approx 2.5'$ from the core (the spatial limit of their spectral studies). The steepening of the spectrum with core distance on arcminute scales also confirms earlier results by \cite{Jaegers_1987} and \cite{Andernach_1992}.
This spectral behaviour is typical of FRI radio galaxies \citep[see also, e.g., the case of 3C~31;][]{Heesen_2018}, where particle acceleration processes are thought to occur closer to the core, and the accelerated particles radiatively age as the radio source propagates outwards.

A comparison of the three maps of Fig.~\ref{fig:spectral_combined} and of the corresponding radial profiles of Fig.~\ref{fig:SM_distribution} shows that,  globally, the radial evolution of the spectral index is different in different frequency ranges.
More in detail, the radial profiles are consistent with one another, at the 1$\sigma$ level,  within $\approx 50\arcsec$ from the core (i.e., within regions N1 and S1): thus, in this region,
%Beyond those regions, the radial profiles display a positive vertical offset, on arcminute scale, even though fluctuations are observed on smaller scales: the higher-frequency spectral indices are always higher than the lower-frequency indices.
%This behaviour 
%This indicates that, within these regions, 
the average source spectrum is consistent,  within the uncertainties, with a single power law (with  $\alpha \simeq 0.6$) over the full, 145-8485 MHz frequency range.
%out to $\approx 50\arcsec$ from the core.
However, beyond $\approx 50\arcsec$, the radial profiles display a positive vertical offset, on arcminute scale, even though fluctuations are observed on smaller scales: the higher-frequency spectral indices are always higher than the lower-frequency indices.
This means that the spectral shape deviates from a single power law, and a spectral curvature appears that makes the spectrum steeper at higher frequencies.
Furthermore, the increase of the average spectral index with the distance to the radio core appears to be faster at higher frequencies, indicating that the spectral curvature increases, and the break frequency decreases, with the distance to the core.

\begin{figure}[t]
  \resizebox{\hsize}{!}{\includegraphics{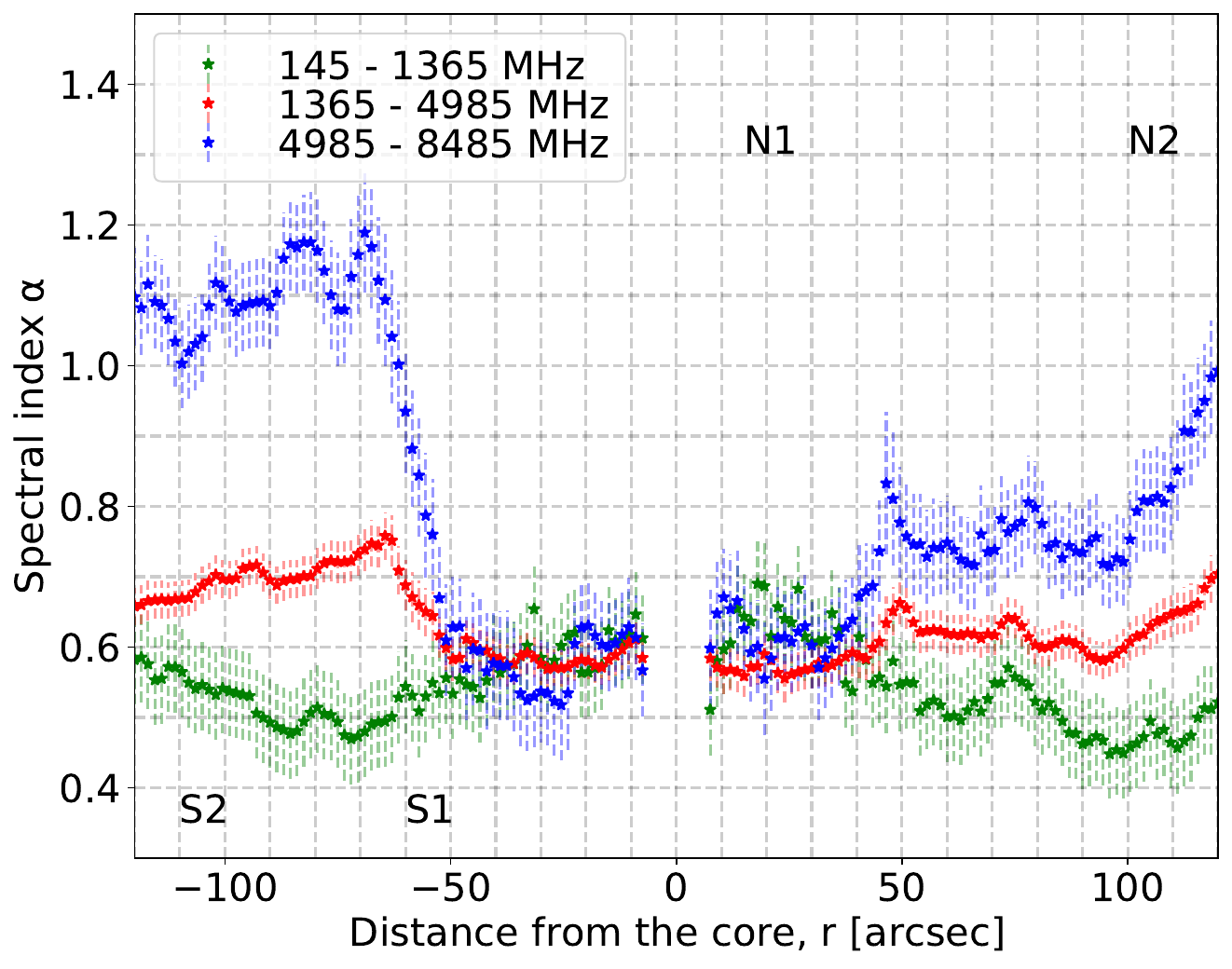}}
  \caption{Same as Fig.~\ref{fig:SM_distribution}, but for the inner $\approx 240\arcsec$ region of the source.}
  \label{fig:SM_distribution_zoom}
\end{figure}

The spectral evolution with core distance is not symmetric about the core at all scales and in all frequency ranges.
The onset of the spectral curvature appears to become significant (at $>1\sigma$)  at about the same distance ($r\approx 50\arcsec$) from the radio core in the two jet directions, as it can be better appreciated in  Fig.~\ref{fig:SM_distribution_zoom}, an enlargement of Fig.~\ref{fig:SM_distribution} corresponding to the region centred on the radio core and extending from N2 to S2.
However, at $r\gtrsim 50\arcsec$,
different spectral behaviours are observed in the two jet directions in different frequency ranges. 
Specifically, in the 1365-8485 MHz range (red and blue curves), 
a spectral 
%curvature 
steepening with core distance is clearly detected %at $r\gtrsim 50\arcsec$  
both in the northern and in the southern jet. 
In the $145-1365$ MHz range (green curve) the spectrum flattens, with respect to the inner jet region, between $r\approx 50\arcsec$ and $r \approx 100 \arcsec$ in the northern jet, and between $r\approx 50\arcsec$ and $r \approx 80 \arcsec$ in the southern one: this initial flattening determines the onset of a spectral break in the frequency range 145-4985 MHz. 
At larger $r$'s, as shown in Fig.~\ref{fig:SM_distribution}, the $145-1365$ MHz spectrum gradually steepens with distance on both sides, although with asymmetric behaviours: on the northern side, the steepening continues out to region N4 ($r\approx 400 \arcsec$), where $\alpha \simeq 0.7$, while on the southern one the steepening proceeds out to region S3 ($r\approx 300\arcsec$), and then the spectral index fluctuates about a constant value $\alpha \simeq 0.75$. A flattening of the spectrum beyond region S3 was also observed by \cite{Jaegers_1987} at 600-1400 MHz and by \cite{Andernach_1992} at 400-2700 MHz.

Overall, in all the frequency ranges considered in this work, a North-South asymmetry in the distribution of $\alpha$ can be appreciated both visually in the spectral index maps and in the radial spectral index profiles on different spatial scales. 
In particular, in the $4985-8485$ MHz frequency range, 
the southern, inner lobe (region S2) appears to have a significantly steeper spectrum than the northern, inner lobe (region N1), as can be seen in the upper, right panel of Fig.~\ref{fig:spectral_combined} and in the blue profile in 
Fig.~\ref{fig:SM_distribution_zoom}; in the $145-1365$ MHz frequency  range, the southern, outer lobe (region S4)  displays a significantly steeper spectrum than the northern plume (regions N4-N5), as can be seen in the upper, left panel of Fig.~\ref{fig:spectral_combined} and in the green profile in Fig.~\ref{fig:SM_distribution_zoom}.

Besides showing a different average values of the spectral  index, the northern tail and southern lobe also show a different spatial distribution of $\alpha$ at both angular resolutions, as can be partly seen in the upper, left panel of Fig.~\ref{fig:spectral_combined} and, more clearly, in Fig.~\ref{fig:Spec_index_low} (left panel).
The northern, narrow tail of the source (regions N4-N5), which exhibits a low 145 MHz surface brightness everywhere, displays a distribution of the $145-1365$~MHz spectral index significantly patchy, whose average value is $\alpha \simeq 0.75$ (with $\sigma_{\alpha} \simeq 0.10$) at $20\arcsec \times 20\arcsec$ resolution. 
Conversely, the southern lobe of the source (region S4), characterised by a high 145 MHz surface brightness in the inner region that slowly falls off towards the outskirts of the lobe, has a well-defined spectral index distribution: at $20\arcsec \times 20\arcsec$ resolution, we clearly see that the central, brighter region of the lobe has a significantly flatter spectrum ($\alpha \simeq 0.7-0.9$, with $\sigma_{\alpha} \simeq 0.06-0.07$) 
than the surrounding, fainter regions ($\alpha \simeq 0.9-1.3$, with $\sigma_{\alpha} \simeq 0.12$), 
yielding an overall spectrum which is steeper 
%($\alpha \simeq 0.75${\color{red}[CHECK WITH LUCA]}) 
than that of the northern tail. This result confirms previous findings on the North-South spectral asymmetry by \citet{Katz-Stone_Rudnick_1997} and \citet{Feretti_1999}, also suggested by earlier, lower-resolution studies by \cite{Jaegers_1987} and \cite{Andernach_1992}.

\begin{figure}
  \resizebox{\hsize}{!}{\includegraphics{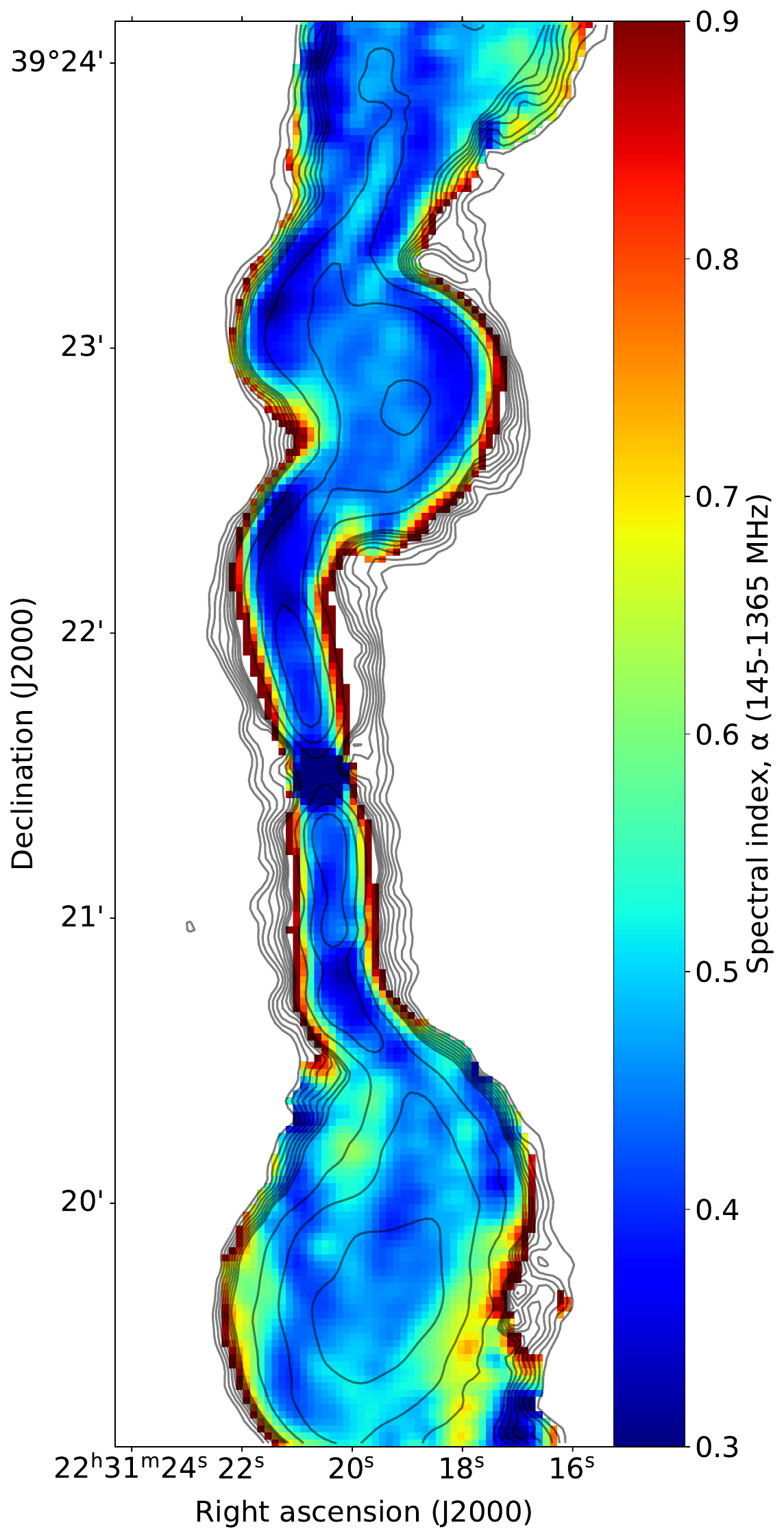}}
  \caption{Enlargement of the inner $\approx 5'$ region of the $145-1365$ MHz spectral index map at the angular resolution of $6.0\arcsec \times 6.0\arcsec$ (upper, left panel of Fig.~\ref{fig:spectral_combined}). The dark-blue patches, representing the flattest spectra ($\alpha \simeq 0.32-0.35$, $\sigma_{\alpha} = 0.063$), are visible about the jet turns. For the uncertainties, see the lower left panel of Fig.~\ref{fig:spectral_combined}.}
  \label{fig:SM_zoom}
\end{figure}

\subsubsection{Transverse variations of the spectral index}
%{Spectral index transverse profiles}
\label{sec:transverse_profiles}

Besides showing an evolution of the average spectral index with the distance from the core, 
%and, at any given distance, with the frequency range,  
our spectral index maps also reveal  significant substructures in the distribution of $\alpha$ along the East-West direction. 
This transverse substructure is clearly detected, in all the explored frequency ranges, from the inner lobes (regions N2 and S2) outwards.
However, in the frequency range $145-1365$ MHz, a transverse structure in $\alpha$ is also detected both in the northern and in the southern inner jet, specifically in the nearly straight and collimated structures that extends out to $\approx 1'$ from the radio core (regions N1 and S1) . 
As shown in the higher angular resolution, $145-1365$ MHz  spectral index map of Fig.~\ref{fig:spectral_combined} (upper, left panel) and in its $\approx 5'$ enlargement of Fig.~\ref{fig:SM_zoom}, as well as in the lower angular resolution maps of Fig.~\ref{fig:Spec_index_low} (left panel), the higher surface brightness part of each of the inner jets, closer to the jet axis, is characterized by a flat spectrum (blue colour, $\alpha \simeq 0.4$), and appears to be surrounded by layers whose spectra are significantly steeper (green-red colour, $\alpha \simeq 0.6-1.1$).

In both jets, about the first turn, at a distance of $\approx 1'$ from the core, the spectral index distribution starts to become more complex: the steep spectrum regions seem to be responsible for the widening of the source, and appear to be progressively mixed to a less collimated, although continuous, flat spectrum structure that extends with no interruptions out to  $\approx 3'$ from the core (regions N2-N3) for the northern jet, and out to $\approx 2.5'$ from the core (region S2) for the southern jet.

The evidence for a flat spectrum jet spine surrounded by a steeper spectrum sheath in the inner, $\approx 1\arcmin$ jets of 3C~449 is a unique feature of our $145-1365$ MHz spectral index maps, which combine LOFAR and VLA data.
In the remaining explored frequency ranges, from 1365 to 8485 MHz (see middle and right panels of Fig.~\ref{fig:spectral_combined}), in the first $\approx 1'$ of the jet we do not clearly detect a flatter spectrum jet surrounded by a steeper spectrum layer.
However, beyond $\approx 1'$ from the core,  we do see that steeper spectrum regions appear to enshroud a flatter spectrum jet, in both jets and in all of the three spectral index maps of Fig.~\ref{fig:spectral_combined}: the steep spectrum regions widen with the distance from the core, and start to become patchy about the edge of the collimated jet, namely in regions N4-N5 for the northern jet, and in region S3 for the southern jet; beyond these regions, there is no clear structure in the distribution of the spectral index in the northern plume, whereas the southern lobe displays a flatter core surrounded by a steeper shell, as mentioned in Sect.~\ref{sec:radial_profiles}.

\begin{figure*}[h]
    \centering
\begin{multicols}{2}
   \includegraphics[width=\linewidth]{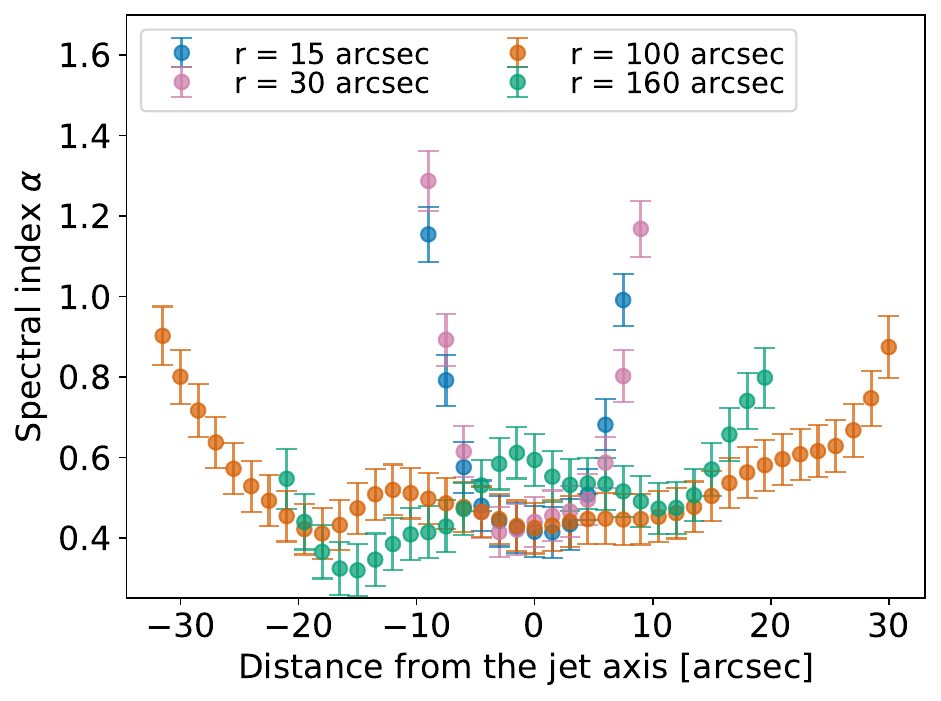}\par
   \includegraphics[width=\linewidth]{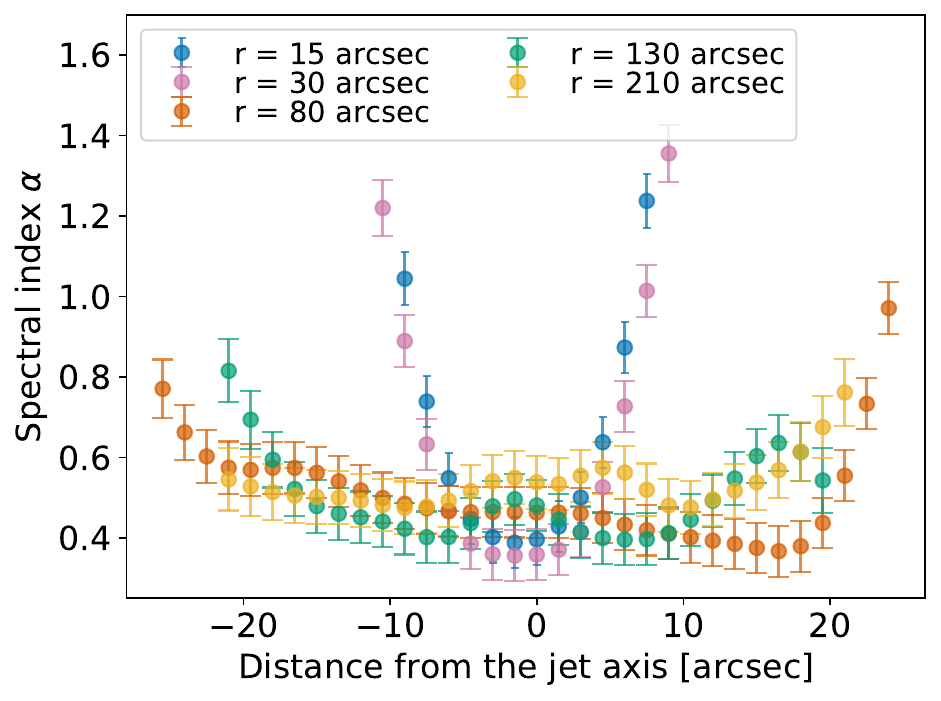}\par
\end{multicols}
  \caption{Transverse profiles of the spectral index between 145 and 1365 MHz, derived from the $6\arcsec \times 6\arcsec$ spectral index map: each profile is evaluated over a slice arranged along the East-West direction, roughly perpendicular to the jet axis;  the slice thickness is 1 pixel (i.e., $1.5\arcsec$). All the slices are shown in the upper, left panel of Fig.~\ref{fig:spectral_combined}. Left panel: southern jet. Right panel: northern jet.} 
  \label{fig:transverse_profiles}
  \end{figure*}

The presence of a flat-spectrum jet spine surrounded by a steep-spectrum sheath in the inner jet of 3C~449 can be better visualised by the transverse profiles of the spectral index, shown in Fig.~\ref{fig:transverse_profiles}.
These profiles were constructed for a number of representative source slices arranged in the East-West direction on the $145-1365$ MHz spectral index map at $6\arcsec \times 6\arcsec$ angular resolution, covering both the northern and the southern jets.  
The slices are highlighted with red segments in the upper, left panel of Fig.~\ref{fig:spectral_combined}.

In Fig.~\ref{fig:transverse_profiles}, the left panel shows the southern jet and the right panel shows the northern jet. The transverse spectral index profiles of slices selected at different distances from the core are represented with different colours. 
Out to $\approx 100\arcsec$ from the core, in each profile the spectral index is lower (i.e., the spectrum is flatter) about the slice center (close to the jet axis), where $\alpha \simeq 0.4-0.6$,  and becomes larger and larger (i.e., the spectrum steepens) as one moves towards the eastern and western edges of the slice (far from the jet axis), where the spectral index reaches values of $\alpha \simeq 1.3$ in the jet regions closer to the core. 

The steeper spectrum sheath is narrower (a few arcseconds across) and displays a faster rise of the spectral index in slices located at smaller distances from the radio core; it widens and displays a more gradual rise of the spectral index as the slice distance from the radio core increases.
About $\approx 3\arcmin$ from the core in the northern jet, and about $\approx 2.5\arcmin$ in the southern jet, the shape of the profiles becomes more complex, and the separation between a flat jet spine and a steep sheath structure is no longer clear. 
In these regions, the central, flatter-spectrum structure fades, and mixing between the flatter-spectrum and steeper-spectrum regions seems to occur.

The evidence we found for a steep spectrum sheath embedding the flatter spectrum jet beyond $\approx 1'$ from the core confirms previous findings by \citet{Katz-Stone_Rudnick_1997} in the 330-4835 MHz frequency range for both jets; it also confirms the results obtained by \citet{Feretti_1999} in the 4985-8400 MHz frequency range for the southern jet, but not for the northern jet, where these authors do not identify a spine-sheath structure.
On the other hand, our results on the transverse spectral index structure of the inner, $\approx 1'$ jet in the $145-1365$ MHz range were not reported previously.
We discuss the possible origin of the spine-sheath spectral structure in Sect.~\ref{sec:spine-sheath}.

\subsubsection{Substructures with very flat spectrum}
\label{sec:flat_spectra}

A noteworthy aspect of our spectral index maps relates to the unusually low values that the $145-1365$ MHz spectral index assumes in some regions of the source.  
Both the higher angular resolution spectral index maps (upper, left panel of Fig.~\ref{fig:spectral_combined}, and its enlargement of Fig.~\ref{fig:SM_zoom}) and the lower angular resolution map (left panel of Fig.~\ref{fig:Spec_index_low}) show that, in this frequency range, the spectrum is very flat, with $\alpha \simeq 0.35-0.45$, in significant portions of jet regions N1, N2, N3, S1, and S2.
Furthermore, a few isolated substructures display an even flatter spectrum, with $\alpha \simeq 0.30-0.35$; they appear about the eastern jet turns of the northern, inner jet (in regions N1 and N2) and the western turn of the southern, inner jet (in region S1), and have a linear size of $\approx 10\arcsec$ ($\approx 3.6$~kpc).
The uncertainty on $\alpha$  is $\sigma_{\alpha} \simeq 0.06$ across the entire jet region, making all the $\alpha$'s we measured be consistent with the ``universal'' spectral index value of 0.5 (expected from diffusive shock acceleration of particles by non-relativistic, strong shocks; see Sect.~\ref{sec:flat_index}) at the $2-3\sigma$ level.
Some of them have a counterpart in the $1365-4985$ MHz and/or the $4985-8485$ MHz frequency range. However, at these frequencies, their spectrum always has an index $\alpha > 0.5$, consistent with the spectral index of the adjacent regions at 1$\sigma$ level; therefore, the spectral flattening of these regions at $1365-4985$ MHz has a lower significance than the flattening of the same regions at $145-1365$ MHz.
Specifically,  the $1365-4985$ MHz map (Fig.\ \ref{fig:spectral_combined}, middle panels) shows a counterpart of all of the three flattest substructures detected in the $145-1365$ MHz frequency range: the substructures in regions N1 and S1 display $\alpha \simeq 0.54$ (against $\alpha \simeq 0.57$ for the adjacent region), whereas the substructure in region N2 shows $\alpha \simeq 0.57$ (against $\alpha \simeq 0.63$ for the adjacent region). 
On the other hand, in the $4985-8485$ MHz spectral index map, only the flat substructure of region N2 has a counterpart, with $\alpha \simeq 0.57$ (lower than the adjacent region, with $\alpha \simeq 0.65$); no counterpart is detected for the flat substructures in N1 and S1.
Interestingly, the flatter-spectrum regions do not display an enhanced surface brightness in any of the observing frequencies (see Fig.\,\ref{fig:LOFAR_maps} and Appendix \ref{app:vlamaps}.)

We note that the flatter-spectrum regions that we identify
were not found neither in the $2.5\arcsec$ resolution spectral index
maps of the source at 4985-8485 MHz by \cite{Feretti_1999} nor
in the $3.6\arcsec$ resolution spectral index map at 1445-4835 MHz by 
\cite{Katz-Stone_Rudnick_1997}. No comparison with previous
spectral images could be performed at 145-1365 MHz, due
to the lack of spectral studies in this frequency range prior
to ours.

We discuss the possible origin of the unusually low values of the spectral index detected in the $145-1365$ MHz frequency range, as well as of the flattest substructures of regions N1, N2, and S1 in the different frequency ranges in Sect.~\ref{sec:flat_index}.
%Sect.~\ref{sec:particle_population}.}

%------------
\section{Spectral analysis and implications} \label{sec:spectral_age}
%------------
The spatial distribution of radio spectral indices in radio galaxies is traditionally interpreted as the sign of a radiative ageing of a population of emitting electrons with a power-law energy distribution. In this scenario, a spatial distribution of spectral ages across the source can be inferred either from the spatial distribution of spectral break frequencies by using analytical formulae \citep[e.g.,][]{van_der_Laan_1969} or from the numerical modelling of the observed radio spectra as the result of the time evolution of an injected spectrum due to plasma radiative losses \citep{Jaffe_Perola_1973,Kardashev_1962,Harwood_2013,Harwood_2015}. In both approaches, the assumption of a uniform  magnetic field in the whole source is common \citep[e.g.,][]{Parma1999,Brienza_2020}, even though for a few sources  spectral ages and/or break frequencies were estimated in different source points or regions, for which different magnetic field intensities were estimated \citep[e.g.,][]{Andernach_1992,Heesen_2018,kukreti_2022}. 
Alternative scenarios call into question non-uniform magnetic fields with complex structure, electron energy distributions that deviate from a power-law, adiabatic expansion of the jet, or a combination of these effects \citep{Eilek_1996,Eilek_1997,Katz-Stone_Rudnick_1997}. 

Under the traditional ageing assumption, we mapped the spectral age of 3C~449 by carrying out a spectral analysis of our radio maps by means of the Broadband Radio Astronomy ToolS\footnote{\url{https://github.com/JeremyHarwood/BRATS}  \citep[BRATS;][]{Harwood_2013, Harwood_2015}} software package. BRATS provides, among other tools, the implementation of a variety of spectral ageing models. 
Assuming that the radiative losses  dominate over the expansion losses, these models compute, for a spatially resolved radio source,  the temporal evolution of a power-law synchrotron spectrum as a function of position in the source, accounting for the radiative losses of the plasma due to both synchrotron emission and inverse-Compton scattering of the cosmic microwave background (CMB) off plasma electrons in a magnetic field that does not vary with time. From the spectral shape at each position, the models provide the time since the particles located in that position were last accelerated; from the particle radiative ages,  the spectral age of the source can be inferred.
Among the models implemented in BRATS, we only considered models that assume a single injection of accelerated particles in the jet at a given epoch.
These models are suitable for spatially resolved spectral studies: on small scales, particles can be considered as affected by the same acceleration event.

We considered all the single injection models implemented in BRATS, namely Jaffe-Perola's model \citep{Jaffe_Perola_1973}, hereafter referred to as JP's model, Kardashev-Pacholczyk's model \citep{Kardashev_1962,Pacholczyk_1970}, hereafter referred to as KP model, and Tribble's model \citep{Tribble_1991, Tribble_1993}.
Both JP's and KP's models assume a uniform magnetic field distribution along the source.
While in JP's model each electron is subject to multiple scattering events, that randomise its pitch angle (i.e., the angle between the electron velocity vector and the magnetic field), in KP's model each electron has a constant pitch angle, implying that more higher-energy electrons can be found at small pitch angles.
As a result, JP's model spectrum is steeper than the KP counterpart at high frequencies, due to the ability of high-energy electrons at small pitch angles to radiate at higher frequencies \citep{Katz-Stone_Rudnick_1997,Harwood_2013,Brienza_2020}.
On the other hand, Tribble's model assumes the magnetic field to be turbulent, and describes it as a Gaussian random field; in the weak-field, strong-diffusion case, the magnetic field distribution is drawn from a Maxwell-Boltzmann distribution within each volume element \citep{Tribble_1991,Hardcastle_2013}.

\begin{figure*}
\begin{multicols}{2}
   \includegraphics[width=1.025\linewidth]{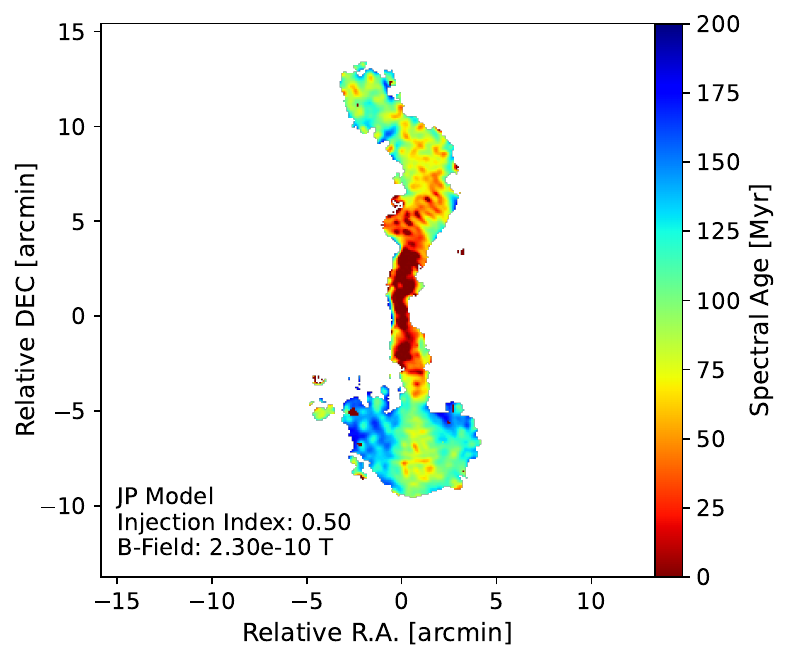}\par
   \includegraphics[width=1.03\linewidth]{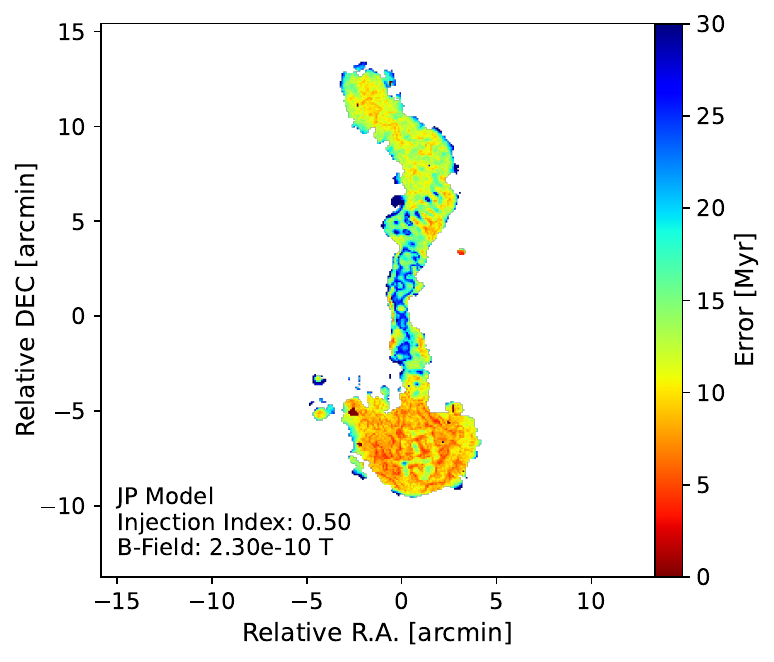}\par
\end{multicols}
\begin{multicols}{2}
   \includegraphics[width=\linewidth]{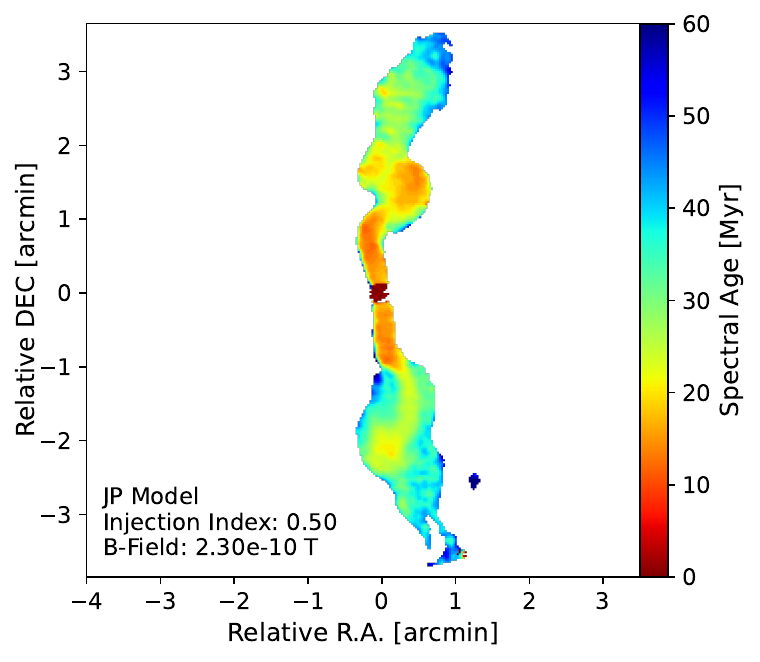}\par
   \includegraphics[width=\linewidth]{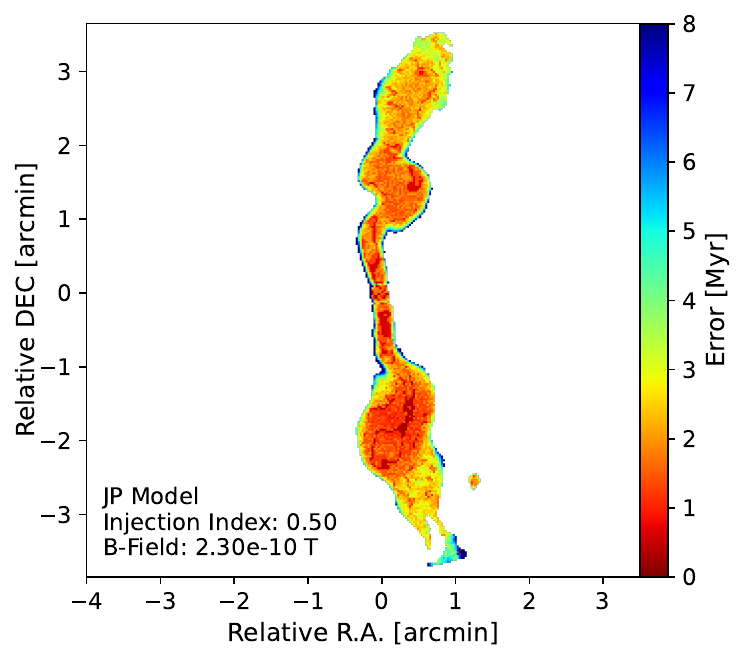}\par
\end{multicols}
  \caption{Spectral ageing maps obtained by fitting JP's model to the intensity maps, and corresponding maps of the uncertainty. Upper, left panel: map of the spectral age of 3C~449 between 145, 1365, and 1485 MHz, with the angular resolution of $20.0\arcsec \times 20.0\arcsec$, by means of the BRATS software package. 
  %Overall, the spectral age increases with the distance to the radio core. 
  Upper, right panel: map of the upper error on the spectral age shown in the left panel. Lower, left panel: map of the spectral age of 3C~449 between 145, 1365, 4985, and 8485 MHz, with angular resolution of $6.0\arcsec \times 6.0\arcsec$. Lower, right panel: map of the upper error on the spectral age shown in the left panel.
  %The parameters of JP's model are reported in the legend.
  In all the maps, the assumed magnetic field intensity is $B=B_{\rm eq}=2.3\, \mu$G.}
  \label{fig:BRATS_maps}
\end{figure*}

\begin{table*}[ht]
\caption{Results of the model fitting for the spatially resolved spectral analysis performed on the intensity maps at 145-1485 MHz, with $20\arcsec\times 20\arcsec$ angular resolution (rows 1--3), and on the intensity maps at 145-8485 MHz, with $6.0\arcsec\times 6.0\arcsec$ angular resolution (rows 4--6).}              % title of Table
\label{tab:age_statistics}      % is used to refer this table in the text
\centering                                      % used for centering table
\begin{tabular}{c c c c c c c c c c c c }          % centered columns (4 columns)
\hline\hline       % inserts double horizontal lines
Frequency range & Model & B & $\alpha_{\rm inj}$ & $\chi^2_{\rm red}$ &   \multicolumn{5}{c}{Confidence bins} & Rejected & Median\\

(MHz) &  & ($\mu \mathrm{G}$) & & & < 68\% & 68-90\% & 90-95\%& 95-99\% & 99\% & &  confidence \\ 
\hline
%     ~~ & ~~ & ~~ & ~~ & ~~ & ~~ & ~~ & ~~ & ~~ & ~~ & ~~ \\ 
$145-1485$     & JP & 2.3 & 0.50 & 3.82 & 43\% & 21\% & 7\% & 12\% & 17\% & No &  < 68\% \\  
$145-1485$     & KP & 2.3 & 0.50 & 3.83 & 43\% & 21\% & 7\% & 12\% & 17\% & No &  < 68\% \\
$145-1485$     & Tribble & 2.3 & 0.50 & 3.83 & 43\% & 21\% & 7\% & 12\% & 17\% & No &  < 68\% \\
%\hline
     ~~ & ~~ & ~~ & ~~ & ~~ & ~~ & ~~ & ~~ & ~~ & ~~ & ~~ \\ 
$145-8485$     & JP & 2.3 & 0.50 & 21.30 & 38\% & 27\% & 10\% & 14\% & 11\% & No &  < 68\% \\  
$145-8485$     & KP & 2.3 & 0.50 & 21.42 & 36\% & 28\% & 10\% & 15\% & 11\% & No & < 68\% \\
$145-8485$     & Tribble & 2.3 & 0.50 & 21.44 & 36\% & 28\% & 10\% & 15\% & 11\% & No &  < 68\% \\
\hline \hline
\end{tabular}
\begin{flushleft} 
\textbf{Notes.} Column 1: Frequency range of the intensity maps used for the spectral analysis; Column 2: Model applied for the spectral analysis; Column 3: Equipartition magnetic field intensity; Column 4: Injection index that best describes the source (i.e., that minimises the mean $\chi^2_{\rm red}$ over the source); Column 5: Value of the $\chi^2_{\rm red}$ that corresponds to the injection index of Column 4; Columns $6-10$: Fraction of adaptive regions that falls in each confidence bin; Column 11: Rejection or non-rejection of the model; Column 12: Median confidence level at which the model cannot be rejected over the entire source.
\end{flushleft}
\end{table*}
We applied JP's, KP's, and Tribble's models to both higher ($6\arcsec \times 6\arcsec$) and lower ($20\arcsec \times 20\arcsec$) angular resolution intensity maps; therefore, we obtained two spectral age maps per model.
The maps obtained with JP's model are shown in Fig.~\ref{fig:BRATS_maps}; those obtained with KP's and Tribble's models are reported in Appendix \ref{app:tribble}.
%Fig.~\ref{fig:sa_Tribble}.

In all of these figures, the first map (upper, left panel), obtained from the intensity maps at 145, 1365, and 1485 MHz, has an angular resolution of $20\arcsec \times 20\arcsec$, and an extension of $\approx 20\arcmin$ ($\approx 430$ kpc); the second map (lower, left panel), obtained from the intensity maps at 145, 1365, 4985, and 8485 MHz, has an angular resolution of $6\arcsec \times 6\arcsec$ and an extension of $\approx 7\arcmin$ ($\approx 150$ kpc).
The lower resolution, $20\arcsec \times 20\arcsec$ spectral age map was obtained from intensity maps at three frequencies only.
This restriction may introduce uncertainties in the spectral age due to the sparse sampling of the radio spectrum, but preserves the large scale ($\approx 20\arcmin$) of the map.
Indeed, including in the analysis our additional intensity maps at 4985 and 8485 MHz, whose field of view is limited to $\approx 10\arcsec$ and $\approx 7\arcsec$, respectively,
would significantly reduce the size of our spectral age map.

The spatially resolved spectral analysis performed with BRATS can be summarized as follows.
With an assumed constant magnetic field intensity,
BRATS initiates the spectral modelling with a power-law electron energy distribution, $N(E) = N_0 E^{-p_{\rm inj}}$, that generates a synchrotron spectrum $S(\nu) = C E^{-\alpha_{\rm inj}}$, with $\alpha_{\rm inj}=(p_{\rm inj}-1)/2$ and $C$ a constant of proportionality  \citep{Harwood_2013}. 
To carry out the spectral analysis, BRATS divides the source in a number of ``regions''. 
%Each region contains $\sim 2$ pixels, and has a size of $10\arcsec \times 10\arcsec$. 
In each region, the initial spectrum evolves with time, developing a spectral break and a steepening of its higher-frequency portion, according to the spectral ageing model chosen for the analysis. The time $\tau$ at which the chosen model best fits the spectrum of a given region is selected as the spectral age of that region according to that model. The final result is thus a map of the spectral age of the source for each model considered in the analysis. 

To model the spatially resolved spectrum, we employed  JP's, KP's, and Tribble's models assuming an energy injection index $\alpha_\mathrm{inj} = 0.50$ (i.e., a particle injection index $p_\mathrm{inj} = 2.0$).
This value of the injection index is the expected lower limit for particles accelerated through the diffusive shock acceleration mechanism in strong, non-relativistic shocks; it is consistent at 3$\sigma$ level with the flattest spectral index that we measure in the source (see Sect.~\ref{sec:spec_index_maps}).

A key ingredient of JP's, KP's, and Tribble's models
is the intensity of the magnetic field, $B$, in the region of interest. 
The most recent estimates of the magnetic field of 3C~449 are those made by  \citet{Hardcastle_1998} and \citet{Croston_2003}  for the southern lobe of the source.
Specifically, these authors found that the equipartition magnetic field evaluated under the assumption that only the relativistic, synchrotron-emitting electrons contribute to the internal lobe pressure, $B_{\rm eq}=2.3\,\mu$G, renders the lobe underpressured with respect to the ambient medium. To achieve pressure balance, the authors propose that (i) either magnetic domination holds in the southern lobe, and $B=12\,\mu$G, or (ii) particle domination holds,  
%with a dominant contribution to the lobe pressure provided by non-radiating particles, 
and $B=0.2\,\mu$G.  However, from the measured upper limit to the inverse-Compton X-ray flux produced by CMB photons scattered off synchrotron-emitting electrons in the lobe region, the authors derive a lower limit to the magnetic field, $B \ge 0.8\,\mu$G. This limit seems to rule out the above particle-dominated scenario case (with $B=0.2\, \mu$G), unless the particles contributing to the pressure are non-relativistic electrons, relativistic protons, or entrained and heated thermal material, which do not contribute to the X-ray emission.
%not contributing to the observed X-ray emission.}

Previous estimates of the equipartition magnetic field intensity by \citet{Perley_1979} and \cite{Andernach_1992} in selected points of the source, by assuming equal energies in relativistic electrons and protons, 
%
%%%%%($\kappa=1$);
%
showed that $B_{\rm eq}$ is not constant across the source, but decreases from the radio core outwards in a similar fashion in the northern and southern source. 
Specifically,  \citet{Perley_1979}, in their higher-resolution images, found a symmetric decrease  of $B_{\rm eq}$ by $\sim 30-40\%$ (from $10.6-11.3 \, \mu$G to $7-7.3\, \mu$G)  from the core out to the inner lobes, at $\sim 3'$, highlighting that $B \propto x^{-1}$, with $x$ the cross sectional radius of the jet;  \citet{Andernach_1992}, from their lower-resolution images, found a fairly symmetric decreasing trend of the magnetic field intensity between $3'$ and $9'$ of the core (where $B_{\rm eq}$ drops from $\simeq 2-2.2 \, \mu$G down to $\simeq 1-1.2\, \mu$G,\footnote{The $B_{\rm eq}$ values from \citet{Andernach_1992} are reported here with the appropriate correction to the cosmology adopted in this work.}), followed by a further decrease in the northern source, where $B_{\rm eq}\simeq 0.63\, \mu$G at $15'$ north of the core, at the edge of the northern tail. A slight asymmetry emerges on large scales, with the northern tail having a magnetic field intensity a factor $\lesssim 2$ lower than the southern lobe. 
The values of $B_{\rm eq}$ estimated by \citet{Perley_1979} and \cite{Andernach_1992} at $3'$ differ by a factor $\gtrsim 3$.

For our analysis, we first assumed that equipartition holds for the full source, and adopted the equipartition magnetic field strength estimated for the southern lobe by \cite{Croston_2003}, $B_{\rm eq}=2.3 \, \mu$G, as the reference value for the isotropic intensity (in JP's and KP's model) and for the mean intensity (in Tribble's model) of the magnetic field in each source region.

Even though a careful mapping of the spectral age across the source would certainly require the estimate of $B_{\rm eq}$ at different core distances, adopting the southern lobe's magnetic field value $B_{\rm eq}=2.3\, \mu$G all over the source enables us to estimate the spectral age of the southern lobe, and to provide an upper limit to the spectral age both in the northern tail, where $B_{\rm eq}$ might be lower than in the southern lobe, and in the jets, where $B_{\rm eq}$ is likely higher than that, according to the trends derived by \cite{Perley_1979} and \cite{Andernach_1992}; we will come back to this aspect at the end of this Section.
In the following, we describe the results of this analysis.

For comparison purposes, we then repeated the same analysis in a scenario of magnetic domination that guarantees the lobe-ambient  pressure balance, by adopting $B=12 \, \mu$G \citep{Croston_2003}. 
These results are reported in Appendix~\ref{app:12muG}.

Table \ref{tab:age_statistics} shows the results of the fitting procedures with JP's, KP's, and Tribble's model, with the equipartition magnetic field, $B_{\rm eq}=2.3\, \mu$G.
For each region, BRATS performs a $\chi^2$ test to evaluate the goodness of the spectral modelling and associates a confidence level to each region's $\chi^2$ value. The cut-off for the model rejection is set to a confidence level $\geq 95\%$.
If more than half of the regions fall in the rejection zone, the model over the source is classed as ``poor fit'' and is rejected \citep{Harwood_2013}.
In our case, with one degree of freedom, the median of the confidence levels over the entire source falls in the < 68\% confidence-level bin for both models: therefore, the two models provide a description of the source spectrum that cannot be rejected at $>95\%$  confidence level.

The upper panels of Fig.~\ref{fig:BRATS_maps} shows the results that we obtained 
%distribution of the spectral age of 3C~449 that we obtained from 
by fitting JP's model to the intensity maps at 145, 1365 and 1485 MHz, with an angular resolution of $20\arcsec \times 20\arcsec$: the left panel displays the map of the spectral age; the right panel shows the corresponding map of the upper error\footnote{The error on the spectral age is asymmetric, and we report here the upper error, following the suggestion by \citet{Harwood_2013}.}. 
The spectral age is on average $\tau \lesssim 20$ Myr in the internal part of the jet flow out to $\approx 3\arcmin$ from the radio core in the northern jet (i.e., out to region N3), and out to  
$1\arcmin$ from the radio core in the southern jet; similar ages appear again in the internal part of region S2, at $\simeq 1.5-2.5 \arcmin$ from the core.
The spectral age then increases to  $\tau \simeq 30-50$ Myr in the inner flow on larger scales (i.e.,  in regions N3 and S2-S3). 
This internal, lower age flow appears to be surrounded by a layer of higher age,  $\tau \simeq 70-110$ Myr, which becomes the dominant component beyond those scales.

The northern tail and the southern lobe show a different spectral age distribution, confirming the North-South asymmetry that we detected in the spectral index maps and in the corresponding profiles (see Sect.~\ref{sec:spec_index}).
The northern tail shows a spectral age smoothly growing with the core distance, from $\tau \simeq 50$ Myr right beyond region N4 to $\tau \simeq 90-150$ Myr at the edge of the detected emission.
In the southern lobe, a central region where $\tau \simeq 70-110$ Myr is surrounded by an older shell whose age is as high as $\tau \simeq 150-200$ Myr.

The lower panels of Fig.~\ref{fig:BRATS_maps} show the results of the spectral age analysis that we performed by fitting JP's model to the $6.0\arcsec \times 6.0\arcsec$ angular resolution intensity maps at  145, 1365, 4985, and 8485 MHz: the left panel shows the map of the spectral age; the right panel shows the corresponding upper error map.
The full extension of this map is smaller ($\approx 7 \arcmin$, from N3  to S3) than that of the map on the upper panel, due to the smaller extensions of the 4985 and 8485 MHz surface brightness maps.

The obtained age distribution displays a radial dependence and a transverse gradient similar to those of the lower-resolution map, shown in the upper panel, on comparable scales. 
%, in terms of radial dependence of the spectral age. 
However, differences in the values of the spectral age emerge, owing to the larger set of frequencies used for the analysis and the different angular resolution of the maps.
For the northern jet, the internal part of the plasma flow shows ages  $\tau \simeq 10-20$ Myr out to $\approx 2'$ from the radio core (i.e., out to region N2); for the southern jet, the internal part shows ages $\tau \simeq 10-20$ Myr out to $\approx 1'$ from the radio core (i.e., out to region S1), and then ages $\tau \simeq 20$ Myr out to  $\approx 2.5'$ from the core (i.e., out to region S2). 
A layer of older particles, with $\tau \simeq 30-50$ Myr, appears to enshroud the younger, internal jets; the older plasma is the dominant component of both jets beyond regions N2 and S2; in the northern jet, region N3 still shows some internal, younger structure, however with a patchy distribution of ages.

Summarizing, we consider as the spectral age of the northern source side the age of the oldest structure in the northern tail, corresponding to $\tau \simeq 150$~Myr. 
Similarly, we consider as the spectral age of the southern source side the oldest structure of the southern lobe, corresponding to $\tau\simeq 200$ Myr.  
The spectral age of the full source can thus be assumed as the higher of the spectral ages of the northern and southern source parts, namely $200$ Myr.
We note that the spectral ages at the northern edge of the tail (N5) and at the southern edge of the lobe (S4), which are relevant when considering the source expansion (see Sect.~\ref{sec:mach_number}), are both equal to $\tau_{\rm sp} \simeq 150$ Myr.
By applying KP's and Tribble's model to the same higher and lower angular resolution intensity maps, we obtained results consistent with those obtained with the JP modelling,
within the uncertainties. We show the corresponding spectral age maps in Appendix \ref{app:tribble}.

We recall that, for a given break frequency,  the spectral age in units of Myr depends  on the magnetic field intensity as $\tau\propto B^{1/2}/(B^2+B_{\mathrm CMB}^2)$ \citep{van_der_Laan_1969,Leahy_1991,Parma1999}, where $B$ is the magnetic field strength in the source and $B_{\rm CMB} = 3.18\times(1+z)^2 \, \mu$G  is the equivalent magnetic field intensity of the cosmic microwave background radiation, which in turns depend on the redshift $z$ of the source. 
While we obtained a spectral age $\tau \simeq 150$ Myr with the equipartition magnetic field strength $B_{\rm eq} = 2.3\, \mu$G,  the spectral age significantly decreases in the out-of-equipartition scenarios by \citet{Croston_2003} mentioned above. 
Specifically, we obtain $\tau_{\rm sp} \simeq 35$~Myr in a magnetic domination scenario with $B= 12\,\mu$G (see Appendix\,\ref{app:12muG}), while we estimate  $\tau_{\rm sp} \lesssim 124$~Myr for the inverse-Compton limit $B\ge 0.8\, \mu$G and $\tau_{\rm sp} \simeq 65$~Myr for the particle-dominated scenario with $B=0.2\, \mu$G.
The results we find are expected from the fact that the $B/B_{\rm CMB}$ ratio determines the function $\tau (B)$;
for 3C~449, with $B=B_{\rm eq}=2.3\, \mu$G, the spectral age has its highest possible value at $B\approx B_{\rm eq}$ \citep[see][their Fig.~4]{Parma1999}. For $B\lesssim B_{\rm eq}$ and $B\gtrsim B_{\rm eq}$, the inferred age is thus lower.
This also implies that possible north-south asymmetries in the magnetic field intensities at the source edges, with the northern tail having a magnetic field lower than that of the  southern lobe by a factor $\lesssim 2$ \citep[as suggested by][]{Andernach_1992}, would result in a $\lesssim 20\%$ lower spectral age for the northern tail. 
The spectral age $\tau_{\rm sp}\simeq 150$ Myr is thus a fairly robust upper limit to the source spectral age.
Finally, if the magnetic field intensity had a decreasing trend from the core outwards, reaching the lowest value $B_{\rm eq}=2.3\, \mu$G in the lobes, the spectral age of the higher-field regions, as the jets and the inner lobes, would be even lower than those we determined by associating the lobe's value to the entire source (see, e.g., Appendix \ref{app:12muG}); the increasing trend of the spectral age of the particles from the core outwards would thus be even steeper than observed in our spectral age maps. 

The spectral age we estimated for the source assuming $B_{\rm eq}=2.3\, \mu$G, $\tau_{\rm sp}\simeq 150$~Myr, is higher than the previous estimates of the equipartition spectral age of the source by \citet{Andernach_1992} and \citet{Parma1999}.
\citet{Andernach_1992} estimated $\tau \simeq 46\,h^{-3/7}$~Myr and $\tau \simeq 34\, h^{-3/7}$~Myr 
%at the northern and southern edge of the source, 
 at distances from the core of $15\arcsec$ to the north   (i.e., beyond the northern edge of the source in our maps) and $9'$ to the south, respectively, which correspond to $\tau \simeq 54$ Myr and $\tau \simeq 40$ Myr, respectively, with the value of $H_0$ we adopted in this work. \citet{Parma1999} estimated a spectral age $\tau \simeq 74$ Myr, which corresponds to {$\tau \simeq 87$ Myr} with our value of $H_0$.
The discrepancy between their age estimates at equipartition and ours might arise from the fact that both groups made use of equipartition magnetic fields lower than ours (thus much lower than $B_{\rm CMB}$), and of radio data at higher frequencies and lower angular resolution than ours; furthermore, in \citet{Parma1999}, the authors estimated the break frequency by applying the JP model to average radio spectra sampled at two frequencies only. 

Our estimate of the spectral age at the source edges is broadly consistent with the estimate of the spectral age of the FRI radio galaxy 3C~31 by  \citet{Heesen_2018}. With a linear size of 1.1 Mpc, 3C~31  is twice as large as 3C~449; these authors estimate its spectral age as $\tau \approx 200$ Myr, $\approx 25\%$ higher than the age of 3C~449. 
 
%------------
\section{Discussion}
\label{sec:discussion}
%------------

%------------
 \subsection{Spectral age and average expansion speed}
\label{sec:mach_number}
%------------

The spectral age 
%$\tau \sim 22-131$ Myr 
that we estimated at the edges of the northern tail and the southern lobe in Sect.~\ref{sec:spectral_age}, $\tau_{\rm sp}$,  can be used to constrain the time-averaged expansion speed of the radio-emitting plasma through the external medium.
This estimate requires the assumption that the above spectral age corresponds to the dynamical age, $\tau_{\rm dyn}$, of the source, i.e.,\ $\tau_{\rm sp}=\tau_{\rm dyn}$.
This assumption holds if the particles in the plasma flow were not subject to any in-situ reacceleration or adiabatic loss process during their propagation. In-situ reacceleration would tend to make the electron population younger than it actually is, while adiabatic losses have the opposite effect, because they shift the entire spectrum (and so the break frequencies) to lower frequencies \citep{Katz-Stone_Rudnick_1997}.

The dynamical age represents the time that has elapsed since the particle flow was initiated and the radio jet started to propagate through the ambient medium.
An estimate of the dynamical age, $\tau_{\rm dyn}$, can be done by assuming that each source side has grown to a size $D$ by expanding through the ambient medium with average expansion speed $v_{\rm exp}$:
\begin{equation}
    \tau_{\rm dyn} = \frac{D}{v_{\rm exp}} \, .
    \end{equation}
The dynamical age can be re-written as a function of the sound-crossing time, $\tau_{\rm cs} = D/c_{\rm s}$ \citep{Birzan_2008,Wykes_2013}, and of the average Mach number, $M$, as
\begin{equation}
    \tau_{\rm dyn}  = \frac{D}{M \cdot c_{\rm s}} = \frac{\tau_{\rm cs}}{M} \, ,
\end{equation}
where $c_{\rm s}$ is the sound speed in the X-ray emitting ambient medium, computed as 
\begin{equation}
    c_{\rm s} = \sqrt{\frac{\gamma kT}{\mu m_\mathrm{AMU}}} \, , 
\end{equation}
with $\gamma$ the adiabatic index, $k$ the Boltzmann constant, $T$ the temperature of the gaseous medium, $\mu$ its mean molecular mass, and $m_{\rm AMU}$ the atomic mass unit.
For 3C~449, adopting a mean temperature $T=1.14\times 10^7$~K (i.e., $kT= 0.98 \, \mathrm{keV}$), derived by \citet{Croston_2003} for the X-ray emitting intergalactic medium of the host galaxy group within a region of radius $\simeq 420\arcsec$ ($\simeq 150$ kpc), and assuming  $\gamma= 4/3$, the mean sound speed in the medium is $c_{\rm s} = 458$ km/s.
Because of the spatial asymmetry of the source, the sound crossing time is slightly different for the two source sides.
Specifically, the estimates of 290 kpc as the distance travelled by the plasma from the core to the edge of the northern tail, and of 200 kpc as the distance travelled by the plasma from the core to the edge of the southern lobe yield sound crossing times $\tau_{\rm cs, N} = 619$ Myr and $\tau_{\rm cs, S} = 426$ Myr for the northern and southern source parts, respectively.\footnote{We estimated the extension of the northern jet as the sum of the distance of the core to the beginning of region N5, computed along the N-S direction, and the extent of region N5 itself, computed along its symmetry axis, roughly aligned to the NE-SW direction; our estimate agrees with the estimate by \citet{Croston_2003}. 
Similarly, we estimated the extension of the southern jet as the distance of the core to the edge of the southern lobe, computed along the N-S direction.}

In these estimates, the inclination of the source to the plane of the sky, which is $<15\degree$ according to \cite{Feretti_1999}, is neglected. 
If the inclination to the plane of the sky of the lobes of 3C~449 were higher than the inclination of the inner jets, the actual sound-crossing time would be higher.
The estimation for the northern jet agrees with the dynamical age of >500 Myr found by \cite{Croston_2003} with the same jet length of 290 kpc.

The equivalence of spectral and dynamical age,  $\tau_{\rm sp}=\tau_{\rm dyn}$, with the spectral age $\tau_{\rm sp} \sim 150$ Myr derived by assuming equipartition (see Sect.~\ref{sec:spectral_age}), yields the average expansion Mach number of the radio source over its lifetime:
\begin{equation}
%    M = \frac{\tau_{cs}}{\tau_{\rm sp}} = 5.5 - 7.1 \, ,
    M = \frac{\tau_{\rm cs}}{\tau_{\rm sp}}.
\end{equation}
We obtained $M_{\rm N} = 4.1$ for the northern part of the source, and $M_{\rm S} = 2.8$  for the southern part, implying supersonic expansion speeds $v_{\rm exp, N}= 6 \times 10^{-3} \, c$ and $v_{\rm exp, S}= 4 \times 10^{-3} \, c$ towards the north and towards the south, respectively.
%for the northern and southern parts,

As shown in Sect.~\ref{sec:spectral_age},  in the out-of-equipartition scenarios the spectral ages are always lower than the equipartition spectral age, implying supersonic flows with Mach numbers even higher than those found for the equipartition scenario.
Specifically, the magnetic field $B=12 \, \mu$G, that yields $\tau_{\rm sp}\simeq 35$~Myr (see Sect.~\ref{sec:spectral_age} and Appendix \ref{app:12muG}), implies $M_{\rm N}\simeq 17.6$ and $M_{\rm S}\simeq 12.1$; the magnetic field limit  $B\ge 0.8 \, \mu$G, that yields $\tau_{\rm sp}\lesssim 124$~Myr, implies  $M_{\rm N}\gtrsim 5.0$ and $M_{\rm S}\gtrsim 3.4$, while $B=0.2 \, \mu$G, that  yields $\tau_{\rm sp}\simeq 65$~Myr, implies $M_{\rm N}\simeq 9.5$ and $M_{\rm S}\simeq 6.5$.

Observationally, no signature of the strong shocks that are expected for highly supersonic flows was detected in studies of the X-ray emitting gaseous environment of 3C~449 on spatial scales comparable to the full radio source, beyond which the shock is expected to be \citep[][]{Hardcastle_1998, Croston_2003}. 
However, we acknowledge that the sensitivity of ROSAT was not sufficient, and that of XMM-Newton (the only instrument currently covering the appropriate spatial scales for 3C~449) may not be sufficient to detect a shock front in the very outer regions of the host galaxy group; therefore, the lack of detection of shock signatures alone does not guarantee that the plasma flow is subsonic.
On the other hand, the pressure-balance arguments by \cite{Croston_2003} (see Sect.~\ref{sec:spectral_age}) do not support highly overpressured lobes, required for shock generation. Therefore, it is possible that the source is currently expanding with subsonic speed. In this case, $\tau_{\rm dyn}>\tau_{\rm cs}$, and the dynamical  age would exceed the spectral age by a factor of $\gtrsim 3-4$.

Similar results were obtained for 3C~31: \cite{Heesen_2018} inferred a time-averaged Mach number $M\approx 5$ from the advection time scale of the oldest visible plasma of the source, assuming no {\it in situ} particle acceleration on large scales.
However, as claimed by the authors, in 3C~31 a supersonic flow is not in agreement with ram pressure balance arguments, which instead suggest a subsonic flow, at least at the current epoch. 

For FRI radio galaxies, subsonic flows on large scales are also suggested by recent 3D MHD numerical simulations: \cite{Massaglia_2019, Massaglia_2022} showed that radio galaxies with FRI morphology can be generated by low-power, magnetised jets that propagate in a stratified medium; even though the jet is initially supersonic, a transition to subsonic flow occurs as soon as non-axisymmetric modes develop, which cause the jet disruption and the formation of distorted plumes; the jet head velocity is thus subsonic for a significant fraction of the source lifetime.

In the equipartition scenario, dynamical ages, even when accurately estimated by accounting for the proper density profile of the ambient medium, are actually often found to exceed the spectral ages in large-scale radio galaxies.
For instance, in a sample of low-luminosity radio sources of the FRI and FRII type, the correlation between dynamical and spectral ages found by \cite{Parma1999} indicates that the dynamical ages evaluated from ram-pressure arguments are in general larger than the spectral ages derived from a model of aged power-law electron distribution in a uniform magnetic field by a factor $\approx 2-4$,  depending on the value of the index $\beta$ in the $\beta$-model used to describe the density profile of the external medium. 
The authors show that the spectral age can increase to become comparable to the dynamical age only when $B_{\rm eq}/B_{\rm CMB}\gtrsim 2$ and $B<B_{\rm eq}$ (see their Fig.~4). 
However, this is not the case for 3C~449, where $B_{\rm eq}/B_{\rm CMB}\simeq 0.7$, and any deviation from equipartition yields spectral ages lower than the equipartition spectral age, as we illustrated in Sect.~\ref{sec:spectral_age}, preventing the solution of possible discrepancies between dynamical and spectral ages.

Summarizing, in a scenario where 3C~449 is expanding with subsonic speed and the magnetic field is constant in time and has either a uniformly distributed intensity across the source (as assumed in Sect.~\ref{sec:spectral_age}) or a gradually decreasing intensity with the distance to the radio core, as suggested by \citet{Perley_1979} and \citet{Andernach_1992} (see also \citet{Heesen_2018} for the case of 3C~31), and the spectral age of the source is evaluated from the lobe's magnetic field, a discrepancy between the spectral age and the dynamical age,  with $\tau_{\rm dyn} \gtrsim (3-4)\, \tau_{\rm sp}$, emerges: this discrepancy may in principle be solved by assuming that reacceleration processes take place in the source, naturally lowering the spectral age of the radiating particles. However, evidence of acceleration processes is compelling only in the inner $\lesssim 20$ kpc in both jets; therefore, reacceleration in these regions would probably not significantly impact the spectral age of the particles that have travelled out to the edge of the radio source, at $\approx 200$~kpc from the core. 

Alternatively, as mentioned in Sect.~\ref{sec:spectral_age}, the discrepancy between spectral and dynamical age 
might be solved by interpreting the steepening of the radio spectrum across the source as due to an inhomogeneous magnetic field with more complex structure, rather than to synchrotron ageing  \citep{Eilek_1996,Eilek_1997,Katz-Stone_Rudnick_1997}. In this scenario, the radiating particles might be much older than they look from the break frequency of the radio spectrum, and our spectral ages would no longer be meaningful. 
Exploring this scenarios is, however, beyond the scope of this paper.

%------------
 \subsection{Particle populations and acceleration mechanisms} 
 \label{sec:particle_population}
 %------------

Our LOFAR-VLA spectral index maps, shown in Sect.\ \ref{sec:maps}, 
enabled us to explore the properties of the particle population in 3C~449.
On the one hand, as mentioned in Sect.~\ref{sec:radial_profiles}, the radial evolution of the spectral index reveals a particle population that,  on scales larger than $\approx 18$~kpc, on average progressively loses energy mostly at higher energies as the distance from the radio core increases, without major reacceleration events, as expected for FR~I sources \citep[see, e.g., the case for 3C\,31,][]{Heesen_2018}.
On the other hand, the small-scale inhomogeneity of the spectral index distribution (see Sect.~\ref{sec:flat_spectra}), and the transverse spectral index profiles, which suggest a spine-sheath structure in the jet (see Sect.~\ref{sec:transverse_profiles}), 
indicate that, even on scales larger than $\approx 18$~kpc, energy losses are not exactly as expected in a model where a particle population acquires a power-law energy distribution in an acceleration event, and then radiates in a uniform, constant magnetic field while propagating downstream. 
Reacceleration processes may be at work, at least locally, and/or the magnetic field may show complex spatial variation across the jet.

%------------
 \subsubsection{Flat-spectrum substructures}
 \label{sec:flat_index}
 %------------

As far as the small-scale inhomogeneities in the spectral index distribution are concerned, as mentioned in Sect.~\ref{sec:flat_spectra} and shown in Fig.\,\ref{fig:SM_zoom}, the values of the $145-1365$ MHz spectral index are unusually low in significant, coherent portions of jet regions N1, N2, N3, S1, and S2, where $\alpha = 0.35-0.45$, and are even lower in a few, isolated substructures about the eastern jet turns of the northern, inner jet (in N1 and N2) and the western turn of the southern, inner jet (in S1), where $\alpha = 0.30-0.35$.
The very low values of $\alpha$ challenge the standard model of diffusive shock acceleration of particles by non-relativistic, strong shocks, which produces spectra with a ``universal'' index, $\alpha \geq 0.5$ \citep{Bell_I_1978}; however, the uncertainty on the spectral indices we measure, $\sigma_{\alpha}\simeq 0.06$, means that the indices are consistent with the theoretical lower limit $\alpha=0.5$ at the 2-3$\sigma$ level. 
Should these low values of spectral index be confirmed by future observations in the frequency range of a few hundred MHz, they may be indicative of acceleration of particles by processes able to generate very flat spectra: for instance, diffusive shock acceleration of particles by strong, relativistic shocks 
%in cold material 
or weak (i.e., barely supersonic) shocks in a hot material can produce spectral indices $\alpha=0.3-0.5$ \citep[e.g.,][]{Peacock_1981}; alternatively, very flat spectra may be the result of multiple-shock acceleration mechanisms, able to yield weakly inverted spectra with $\alpha \approx 0$  above the cutoff frequency and $\alpha \approx -0.5$ below the cutoff \citep[e.g.,][]{Schneider_1993,Melrose_1997}.
An alternative possibility is that the flattening of the spectrum reflects a local low-frequency spectral turnover in the frequency range of a few hundred MHz, resulting from synchrotron self-absorption.
However, current observations do not enable us to make any of these claims.  

Regardless of whether the spectral index is larger or lower than 0.5, the coherent, flatter spectrum portions of regions N1, N2, N3, S1, and S2, surrounded by steeper spectrum areas (see Fig.\,\ref{fig:SM_zoom}) call for a physical interpretation.
If the flattening were intrinsic, these regions would be the site of local particle reacceleration; in the spectral ageing  maps (see Fig.\, \ref{fig:BRATS_maps}), they clearly appear radiatively younger than their surroundings. 

We notice that these flat-spectrum regions are predominantly located in the same positions of the jet bends.
As a consequence, as shown in \cite{Davelaar2020}, 
%one possibility is that 
the re-heating of the particles may be due to kink instabilities, which are expected to arise when a jet undergoes changes in its direction of propagation.
Alternatively, the presence of these flat-spectrum areas could be attributed to the formation of secondary hotspots, a phenomenon commonly associated with jet deflection \citep{Scheuer1982, Hardee1990, Cox1991, Smith1984, Lonsdale1986, Horton2023}.
However, most models predict that these secondary hotspots form as a result of interactions between the jet terminal head and the external medium, and they are typically expected to persist over timescales of several million years \citep[see, e.g.,][]{Horton2023}.
This timescale is inconsistent with the age of tens to hundred million years we estimated for 3C\,449 (see Sect.~\ref{sec:spectral_age}). 
If secondary hotspots formed earlier in the jet’s evolution, specifically during its initial expansion into the external medium, they would likely have dissipated by now.
% PER LUCA: NON RITROVO LO STREAM SPLITTING SULLA MAPPA DI SB! FORSE SI VEDE NELLA MAPPA DI ALFA, MA NON SO SE SIA LA STESSA COSA.
Alternatively, \citet{Horton2023} suggest that these secondary hotspots may form as a consequence of the jet stream splitting into two or more parts following the jet interacting with the lobe boundary. 
The scenario shown in Fig.~5 of their paper for the timestep 131 resembles the situation seen in 3C\,449 in region N2.   
The question is whether such stream splitting happens within an already existing jet.
More broadly, the presence of such regions, combined with the apparent change in jet direction, raises the question of whether the jets are undergoing precession, as suggested by \cite{gower_1982} \citep[see, however,][for a different interpretation of the wiggles]{Lal_2013} . 

A distribution of spectral index similar to that observed in the flat spectrum regions of 3C~449 was reported in the comparable, $144-612$~MHz frequency range for the barbell-shaped, giant radio galaxy J223301$+$131502 \citep{Dabhade_2022}.
Specifically, in the northern portion of a $\sim 100$ kpc kink structure (see Fig.~6 of their paper), these authors found a spectral index which is as flat as $\alpha=0.36 \pm 0.08$ in the outer part of the kink and shows a positive gradient across the instability; they interpreted the kink as the site of particle acceleration due to a strong shock.
This situation strikingly resembles what we see in region N2 of 3C~449.

%------------
 \subsubsection{Spine-sheath spectral structure} 
 \label{sec:spine-sheath}
 %------------

As reported in Sect.\,\ref{sec:transverse_profiles}, and clearly visible in Fig.\,\ref{fig:SM_zoom}, the LOFAR–VLA spectral index map exhibits a significant transverse gradient in spectral index values.
This gradient is very pronounced in the inner $\approx 10$ kpc, where the spectral index increases from $\alpha \simeq 0.4$ in the jet spine to $\alpha \simeq 1.2$ toward the outer sheath, and becomes less pronounced as the distance from the core increases (see Fig.~\ref{fig:transverse_profiles}).
The origin of this steeper spectrum structure remains uncertain.
On one hand, we note that our LOFAR-VLA  spectral index maps were created by combining data acquired by two different radio telescope arrays, which have significantly different configurations, resulting in distinct $(u,v)$-coverages that could introduce artificial gradients in the spectral index map. 
%Future observations performed at hundreds of MHz will be essential to confirm the presence of a genuine spine-sheath structure.
%
%On the other hand, even though on larger scales, on larger scales, the spine-sheath structure also appears in the spectral index maps derived from a combination of VLA radio images at different frequencies. 
On the other hand, numerical simulations show that a shear layer enshrouding the jet spine can indeed develop in a plasma jet that propagates through a gaseous ambient medium \citep[e.g.,][]{Loken_1995,Massaglia_2019}: this layer propagates slower than the jet spine \citep[e.g.,][their Fig.~8]{Massaglia_2019}, and is composed of a mixture of particles that originates from the interaction of the spine itself with the backflow generated by the interaction of the jet head with the ambient medium.
However, the spectral properties of the shear layer are not well-constrained yet. 
Recent 3D RMHD numerical simulations of jets that account for the evolution of relativistic radiating particles show that relativistic jets can have a spine populated by younger particles, surrounded by a layer populated by older particles \citep[e.g.,][]{Mukherjee_2021}; however, a direct comparison between simulation and data is still beyond reach.
A transverse spectral index gradient similar to the one we observed in 3C~449 was found in other sources as well, including in the inner $\approx 70$~kpc structure of NGC~4869 \citep{Lal_2020}, in both jets of 3C~130 \citep{Hardcastle1999}, and in the inner southern jet and in both large scale jets of 3C~465 \citep{Bempong-Manful2020}.
Future observations performed at hundreds of MHz will be essential to confirm the presence of a genuine spine-sheath structure.

%------------
 \subsubsection{North-south jet asymmetries} 
 \label{sec:north-south}
 %------------

In the northern tail (region N5), the particle spectrum appears to evolve smoothly with increasing distance, gradually steepening as the jet becomes progressively fainter.
In contrast, the situation in the southern lobe (region S4) is different.
Fig.~\ref{fig:Spec_index_low} (left panel) illustrates that the spectral index in the central part of the lobe traces the continuation of the jet, which appears to terminate at a distance of 39\degree12\arcmin, where it likely ceases to propagate into the group intergalactic medium \citep[e.g.,][]{Hardcastle_1998}.
This morphology is likely due to a larger environmental density of the southern lobe compared to the northern tail, as suggested by \cite{Hardcastle_1998}.
%Asymmetries in the N-S direction are suggested in 
The southern jet may not have sufficient momentum to pierce through the denser surrounding medium, and it terminates, giving rise to a backflow. The redirected plasma
%, along with the jet material, 
expands predominantly in the East-West direction, possibly giving rise to the observed steep-spectrum structure that enshrouds the jet termination.
Indications of interaction between the radio jet and the X-ray emitting external medium are suggested by \cite{Hardcastle_1998} and \cite{Croston_2003}.

%------------
\section{Conclusions}
\label{sec:conclusions}
%------------

%%%%%%%%%%%%%%%%%%%%%%%%%%%%%%%%%%%%%%%
%%%%% new draft for the conclusione
%%%%%%%%%%%%%%%%%%%%%%%%%%%%%%%%%%%%%%%

In this paper, we investigated the extended emission and spectral properties of 3C~449 by means of new LOFAR observations at 145 MHz, with angular resolution of  $20\arcsec \times 20\arcsec$ and $6\arcsec \times 6\arcsec$, and archival VLA data at 1365, 1485, 4985, and 8485 MHz.
Our results can be summarized as follows.
\begin{itemize}
    \item The high sensitivity and angular resolution of the LOFAR data enabled us to recover the full extent of the previously imaged radio emission of 3C~449 with a resolution of 6.0\arcsec$\times$ 6.0\arcsec at 145 MHz, unprecedented for this source. The $20.0\arcsec \times 20.0\arcsec$ map reveals that the radio-emitting plasma bends from a North-East orientation toward the North at a declination of approximately $\delta \sim 39\degree \ 33\arcmin$.
    \item Combining LOFAR and VLA data, we obtained high resolution spectral index maps of 3C~449 in the 145-8485 MHz frequency range.   
The $145-1365$ MHz spectral index maps show that the source spectrum, on average, stays approximately constant ($\alpha \simeq  0.6$) in the inner, $\approx 40-50\arcsec$  ($\lesssim 20$~kpc) jets (regions N1 and N2), and then progressively steepens with distance from the radio core on both source sides, as expected for an FRI radio galaxy. Small regions with significantly flat spectra  ($\alpha \sim 0.30 - 0.35$, $\sigma = 0.063$) are detected in both jets (regions N1, N2, and S1).    
In the $20\arcsec \times 20\arcsec$ spectral index map, in the northern tail (region N5), we detected a patchy spectral index distribution whose average value is $\alpha \simeq 0.75$; in the southern lobe (region S4) we detected, for the first time, a steep-spectrum region with $\alpha \simeq 0.9-1.3$ surrounding a flatter-spectrum one with $\alpha \simeq 0.7 - 0.9$.
	\item Our two spectral index maps at 1365-4985 MHz and 4985-8485 MHz confirm the constancy of the spectral index within $\approx 40-50\arcsec$ from the core and a progressive spectral steepening beyond that distance.
	\item The average source spectrum  is consistent, within the uncertainties, with a single power law (with $\alpha \simeq 0.6$) over the full, 145-8485 MHz frequency range  out to $\approx 50\arcsec$ from the core (regions N1 and S1). This indicates that particle reacceleration processes compensate for particle radiative ageing in these regions. Beyond $\approx 50\arcsec$, a spectral curvature appears that makes the spectrum steeper at higher frequencies, both in the northern and in the southern part of the source.  
The increase of the spectral index  with the distance to the radio core is faster at higher frequencies, implying an increase of the spectral curvature with the distance to the core, in agreement with the expectations for particle synchrotron ageing in absence of significant reacceleration processes, and consistently with the FRI morphology of the source. 
The localized regions with very flat spectra ($\alpha \simeq 0.30-0.35$), that are flatter than the ``universal'' spectrum ($\alpha=0.5$) produced by diffusive shock acceleration by non-relativistic shocks, indicate either the possible presence of reacceleration processes whose nature cannot currently be established, or a spectral turnover due to synchrotron self-absorption. Additional low-frequency observations will be essential to clarify this behaviour.

     \item The $145-1365$~MHz spectral index map reveals, for the first time, a spine-sheath structure in the inner $\approx 1'$ (22~kpc) of both jets: here, a central flat-spectrum spine, with $\alpha \simeq 0.4$, is surrounded by a steeper spectrum layer whose spectral index increases up to  $\alpha \lesssim 1.3$, with the highest values reached in the external part of the sheath of the jet regions closer to the core. 
The spine-sheath structure continues out to $\approx 3'$ (65~kpc) in the northern jet and $\approx 2.5'$ (54~kpc) in the southern jet in the $145-1365$~MHz frequency range; beyond $\approx 1'$ from the core, it is also detected in both jets in the higher frequency ranges explored in this paper, confirming previous findings by \cite{Katz-Stone_Rudnick_1997} for both jets and by \cite{Feretti_1999} for the southern jet. The spine-sheath structure may be the signature of the interaction between jet and ambient medium that emerges in 3D MHD simulations.           

    \item In the standard ageing scenario, we derive the highest radiative age of the emitting particles under the assumption of equipartition and of a constant magnetic field throughout the source. The spectral age maps yields $\tau \lesssim 20$ in the internal jet parts out to $\approx 3\arcmin$ from the radio core in the northern jet, and out to  $1\arcmin$  and about $\simeq 1.5-2.5 \arcmin$ from the radio core in the southern jet.
    This internal, lower age structure is surrounded by a sheath of higher age, $\tau \simeq 70-110$ Myr, which becomes the dominant component beyond those scales.
    In the northern tail, the spectral age smoothly grows with the core distance, reaching $\tau \simeq 90-150$ Myr at the edge of the detected emission.
    In the southern lobe, we detect a central region with $\tau \simeq 70-110$ Myr, which is surrounded by an older region with an age of $\tau \simeq 150-200$ Myr. We estimated $\tau_{\rm sp}\simeq 150$~Myr as the spectral age of the source.

    \item Assuming that the spectral age of 3C~449 corresponds to its actual dynamical age, we found that the average propagation speed of the source over its lifetime is supersonic: the Mach numbers are $M_{\rm N} \sim 4.1$ for the northern jet and $M_{\rm N} \sim 2.8$ for the southern one when equipartition holds, and are higher in out-of-equipartition scenarios.
    This finding is against expectations from 3D RMHD simulations of FRI radio galaxies, where the jets stop to be supersonic early in their propagation. If the jets of 3C~449 were indeed subsonic on average, a discrepancy between dynamical and spectral age would appear, with $\tau_{\rm dyn} \gtrsim (3-4)\tau_{\rm sp}$. Particle reacceleration processes occurring in the jets may in principle lower the spectral age of the source; however, evidence for reacceleration processes is compelling only in the inner $\lesssim 20$~kpc of both jets, and not on larger scales, as required to solve the age discrepancy.

\end{itemize}

In conclusion, the new LOFAR maps at 145~MHz have significantly enhanced our understanding of the spectral index distribution in 3C~449, while also raising new questions on the nature of the source substructures observed in the $1365-8485$ MHz range. 
Addressing these questions will require additional observations at a few hundred MHz.

\begin{acknowledgements}
L.R. is funded by the Deutsche Forschungsgemeinschaft (DFG, German Research Foundation) – project number 443220636. 
L.R. and L.O. acknowledge partial support from the Italian Ministry of Education, University and Research (MIUR) under the Departments of Excellence grant L.232/2016.
The work of L.R. was part of his Master Thesis project at the University of Torino, Italy, in collaboration with the Kapteyn Astronomical Institute, Groningen, The Netherlands. 
L.O. acknowledges partial support from the INFN Grant InDark. M.J.H. acknowledges support from the UK Science and Technology Facilities Council [ST/ST/Y001249/1].
J.H.C. acknowledges support from the UK Science and Technology Facilities Council (ST/X001164/1, ST/J001600/1).
LOFAR is the Low Frequency Array, designed and constructed by ASTRON. It has observing, data processing, and data storage facilities in several countries, which are owned by various parties (each with their own funding sources), and which are collectively operated by the ILT foundation under a joint scientific policy. The ILT resources have benefited from the following recent major funding sources: CNRS-INSU, Observatoire de Paris and Universit\'e d'Orl\'eans, France; BMBF, MIWF-NRW, MPG, Germany; Science Foundation Ireland (SFI), Department of Business, Enterprise and Innovation (DBEI), Ireland; NWO, The Netherlands; The Science and Technology Facilities Council, UK; Ministry of Science and Higher Education, Poland; The Istituto Nazionale di Astrofisica (INAF), Italy.

This research made use of the Dutch national e-infrastructure with support of the SURF Cooperative (e-infra 180169) and the LOFAR e-infra group. The Jülich LOFAR Long Term Archive and the German LOFAR network are both coordinated and operated by the Jülich Supercomputing Centre (JSC), and computing resources on the supercomputer JUWELS at JSC were provided by the Gauss Centre for Supercomputing e.V. (grant CHTB00) through the John von Neumann Institute for Computing (NIC).

This research made use of the University of Hertfordshire high-performance computing facility and the LOFAR-UK computing
facility located at the University of Hertfordshire (\url{https://uhhpc.herts.ac.uk}) and supported by
STFC [ST/P000096/1], and of the Italian LOFAR IT computing infrastructure supported and operated by INAF, and by the Physics
Department of Turin University (under an agreement with Consorzio Interuniversitario per la Fisica Spaziale) at the C3S Supercomputing
Centre, Italy.

This research has made use of NASA’s Astrophysics Data System Bibliographic Services.
This work made use of Astropy:\footnote{http://www.astropy.org} a community-developed core Python package and an ecosystem of tools and resources for astronomy \citep{astropy:2013, astropy:2018, astropy:2022}.
\end{acknowledgements}

\bibliographystyle{aa.bst}
\bibliography{bibliography}

\begin{appendix}

%----------------------------------------------------------
\section{3C\,449 flux densities} \label{app:fluxes}
%----------------------------------------------------------

We report in Table \ref{tab:integrated_spectrum} the total flux densities displayed in the spectrum of Fig.\, \ref{fig:integrated_spectrum}.

\begin{table*}
\caption{Integrated flux densities of 3C 449 at different frequencies.}
\label{tab:integrated_spectrum} 
\centering                   
\begin{tabular}{ccc}         
\hline\hline      
Frequency & Flux density & Reference \\
(MHz) & (Jy) &  ~~   \\
\hline
     86 &  29.90 $\pm$ 1.10  &  \citet{Laing_Peacock_1980}\\
     145 &  14.56 $\pm$ 2.04  &  this work (LOFAR; $6.0 \arcsec\times6.0 \arcsec$)\\ 
     145 &  15.10 $\pm$ 2.11  &  this work (LOFAR; $20.0 \arcsec\times20.0 \arcsec$)\\ 
     178 &  16.20 $\pm$ 1.30  &  \citet{Kuehr_1981}\\
     178 &  11.07 $\pm$ 0.6  &  \citet{Kuehr_1981}\\
     178 &  12.54 $\pm$ 0.63  &  \citet{Laing_Peacock_1980}\\
     408 &  8.08 $\pm$ 0.53  & \citet{Andernach_1992}\\
     408 &  5.11 $\pm$ 0.44  &  \citet{Kuehr_1981}\\
     750 &  5.35 $\pm$ 0.15   &  \citet{Laing_Peacock_1980}\\
     750 &  5.67 $\pm$ 0.16   &  \citet{Kuehr_1981}\\
     750 &  5.4 $\pm$ 0.3   &  \citet{Kuehr_1981}\\
    1365 &  3.63 $\pm$ 0.07   &  this work (VLA; $20.0 \arcsec\times20.0 \arcsec$)\\
    1400 &  3.8 $\pm$ 0.2   &  \citet{Kuehr_1981}\\
    1400 &  3.4 $\pm$ 0.10   &  \cite{Kuehr_1981}\\
    1400 &  3.67 $\pm$ 0.12   &  \cite{Laing_Peacock_1980}\\
    1410 &  3.69 $\pm$ 0.10   & \citet{Andernach_1992}\\
    1485 &  3.42 $\pm$ 0.07   &  this work (VLA; $20.0 \arcsec\times20.0 \arcsec$)\\
    2695 &  2.18 $\pm$ 0.11  & \citet{Andernach_1992}\\
    2695 &  2.5 $\pm$ 0.05   &  \cite{Kuehr_1981}\\
    2695 &  2.48 $\pm$ 0.05   &  \cite{Laing_Peacock_1980}\\
    4750 &  1.33 $\pm$ 0.05   & \citet{Andernach_1992}\\
    4850 &  1.15 $\pm$ 0.17   &  \citet{Becker_1991}\\
    4985 &  1.18 $\pm$ 0.03   &  this work (VLA; $6.0 \arcsec\times6.0 \arcsec$)\\
    5000 &  1.38 $\pm$ 0.14   &  \cite{Kuehr_1981}\\
    5000 &  1.38 $\pm$ 0.07   &  \cite{Laing_Peacock_1980}\\
    8485 &  0.72 $\pm$ 0.02   &  this work (VLA; $6.0 \arcsec\times6.0 \arcsec$)\\
\hline
\end{tabular}
\end{table*}

%----------------------------------------------------------
\section{VLA archival maps} \label{app:vlamaps}
%----------------------------------------------------------

\begin{figure*}
  \centering
\begin{multicols}{3}
   \includegraphics[width=1.10\linewidth]{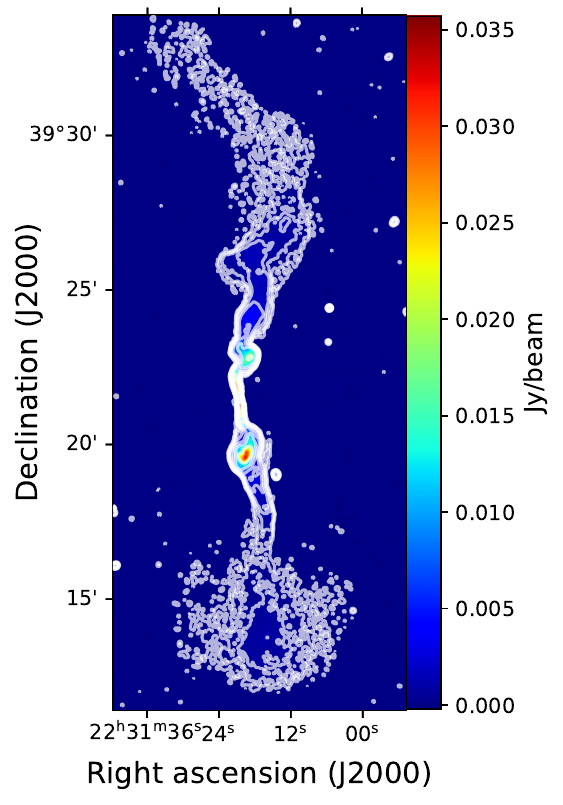}\par
   \includegraphics[width=0.94\linewidth]{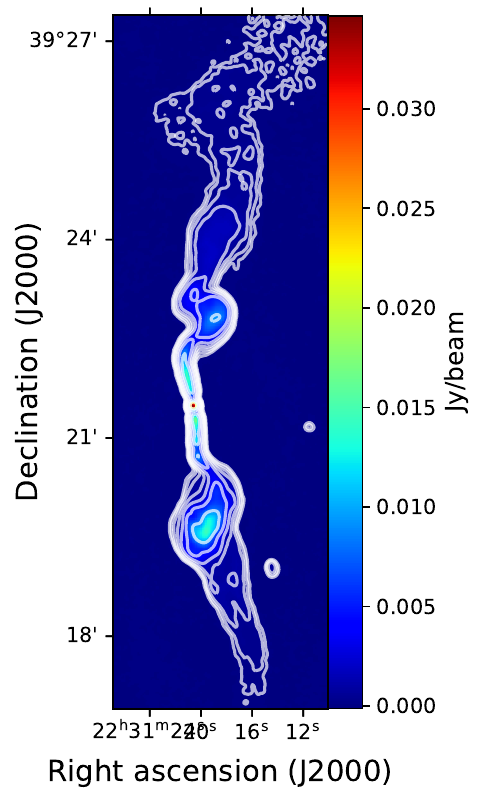}\par
   \includegraphics[width=0.99\linewidth]{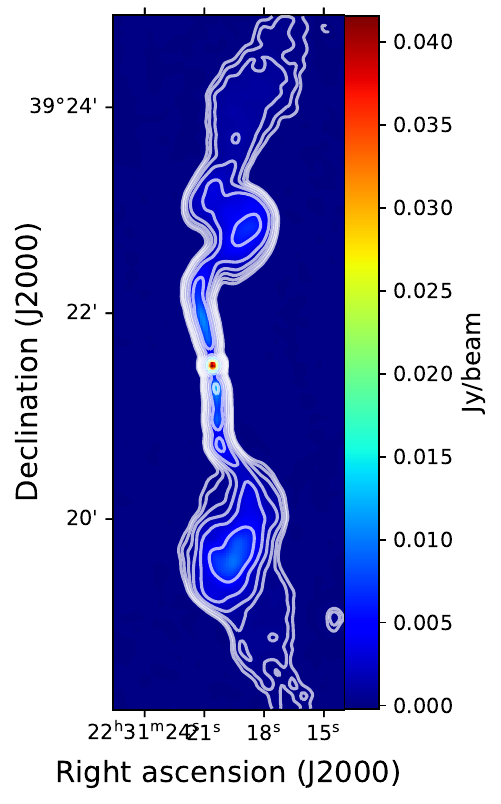}\par
\end{multicols}
\begin{multicols}{2}
   \includegraphics[width=0.85\linewidth]{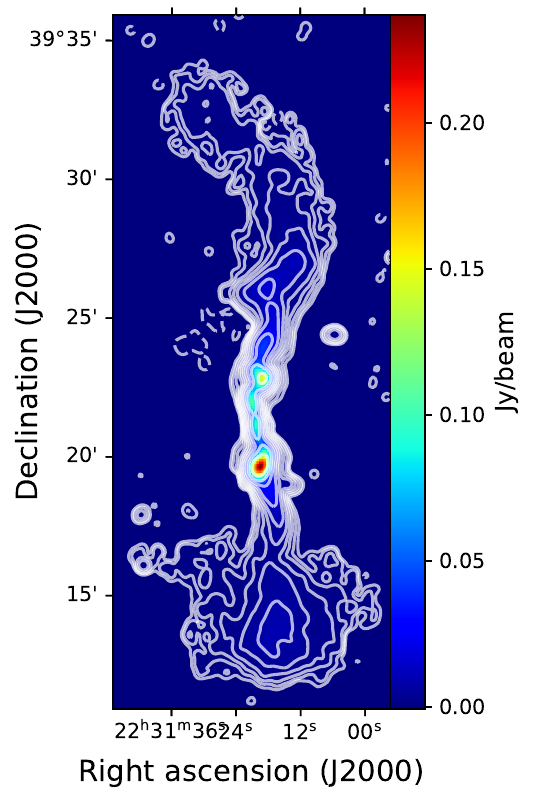}\par
   \includegraphics[width=0.85\linewidth]{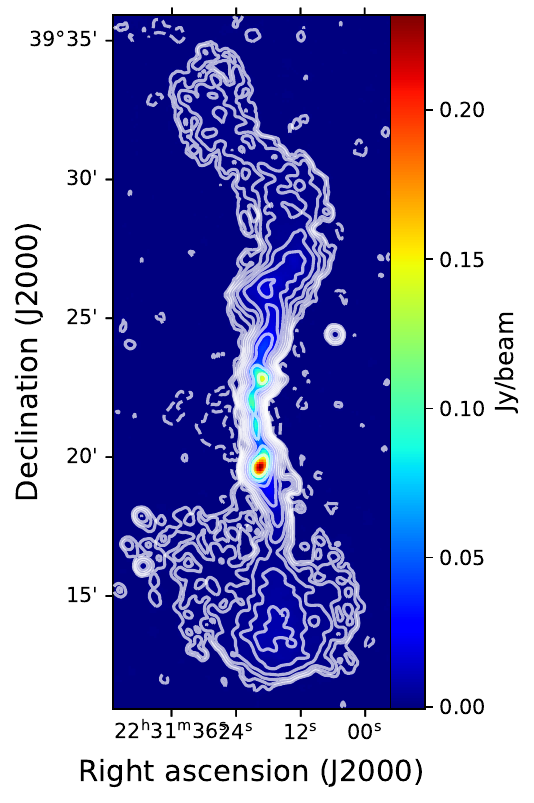}\par
   
\end{multicols}
  \vspace{-0.5cm}
  \caption{VLA maps reanalysed in this work. Upper panels: maps with angular resolution of  $6.0\arcsec \times 6.0\arcsec$. From left to right: 
  i) 1365 MHz VLA map 
  %at $6.0\arcsec \times 6.0\arcsec$ 
  with contour levels [-0.4\%, 0.4\%, 0.8\%, 1.6\%, 3.2\%, 6.4\%, 12.8\%, 25.6\%, 51,2\%] $\times$ the peak of $0.035 \, \mathrm{Jy/beam}$; 
  ii) 4985 MHz VLA map 
  %at $6.0\arcsec \times 6.0\arcsec$ 
  with contour levels [-0.4\%, 0.4\%, 0.8\%, 1.6\%, 3.2\%, 6.4\%, 12.8\%, 25.6\%, 51,2\%] $\times$ the peak of $0.034 \, \mathrm{Jy/beam}$; 
  iii) 8485 MHz VLA map 
  %at $6.0\arcsec \times 6.0\arcsec$ 
  with contour levels [-0.4\%, 0.4\%, 0.8\%, 1.6\%, 3.2\%, 6.4\%, 12.8\%, 25.6\%, 51,2\%] $\times$ the peak of $0.041 \, \mathrm{Jy/beam}$. 
  Lower panels: maps with angular resolution of $20.0\arcsec \times 20.0\arcsec$. From left to right: 
  i) 1365 MHz VLA map 
  %at $20.0\arcsec \times 20.0\arcsec$ 
  with contour levels [-0.1\%, 0.1\%, 0.2\%, 0.4\%, 0.8\%, 1.6\%, 3.2\%, 6.4\%, 12.8\%, 25.6\%, 51,2\%] $\times$ the peak of $0.239 \, \mathrm{Jy/beam}$; 
  ii) 1485 MHz VLA map 
  %at $20.0\arcsec \times 20.0\arcsec$ 
  with contour levels [-0.1\%, 0.1\%, 0.2\%, 0.4\%, 0.8\%, 1.6\%, 3.2\%, 6.4\%, 12.8\%, 25.6\%, 51.2\%] $\times$ the peak of $0.23 \, \mathrm{Jy/beam}$.} 
  \label{fig:VLA_maps}
\end{figure*}

We show in Fig.~\ref{fig:VLA_maps} the VLA archival maps at 1365, 1485, 4985, and 8485 MHz re-analysed in this work.

%----------------------------------------------------------
\section{Spectral ageing maps with KP's and Tribble's models} \label{app:tribble}
%----------------------------------------------------------

\begin{figure*}
\begin{multicols}{2}
   \includegraphics[width=\linewidth]{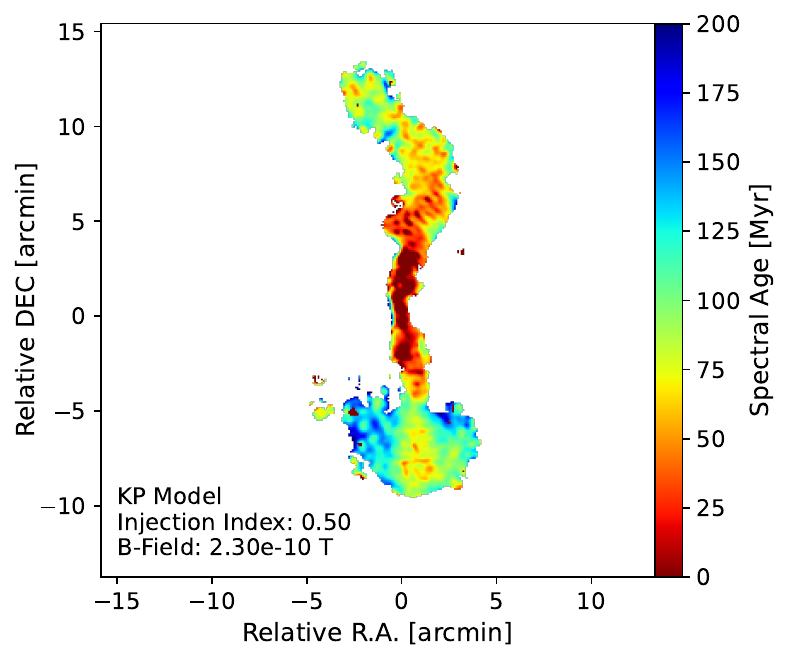} \par
   \includegraphics[width=\linewidth]{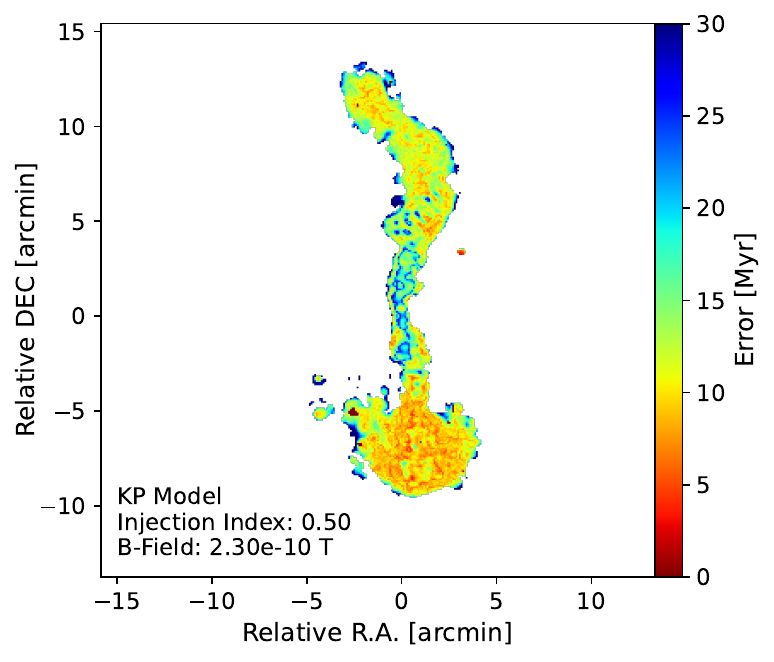}
\end{multicols}
\begin{multicols}{2}
   \includegraphics[width=\linewidth]{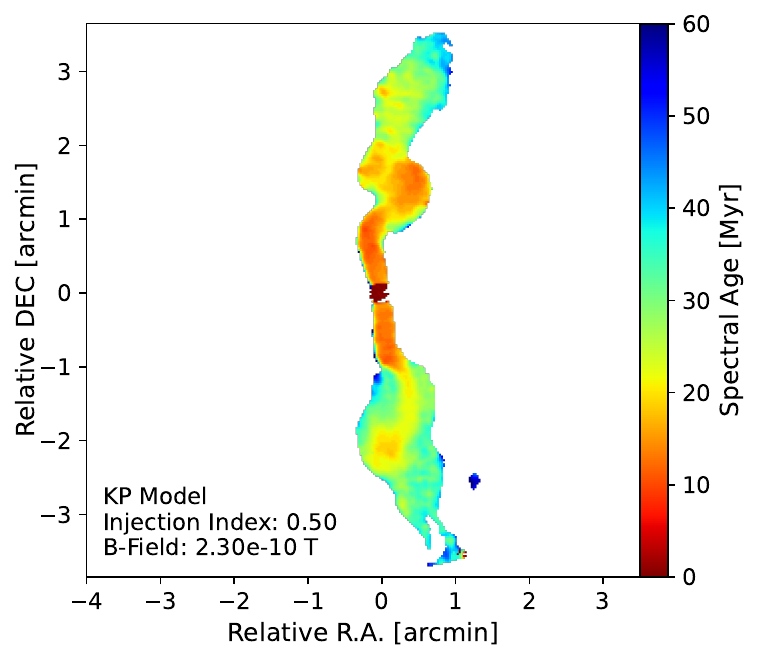} \par
   \includegraphics[width=\linewidth]{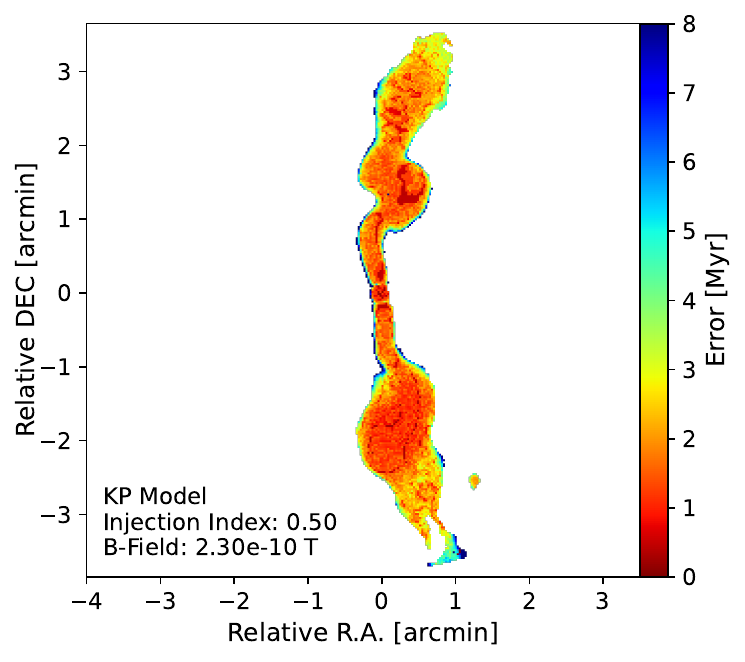}
\end{multicols}
  \caption{Spectral ageing maps obtained by fitting KP's model to the intensity maps, and corresponding maps of the uncertainty.  
 Upper, left panel: map of the spectral age of 3C 449 between 145, 1365, and 1485 MHz, with the angular resolution of $20.0\arcsec \times 20.0\arcsec$, by means of the BRATS software package. The parameters of the model are reported in the legend. Overall, the spectral age increases with the distance to the radio core. Upper, right panel: map of the upper error on the spectral age shown in the left panel. Lower, left panel: map of the spectral age of 3C 449 between 145, 1365, 4985, and 8485 MHz, with angular resolution of $6.0\arcsec \times 6.0\arcsec$. Lower, right panel: map of the upper error on the spectral age shown in the left panel. In all the maps, the magnetic field intensity is  $B=B_{\rm eq}=2.3\, \mu$G.}
  \label{fig:sa_KP}
\end{figure*}

\begin{figure*}
\begin{multicols}{2}
   \includegraphics[width=\linewidth]{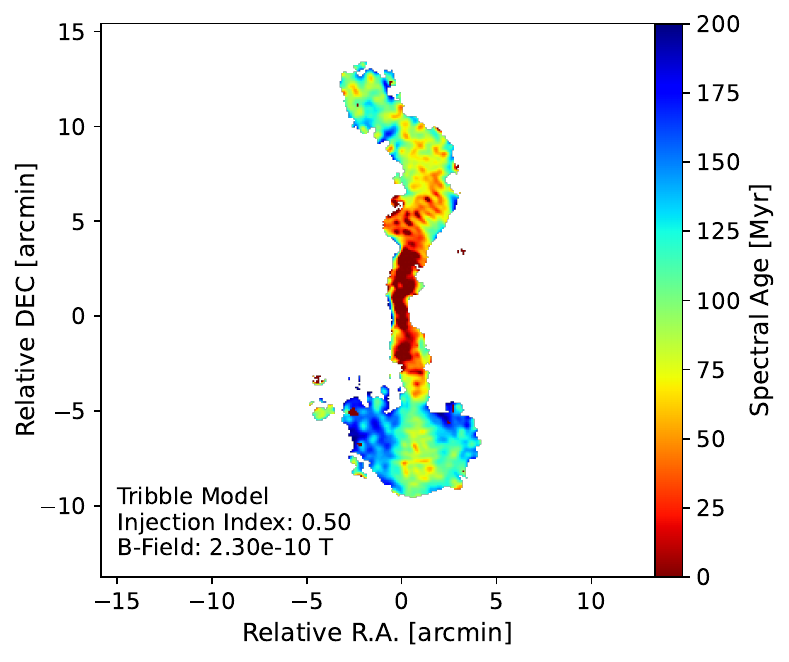} \par
   \includegraphics[width=\linewidth]{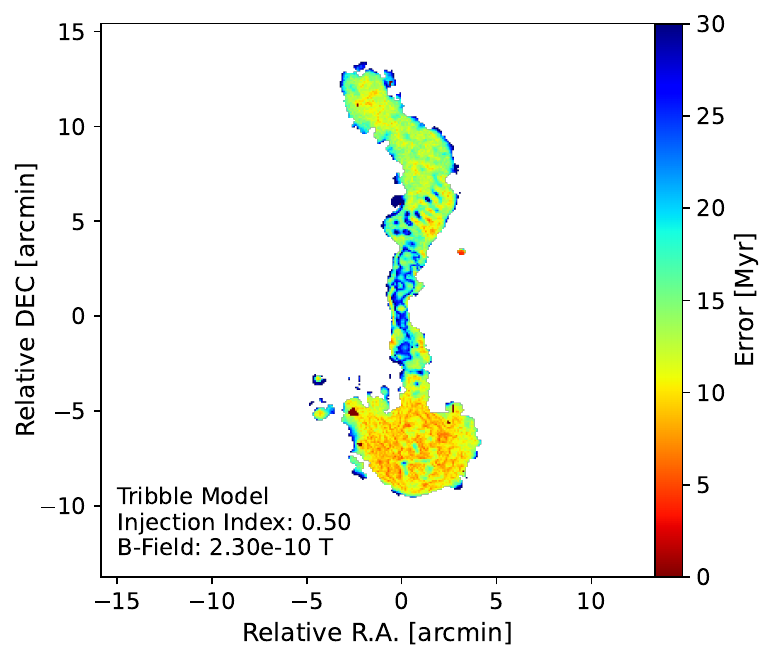}
\end{multicols}
\begin{multicols}{2}
   \includegraphics[width=\linewidth]{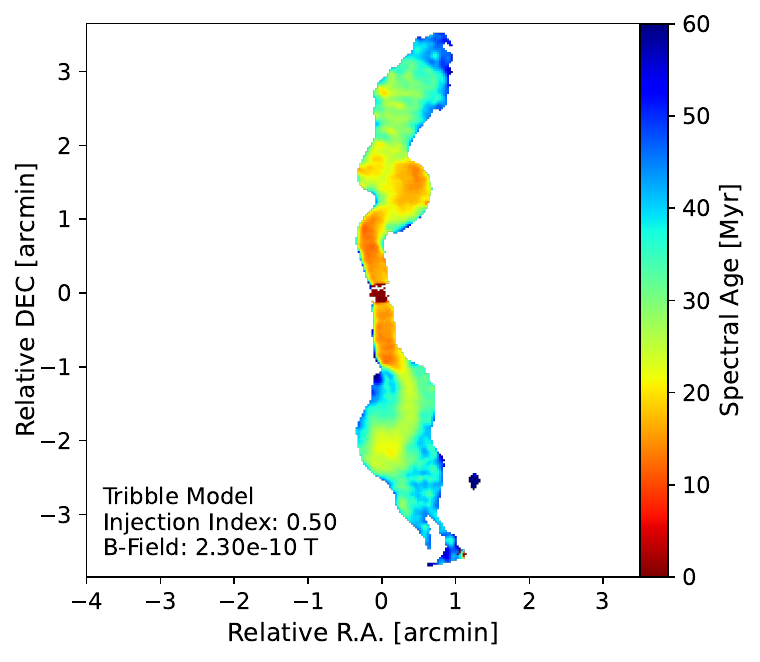} \par
   \includegraphics[width=\linewidth]{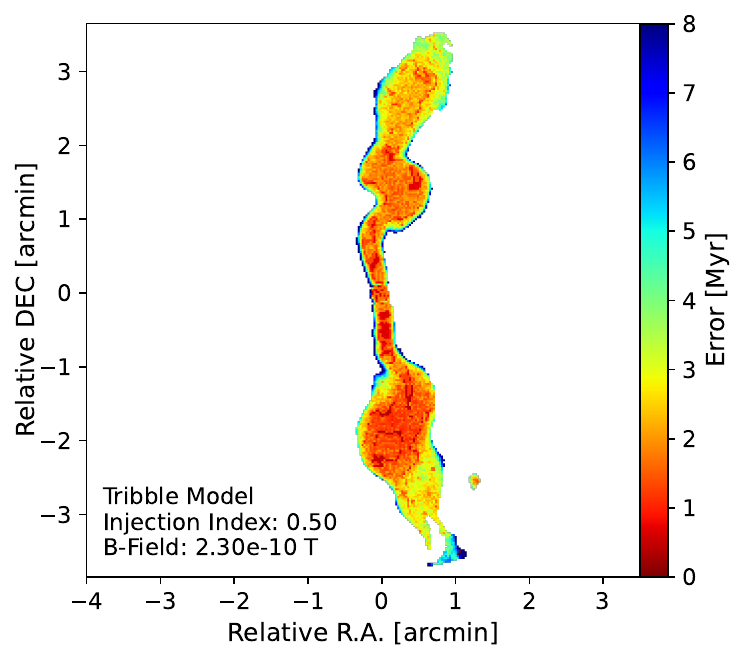}
\end{multicols}
  \caption{Spectral ageing maps obtained by fitting Tribble's model to the intensity maps, and corresponding maps of the uncertainty.  
 Upper, left panel: map of the spectral age of 3C 449 between 145, 1365, and 1485 MHz, with the angular resolution of $20.0\arcsec \times 20.0\arcsec$, by means of the BRATS software package. The parameters of the model are reported in the legend. Overall, the spectral age increases with the distance to the radio core. Upper, right panel: map of the upper error on the spectral age shown in the left panel. Lower, left panel: map of the spectral age of 3C 449 between 145, 1365, 4985, and 8485 MHz, with angular resolution of $6.0\arcsec \times 6.0\arcsec$. Lower, right panel: map of the upper error on the spectral age shown in the left panel. In all the maps, the magnetic field intensity is  $B=B_{\rm eq}=2.3\, \mu$G.}
  \label{fig:sa_Tribble}
\end{figure*}

We show the spectral ageing maps that we obtained by fitting KP's (Fig.~\ref{fig:sa_KP}) and Tribble's (Fig.~\ref{fig:sa_Tribble}) models, as implemented in BRATS, to the intensity maps of 3C~449.
The left panel shows the 145-1485 MHz spectral age map at the angular resolution of $20.0\arcsec \times 20.0\arcsec$;  the right panel shows the 145-8485 MHz spectral age map at the angular resolution of $6.0\arcsec \times 6.0\arcsec$.
The spectral evolution inferred with both KP's and Tribble's model is comparable to that inferred from JP's model within 1$\sigma$ (see Sect.~\ref{sec:spectral_age}, Fig.~\ref{fig:BRATS_maps}).

%----------------------------------------------------------
\section{Spectral ageing maps in a scenario of magnetic dominance}
\label{app:12muG}

\begin{figure*}
\begin{multicols}{2}
   \includegraphics[width=\linewidth]{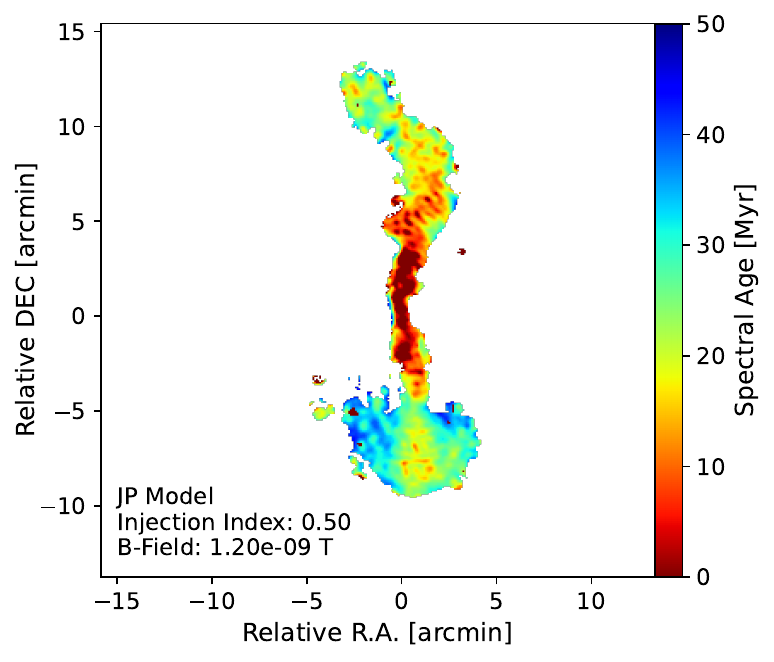} \par
   \includegraphics[width=\linewidth]{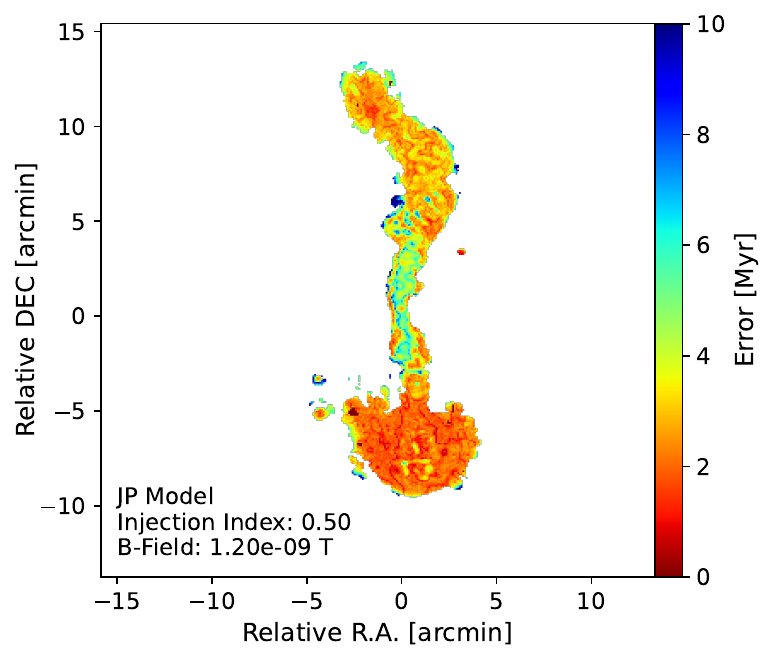}
\end{multicols}
\begin{multicols}{2}
   \includegraphics[width=\linewidth]{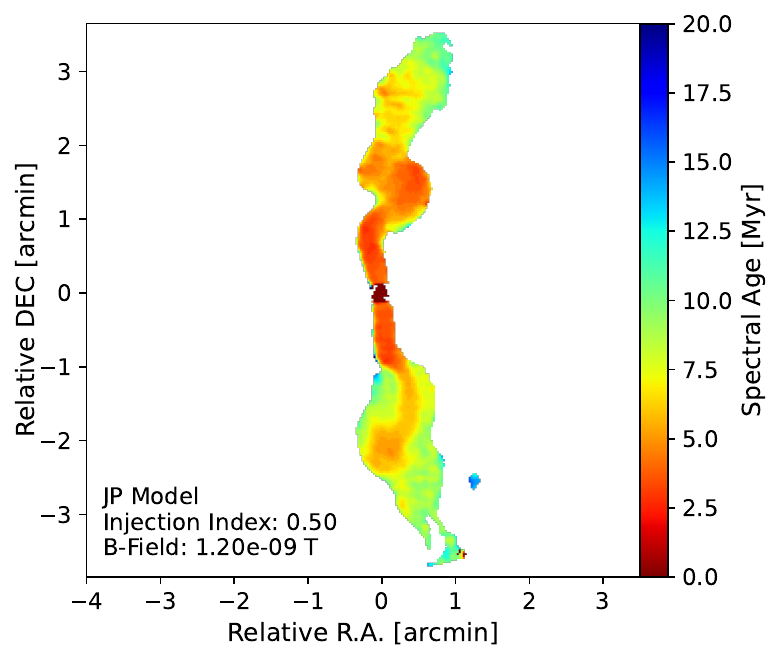} \par
   \includegraphics[width=\linewidth]{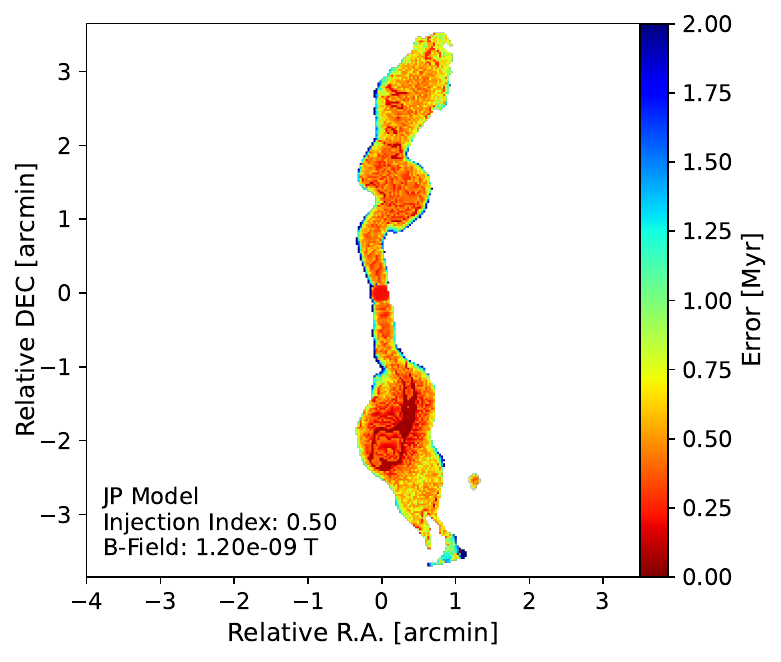}
\end{multicols}
  \caption{Spectral ageing maps obtained by fitting JP's model to the intensity maps, and corresponding maps of the uncertainty, using a magnetic field of $B = 12 \mu \mathrm{G}$.
  Upper, left panel: map of the spectral age of 3C~449 between 145, 1365, and 1485 MHz, with the angular resolution of $20.0\arcsec \times 20.0\arcsec$. Upper, right panel: map of the upper error on the spectral age shown in the left panel.
  Lower, left panel: map of the spectral age of 3C~449 between 145, 1365, 4985, and 8485 MHz, with angular resolution of $6.0\arcsec \times 6.0\arcsec$. Lower, right panel: map of the upper error on the spectral age shown in the left panel.}
  \label{fig:sa_10}
\end{figure*}

We show in Fig.~\ref{fig:sa_10} the spectral ageing maps of 3C~449 obtained by fitting JP's model, as implemented in BRATS, to the intensity maps of 3C~449, by assuming magnetic dominance for the source, with  $B = 12 \, \mu$G  \citep{Croston_2003}. 
%----------------------------------------------------------

\end{appendix}

\end{document}